\documentclass[12pt,a4paper]{article}

\usepackage[utf8]{inputenc}
\usepackage{graphicx}
\usepackage{amsmath}
\usepackage{amssymb}
\usepackage{geometry}
\usepackage{authblk}
\usepackage[colorlinks=true, linkcolor=blue, citecolor=blue, urlcolor=blue]{hyperref}

\usepackage{setspace}
\usepackage{lineno}

\usepackage{natbib}
\newcommand{\apj}{ApJ}

\title{Inside the Eirenesphere: The Interplay of Porosity, Heat Flux and Mineralogy in Exoplanetary Aquable and Habitable Layers}

\author{Santiago A. Orjuela\thanks{Corresponding author: santiagoa.orjuela@udea.edu.co, ORCID: 0009-0005-6902-1939}}

\author{Jorge I. Zuluaga\thanks{jorge.zuluaga@udea.edu.co, ORCID: 0000-0002-6140-3116}}

\affil{SEAP/FACom, Instituto de F\'isica - FCEN, Universidad de Antioquia, Calle 70 No. 52-21, Medellin, Colombia}

\date{}

\begin{document}
\maketitle

\begin{abstract}
The classical circumstellar habitable zone restricts the search for life to planetary surfaces where stellar irradiation sustains liquid water, overlooking vast subsurface environments. 
Here, we present a self-consistent geophysical model that integrates internal radial structure, a parametric mineralogical model, radiogenic heat production, and pressure-dependent crustal porosity to quantify the three-dimensional habitable volume—the eirenesphere—of rocky exoplanets. 
We distinguish between subsurface aquability (thermodynamic water stability) and subsurface habitability, which strictly requires temperatures and pressures within extremophile biological limits, alongside sufficient porosity for fluid circulation. 
To compare planetary capacities, we introduce the Eirenesphere Volumetric Index (EVI), measuring the average subsurface biosphere volume across the habitable zone. Applying this framework, we find that Earth's current state (EVI~$\approx 0.06$ terrestrial oceans) represents only a moderate regime. Instead, mature super-Earths with high geothermal activity provide the most volumetrically extensive environments for deep biospheres. Furthermore, crustal mineralogy exerts a first-order control: thermally insulating felsic crusts sustain significantly larger eirenespheres than primary mafic lithologies. 
Tracking secular cooling reveals that internal habitability is an evolutionary property; young planets host confined biospheres due to steep thermal gradients, whereas mature worlds maximize their habitable volumes over billions of years. 
Finally, we demonstrate that around solar like-stars subsurface habitability persists out to 5--7 au, effectively decoupling life's potential from the surface radiative balance. Extrapolating these findings to the Milky Way reveals a staggering galactic capacity for subsurface life, on the order of billions of terrestrial oceans.
Ultimately, this volumetric framework shifts the astrobiological paradigm from a surface-dependent phenomenon to an intrinsic planetary property, offering a quantifiable metric to prioritize targets.
\end{abstract}

\vspace{0.5cm}

\section{Introduction}

In recent decades, exoplanetary research has shifted from a phase of statistical discovery to an era of exhaustive characterization, allowing us to access physical properties that were previously out of reach. A little over thirty years after the first detections, the astronomical inventory has recently surpassed the astonishing figure of 6,000 confirmed exoplanets. Today, we have precise measurements of mass, radius, and mean density for a growing population of these rocky exoplanets, as well as a detailed understanding of their dynamical architectures and orbital resonances (see e.g., \citealt{Batalha2014, Grimm2018}). This observational progress has been accompanied by unprecedented sophistication in internal structure and thermal evolution models (see e.g., \citealt{Noack2017, Zuluaga2013}). The integration of these geophysical frameworks allows us, for the first time, to rigorously predict the thermodynamic conditions and heat flow in the interiors of distant worlds. This predictive capability, which incorporates secular cooling and radiogenic decay, offers a window into exoplanetary geodynamics that would have been unthinkable barely a decade ago, laying the groundwork for assessing habitability from a three-dimensional, evolutionary perspective.

Against this backdrop, one of the primary objectives of contemporary astrobiology is the detection of environments that could support life. Traditionally, this problem has been approached using the concept of the habitable zone (HZ), defined as the region around a star where a rocky planet can maintain liquid water on its surface over geological timescales (see, e.g., \citealt{Kasting1993,Kopparapu2013,Kopparapu2014}). This approach, which is deeply linked to the surface radiative balance and to simplified atmospheric models, has proven to be a useful tool for classifying exoplanets on the basis of stellar and orbital observables. 

However, the exclusive emphasis on surface conditions imposes an important conceptual limitation: it assumes that habitability is necessarily coupled to direct stellar irradiation. Theoretical and observational studies in the Solar System have shown that shielded subsurface environments can harbor liquid water and energy sources sufficient to sustain biological processes, even when surface conditions are hostile (see e.g., \citealt{Chyba2000, Hand2009, Orcutt2013, tenelanda2021enceladus}). These scenarios motivate an extension of the classical habitability framework into the interiors of planetary bodies. 

In recent years, various studies have explored qualitatively and quantitatively how the internal conditions of rocky exoplanets could broaden the traditional concept of the habitable zone. For example, \citet{McMahon2013} carried out studies in which subsurface habitability (SSHZ, for its initials in English) was explicitly incorporated into the framework of the circumstellar habitable zone by combining geothermal gradients with assumed maximum depths, thus estimating how far the deep biosphere of Earth (and that of other bodies such as Mars) could extend beneath the surface. In addition, \citet{Noack2017} showed that the internal habitability of super-Earths is strongly controlled by their tectonic regime and internal heat flow. Complementing these efforts, \citet{Lingam2018} carried out a detailed analytical study of worlds with subsurface oceans beneath thick ice layers, demonstrating that radiogenic heating alone can keep water in a liquid state and drive the production of prebiotic compounds without any dependence on stellar radiation. Their finding that such internally habitable worlds could outnumber rocky planets in the classical habitable zone by several orders of magnitude highlights the pressing need for volumetric measures of habitability. Moreover, from a geomicrobiological perspective, \citet{Escudero2023} stressed that the hard-rock deep subsurface on Earth harbors an extensive ``dark biosphere'' supported by water--rock interactions, strongly arguing that the concept of planetary habitability should be broadened to explicitly encompass analogous environments in rocky exoplanets and on Mars.

Alternative geodynamical perspectives have also investigated the mechanisms controlling this interior water availability---the fundamental prerequisite for life-supporting environments inside rocky planets. Studies of terrestrial porous media indicate that the exponential decrease of porosity with increasing lithostatic pressure imposes a crucial mechanical limit on the overall volume of deep aquifer systems \citep{Parnell2016}. Concurrently, thermomechanical simulations have shown that differences in geodynamical regimes---especially in planets exhibiting a stagnant-lid configuration---lead to geothermal gradients, heat transport patterns, and degassing behaviors that diverge significantly from those on Earth (see e.g., \citealt{Noack2017, Foley2018, Plesa2018, Cockell2014}). 

While these prior studies provide essential astrobiological and geophysical groundwork, none of them concurrently treat the mechanical evolution of crustal porosity under lithostatic pressure and a detailed mineralogical parameterization as primary variables. It is precisely this combined thermomechanical and compositional treatment that distinguishes the volumetric framework presented here from earlier work. To this end, building upon the physical distinctions noted by previous authors regarding water stability versus biological viability \citep{Lingam2018}, we formally introduce the concept of \textit{subsurface aquability} (SA). We define SA as the thermodynamic state that permits stable subsurface liquid water irrespective of biological constraints, and we explicitly contrast it with \textit{subsurface habitability} (SH), which additionally requires that pressure, temperature, and porosity fall within the known tolerance limits of extremophilic organisms. From this planet-centered standpoint, we reintroduce the notion of the \textit{eirenesphere} (ES), first proposed by Mendez in 2011 and derived from his earlier investigations into spatial distribution of life on planetary bodies \citep{mendez2001planetary}\footnote{The term eirenesphere was introduced by Abel Méndez to designate the region inside a planetary body where temperature and pressure could, in principle, sustain microbial life. Taking its name from Eirene, the Greek goddess of peace, the concept describes a thermodynamic sanctuary buffered from both harsh surface conditions and extreme environments in the deep interior. Méndez drew inspiration from his colleague Irene Schneider; in a conversation about Venus’s habitable zones, she asked whether analogous regions could be determined for other planets such as Jupiter (A. Méndez, personal communication). For further information, consult the Planetary Habitability Laboratory (PHL). at \url{https://phl.upr.edu/projects/eirenespheres}.}, to describe the three-dimensional volume within a rocky exoplanet where these internal conditions are met. Moreover, to quantify the capacity of a given planet to maintain such a deep biosphere, we introduce the VIBE, or \textit{Volumetric Index of Biospheric Extent}, as a measure of the average eirenesphere integrated across the stellar habitable zone.

While the existence of life in the deep subsurface might not directly influence the prioritization of near-term observational targets---given that such biospheres would scarcely generate atmospheric biosignatures detectable across interstellar distances---the quantification of the fraction of planetary volume capable of sustaining life is of fundamental astrobiological interest \citep{Lingam2018}. This approach transcends the search for candidates for remote characterization and becomes a question about the intrinsic habitability of the universe as a whole: an assessment of the true abundance of biological niches that can persist in environments independent of the photosynthetic habitable zone (see e.g., \citealt{Vance2018, Atri2025, lloyd2025intraterrestrials}). In this context, the study of processes such as subsurface radiolysis suggests that the habitable zone could extend significantly beyond the traditional surface thermal limits (see e.g., \citealt{Atri2025, Spohn2026}). In this work, we synthesize these insights into a unified framework that couples geology, internal structure, and thermal evolution, enabling us to rigorously evaluate the spatial extent of these potentially eirenespheres within rocky planets.


This work is organized as follows: \autoref{sec:sshz} introduces the conceptual framework of subsurface habitability, exploring the thermodynamic stability of liquid water in the planetary interior and distinguishing between aquable environments and biologically habitable ones. \autoref{sec:model} describes the internal structure and thermal evolution model adopted for rocky planets, including the sources of radiogenic heat, scaling relations, and the conductive thermal regime. Section 4 details the numerical implementation used to self-consistently solve the coupled physical profiles. In \autoref{sec:criteria}, we define the physical and biological boundaries used to identify habitable regions, which leads to the formal definition of the eirenesphere in \autoref{sec:eirenesphere}. The formulation of the Eirenesphere Volumetric Index (EVI) is developed in \autoref{sec:index}. The main results are presented in \autoref{sec:results}, covering the thermal structure of rocky planets, the geometry of subsurface eirenespheres, the impact of mineralogical diversity, physical regimes of habitability, and the dependence of the EVI on planetary mass and internal heat flux. \autoref{sec:discusion} analyzes the astrobiological implications, metabolic energy constraints, model limitations, and consequences for exoplanetary life detection. Finally, the main conclusions of this work are summarized in \autoref{sec:conclusions}.

\section{Subsurface aquability and habitability}\label{sec:sshz}

The classical concept of habitability has historically been linked to the presence of liquid water on a planet’s surface, regulated primarily by stellar irradiation. However, studies within both the Solar System and the exoplanetary context have revealed that life could persist in shielded environments beneath the surface, even in the absence of benign surface conditions (see e.g. \citealt{Chyba2000, SchulzeMakuch2008, Noack2017, Lingam2018, Escudero2023}).

Subsurface habitability (SH) refers to the possibility that habitable niches may exist within the interiors of planetary bodies\footnote{Hereafter, the term \textit{planet} designates any body with a differentiated interior, a mineralogically complex lithosphere, and active geological processes, regardless of whether it orbits a star or another planet. This geophysical definition, focused on the internal complexity of the body rather than on its orbital dynamics, follows the criterion proposed by \citet{Metzger2022}.}, where the environment is partially or completely isolated from outer space. In these scenarios, the energy required to sustain biological processes does not come directly from stellar radiation, but from internal sources such as radiogenic heat, tidal dissipation, or geochemical reactions (see e.g. \citealt{Hand2009,Orcutt2013}).

Paradigmatic examples within the Solar System include the proposed subsurface oceans on Europa and Enceladus, whose existence is inferred from geophysical and geochemical observations (see e.g. \citealt{Kivelson2000,Sotin2004,Porco2006,Nimmo2009}). These cases show that habitability is not restricted to the classical circumstellar habitable zone, but can extend to much broader regions of planetary space.

\subsection{Thermodynamic stability of liquid water}

A central requirement for habitability, both at the surface and in the subsurface, is the presence of liquid water as a biochemical solvent. Although alternative solvents—such as ammonia, methanol, or hydrogen sulfide—have been considered in the literature as possible media for the chemistry of life (see e.g. \citealt{Bains2004, Stevenson2015}), in this work we restrict ourselves to the case of liquid water, given its relevance as the only known biological example and its central role in universal biochemistry (see e.g. \citealt{Cockell2016}).

In subsurface environments, the stability of liquid water depends mainly on the pressure and temperature conditions imposed by the internal structure of the planetary body. Unlike the surface case—where the atmosphere guarantees a pressure higher than that of the triple point ($P_{\rm tp} = 611.7\,\mathrm{Pa}$, $T_{\rm tp} = 273.16\,\mathrm{K}$) and the radiative balance regulates the temperature within the liquid range—in the planet’s interior the lithostatic pressure increases with depth at rates much higher than those of the atmospheric column, due to the high density of rocky materials.

This increasing pressure expands the stability domain of liquid water in two complementary ways. On the one hand, it shifts the boiling point toward significantly higher temperatures. On the other, and in a less intuitive way, it shifts the melting point of ordinary ice (phase Ih) toward slightly lower temperatures, which can favor the presence of liquid water even at moderate depths on planets with cold crusts (see e.g. \citealt{IAPWS1995, IAPWSRelease2009, IAPWSSeawater2010}). This effect is particularly relevant in thick crusts or beneath ice layers, where mechanical confinement favors the formation of subsurface oceans (see e.g. \citealt{Nimmo2016, Lingam2018}).

\begin{figure*}
    \centering
    \includegraphics[width=0.7\textwidth]{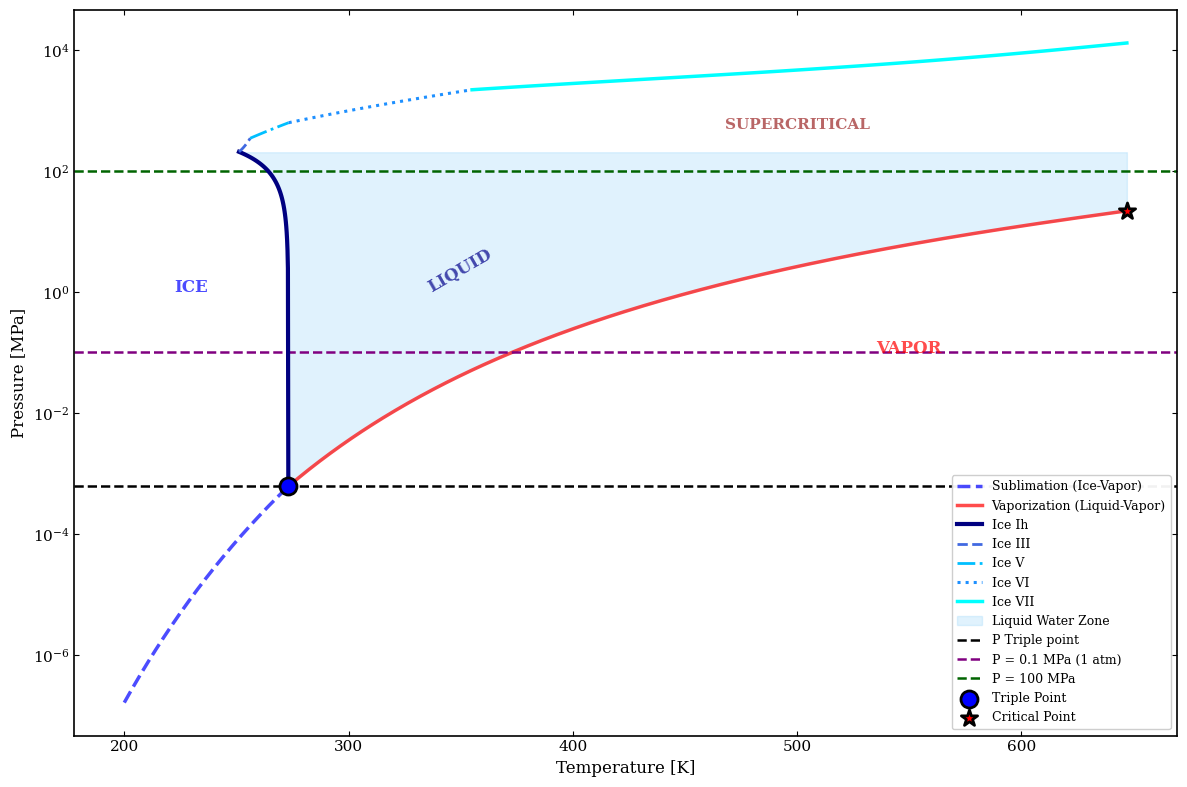}
    \caption{Phase diagram of pure water in pressure–temperature space, showing the main boundaries between ice, liquid, vapor, and supercritical fluid. The shaded blue area marks where liquid water is thermodynamically stable under conditions relevant to rocky planetary interiors. Dashed horizontal lines mark key pressures in astrobiology and planetary geophysics: Earth’s atmospheric pressure (0.1 MPa), the approximate upper biological pressure limit used here (100 MPa), and the water triple-point pressure. The blue circle and red star denote the water triple point and critical point, respectively.}
    \label{fig:phase_water_diagram}
\end{figure*}

The phase diagram of pure water, shown in \autoref{fig:phase_water_diagram}, explicitly illustrates these conditions. The shaded region delimits the stability domain of liquid water in pressure–temperature space, bounded below by the vaporization curve (liquid–vapor) and above by the melting curves corresponding to the different polymorphs of ice (Ih, III, V, VI, and VII). The triple point ($P_{\rm tp}$, $T_{\rm tp}$) marks the thermodynamic minimum below which liquid water cannot exist under any conditions. The critical point ($P_c \approx 22.06\,\mathrm{MPa}$, $T_c \approx 647\,\mathrm{K}$) defines the upper limit beyond which water enters a supercritical state in which the liquid and gaseous phases are thermodynamically indistinguishable \citep{IAPWS1995, Wagner2002}.

The reference lines at $P = 0.1\,\mathrm{MPa}$ ($\approx 1\,\rm{atm}$) and $P = 100\,\mathrm{MPa}$ illustrate, respectively, typical terrestrial surface conditions and the pressures attainable at depths of several kilometers in the crust of a rocky planet (see e.g. \citealt{Wagner2002, Choukroun2010}). In the latter case, the temperature range compatible with liquid water extends from approximately $273\,\mathrm{K}$ to more than $580\,\mathrm{K}$, greatly expanding the thermodynamic space for habitability relative to standard surface conditions. This expansion of the liquid domain with pressure is precisely what makes the planetary interior a potentially habitable environment, even when the planet’s surface is cold, arid, or lacking an atmosphere.

\subsection{Aquability versus habitability}
It is important to make a conceptual distinction between \emph{aquability} and \emph{habitability} in the context of planetary interiors. We introduce the term \emph{aquability} to denote the local thermodynamic conditions—pressure and temperature—that allow the stable existence of liquid water, irrespective of whether additional conditions required for life are met. Aquability is therefore a necessary but not sufficient condition for habitability: a planetary volume can be aquable without being habitable if, for example, the pressure exceeds biological tolerance limits, or if chemical activity is insufficient to sustain metabolisms (see \autoref{sec:habitable_limits} below). This distinction is analogous to that made by \citet{Lingam2018} in separating the conditions for the stability of liquid water from the energetic and geochemical requirements for life on frozen planets. The notion of aquability generalizes and decouples the thermodynamic analysis of water stability from strict biological criteria, and depends explicitly on the internal structure and thermal evolution of the planet (see e.g.\ \citealt{McMahon2013, Unterborn2016, Lingam2018, Riu2022}).

\bigskip

The calculation of habitability conditions requires knowing the pressure and temperature profiles within the planetary lithosphere. In the following section, we present the geophysical model adopted to obtain these profiles from the bulk properties of the planet.

\section{Internal structure model and thermal regime}\label{sec:model}

\subsection{General physical framework}

The thermal regime of the interior of a rocky planet is controlled by the combination of several internal energy sources, including radiogenic heat produced by the decay of long-lived isotopes, primordial heat remaining from planetary accretion and differentiation, and latent heat associated with core crystallization during the early stages of planetary evolution (see e.g. \citealt{Stacey2008, Turcotte2014}). On long geological timescales, relevant for subsurface habitability, radiogenic heat usually dominates the energy budget of the lithospheric interior (see e.g. \citealt{McDonough2020, Ruedas2017}).

Thermal transport in the planetary interior can be described by the general heat diffusion equation,
\begin{equation}
\nabla \cdot \left( k \nabla T \right) + A = \rho c_p \frac{\partial T}{\partial t},
\label{eq:heat_equation}
\end{equation}
where $k$ is the thermal conductivity, $A$ the volumetric internal heat production rate (with units of J s$^{-1}$ m$^{-3}$), $\rho$ the density, and $c_p$ the specific heat at constant pressure. In a quasi-steady state regime—appropriate for the long-term thermal evolution of the lithosphere—the time-dependent term on the right-hand side can be neglected, reducing the equation to a local balance between internal heat production and its transport by conduction. It should be noted that, in the general case, both $A$ and $k$ depend on the density and local temperature of the material.

Solving this equation requires knowing the density, pressure, and temperature profiles throughout the planetary interior, which are mutually coupled. Under the assumption of spherical symmetry and hydrostatic equilibrium, these profiles are obtained from the following system of equations:
\begin{equation}
\frac{dm(r)}{dr} = 4\pi r^2 \rho(r),
\label{eq:masa}
\end{equation}
\begin{equation}
\frac{dP(r)}{dr} = -g(r)\,\rho(r),
\label{eq:hidrostatico}
\end{equation}
where the local gravitational acceleration is obtained from the enclosed mass:
\begin{equation}
g(r) = \frac{G\,m(r)}{r^2}.
\label{eq:gravedad}
\end{equation}
To close the system, the relationship between density and pressure in the deep interior is given by the Adams–Williamson equation,
\begin{equation}
\frac{d\rho(r)}{dr} = -\frac{\rho(r)\,g(r)}{K_s(r)},
\label{eq:adams-williamson}
\end{equation}
where $K_s(r) \equiv -V(\partial P/\partial V)_S$ is the adiabatic bulk modulus. This equation states that, in the absence of compositional changes or phase transitions, the density gradient is controlled by the compressibility of the material and the local gravity. To model planets with compositions different from Earth, this equation is complemented with Birch–Murnaghan or Vinet equations of state for each relevant mineralogical phase (see e.g. \citealt{Sotin2007, Seager2007, Dorn2015}).

The self-consistent solution of \autoref{eq:masa} and \autoref{eq:adams-williamson} provides the radial profiles of density, pressure, and gravity that constitute the structural basis of the thermal model. However, for this model to be predictive, it is necessary to quantify the energy sources that drive heat flow in the planetary interior. The next subsection describes the internal heat sources considered in this work and the parametrization adopted to model their distribution with depth.

\subsection{Sources of Internal Heat}

The internal heat of a rocky planet has three fundamental sources: the decay of long-lived radioactive isotopes, the residual energy accumulated during the processes of planetary accretion and differentiation, and tidal dissipation in gravitationally active systems (see e.g. \citealt{Stacey2008, Turcotte2014, Nimmo2016}).

The relative importance of these sources varies with the age, mass, and dynamical history of the planet. In young worlds, and assuming that the planet is sufficiently distant from its star and from other massive bodies, the dominant source is accretionary heat; whereas in mature planets radiogenic heat and secular heat contribute in comparable proportions (see e.g. \citealt{Jaupart2007, McDonough2020}). In the absence of direct stellar irradiation, these endogenous sources constitute the only mechanism capable of maintaining temperatures compatible with liquid water in the planet’s interior, which makes them the determining factor for subsurface habitability: a planet with a frozen surface or lacking an atmosphere can harbor a significant internal volume where habitable conditions are stably satisfied over geological timescales, provided that the internal heat flux is sufficient (see e.g. \citealt{Artemieva2006, Hasterok2011, Lingam2018}). It should be noted that, although tidal heating can be a determining factor in exomoons and planets in orbital resonance, its effect will not be considered in this work; an analysis of habitability under a significant source of tidal energy is left for future studies.

Among the sources mentioned, radiogenic heating is the one that can be modeled with the greatest precision from the compositional properties of the planet, and the one that dominates the energy budget in mature planets such as Earth \citep{Ruedas2017, McDonough2020}. This heat arises mainly from the decay of four long-lived isotopes: $^{238}$U, $^{235}$U, $^{232}$Th, and $^{40}$K. Since these elements are lithophile and incompatible, they tend to fractionate preferentially into the melt phase during partial melting and magmatic differentiation, becoming concentrated in the upper crust and upper lithospheric mantle \citep{Artemieva2006, Hasterok2011}. The volumetric rate of radiogenic heat production can be expressed as:
\begin{equation}
A = \sum_i \rho\, X_i\, \lambda_i\, E_i,
\label{eq:radiogenic}
\end{equation}
where $X_i$ is the mass abundance of isotope $i$, $\lambda_i$ its radioactive decay constant, and $E_i$ the thermal energy released per decay.

To parameterize the vertical distribution of radiogenic production in the lithosphere, we adopt the exponential model of \citet{Turcotte2014}:
\begin{equation}
A(z) = A_0\,\exp\!\left(-\frac{z}{h_r}\right),
\label{eq:A_exponencial}
\end{equation}
where $z$ is depth, $h_r$ is the radiogenic decay scale —the characteristic depth at which heat production is reduced by a factor of $e$— and $A_0$ is the surface production.

This model was originally introduced to explain the empirically observed linear relationship between surface heat flow and heat production in granitic plutons \citep{Turcotte2014}, and it has a clear physical interpretation: it reflects the upward migration of incompatible elements during crust formation, which generates an exponentially decreasing radiogenic enrichment with depth.

Instead of assigning $A_0$ an independent constant value, we adopt the thermal partitioning approach of HC2011, who, using a layered model as an observational reference, determine that approximately 40\,\% of the total surface heat flow $q_s$—energy released per unit area per unit time—is generated internally within the first $h_r$ of the upper crust. This allows us to express $A_0$ directly in terms of $q_s$:
\begin{equation}
A_0 = \frac{(1 - F)\,q_s}{h_r},
\label{eq:A0}
\end{equation}
where $F \approx 0.6$ is the fraction of the surface heat flow that comes from the mantle and lower crust, and $h_r \sim 10$--$15\,\mathrm{km}$ for the Earth (see e.g. \citealt{Hasterok2011, Turcotte2014}). The decay scale $h_r$ is therefore the parameter that encapsulates all the information about how radiogenic heat production is distributed with depth. Its value depends, in general, on the composition and differentiation history of the planet. In the following subsection we discuss how this parameter, together with other bulk properties of the planet, can be scaled to exoplanets of different mass and radius.

\subsection{Scaling relations for rocky exoplanets}\label{subsec:scaling}

In the exoplanetary context, observations are usually limited to the planet’s mass and radius. The goal of this section is to establish, from these two global quantities, the scaling relations that make it possible to infer three key model parameters: the surface gravity $g$, the pressure at the base of the atmosphere $p_0$—which acts as the upper boundary condition for the interior pressure profile—and the radiogenic decay scale $h_r$. The first two are required to integrate the lithostatic pressure profile from the surface inward; the third parameterizes the vertical distribution of radiogenic heat production discussed in the previous section.

\subsubsection{Mass–radius relation and surface gravity}

For rocky planets with Earth-like composition, the mass–radius relation can be approximated by a power law (see e.g. \citealt{Valencia2006, Zeng2016, Unterborn2016}):

\begin{equation}
\frac{R}{R_\oplus} \simeq \left(\frac{M}{M_\oplus}\right)^{\alpha},
\qquad \alpha \approx 0.25\text{--}0.30.
\label{eq:mass_radius}
\end{equation}

In this work we adopt $\alpha = 0.27$, the central value of the empirical range that best reproduces the observed mass–radius relation for rocky exoplanets with densities consistent with compositions dominated by silicates and iron, according to the statistical fits by \citet{Zeng2016} to the population of confirmed exoplanets with masses and radii measured to better than 20\,\%. The surface gravitational acceleration $g = GM/R^2$ then scales as:

\begin{equation}
\frac{g}{g_\oplus} \simeq \frac{M/M_\oplus}{(R/R_\oplus)^2} 
= \left(\frac{M}{M_\oplus}\right)^{1-2\alpha} 
\simeq \left(\frac{M}{M_\oplus}\right)^{0.46}.
\label{eq:gravedad_scaling}
\end{equation}

\subsubsection{Surface atmospheric pressure.}

The pressure at the base of the atmosphere, $p_0$, constitutes the upper boundary condition for integrating the lithostatic pressure profile. Determining it precisely would require detailed modeling of the composition and mass of the atmosphere, which lies beyond the scope of this work. Instead, we adopt a zeroth-order approximation based on two assumptions:
\begin{enumerate}
    \item The atmosphere is thin compared to the planetary radius ($H_{\rm atm} \ll R$), so that gravity can be considered constant throughout the entire atmospheric column.

    \item The atmospheric mass fraction relative to the total planetary mass, $f_{\rm atm} = M_{\rm atm}/M$, is kept constant when scaling from Earth to other rocky planets.
\end{enumerate}
The limitations inherent to these assumptions are discussed in ~\autoref{sec:discusion}.

Under these assumptions, the surface pressure is obtained by integrating the hydrostatic balance $dp = -\rho g\, dz$ over the entire atmospheric column with constant $g$, which is equivalent to equating $p_0$ to the weight of the atmospheric mass column per unit area:

\begin{equation}
p_0 = \frac{M_{\rm atm}\,g}{4\pi R^2} = \frac{f_{\rm atm}\,M\,g}{4\pi R^2},
\label{eq:p0}
\end{equation}

where $M_{\rm atm}$ is the total mass of the atmosphere. Note that this expression is exact under the thin-atmosphere hypothesis, since in that limit the integral $\int \rho\, dV$ over the atmosphere coincides with $M_{\rm atm}$ regardless of the vertical density profile.

Combining \autoref{eq:p0} with the scalings \autoref{eq:mass_radius} and 
\autoref{eq:gravedad_scaling}, we obtain:

\begin{equation}
\frac{p_0}{p_{0,\oplus}} 
= \frac{M}{M_\oplus}\cdot\frac{g}{g_\oplus}\cdot
\left(\frac{R_\oplus}{R}\right)^2
= \left(\frac{M}{M_\oplus}\right)^{2-4\alpha}
\simeq \left(\frac{M}{M_\oplus}\right)^{0.92}.
\label{eq:p0_scaling}
\end{equation}

This scaling sets the initial value of the pressure–depth profile and directly affects the thermodynamic stability of liquid water in the shallow layers of the lithosphere.

\subsubsection{Radiogenic decay scale.}

The radiogenic decay scale $h_r$, introduced in ~\autoref{eq:A_exponencial}, also depends on the planet’s mass and radius.

On a planet with higher surface gravity, the lithostatic pressure increases more rapidly with depth, compressing the radiogenic layer into a smaller thickness for the same pressure difference. Following the argument of \citet{Turcotte2014}, if $h_r$ is determined by a range of lithostatic pressure $\Delta P = \rho_c\,g\,h_r$ that is approximately constant among planets, then:

\begin{equation}
h_r = \frac{\Delta P}{\rho_c\,g} \propto \frac{1}{g} 
\propto \frac{R^2}{M},
\label{eq:hr_fisico}
\end{equation}

which leads to the scaling:

\begin{equation}
h_r = h_{r,\oplus}\,\frac{(R/R_\oplus)^2}{M/M_\oplus}
    = h_{r,\oplus}\,\left(\frac{M}{M_\oplus}\right)^{2\alpha - 1}
    \simeq h_{r,\oplus}\,\left(\frac{M}{M_\oplus}\right)^{-0.46}.
\label{eq:hr_scaling}
\end{equation}

This result implies that more massive planets have relatively thinner radiogenic layers, due to the greater compaction imposed by their surface gravity, which reduces the volume of crust where heat production is concentrated and, therefore, the near-surface geothermal gradient.

\bigskip

Together with the internal heat sources mentioned in the previous section, this scaling completes part of the set of parameters needed to construct the internal thermal structure model, which is presented in the following section.

\subsection{Internal Thermal Structure Model}

Inside a planet, heat can be transported by conduction or convection. In the deep mantle, where material flows viscously on geological timescales, convection dominates thermal transport. However, in the lithosphere —the region of direct interest for subsurface habitability— the mechanical behavior is rigid and viscosities are high enough to suppress convection, so that heat transport is governed by thermal conduction (see e.g. \citealt{Stacey2008, Turcotte2014}). This approximation justifies the use of the steady-state heat diffusion equation as the basis of the thermal model presented below.

Under these considerations, the geothermal profile of the lithosphere is modeled using a steady-state conductive scheme with internal sources. We adopt a coordinate $z$ increasing with depth, and define the magnitude of the heat flux $q(z)$ as positive when it flows toward the surface. Fourier’s law is then written as:

\begin{equation}
q(z) = k(z,T,P)\,\frac{\mathrm{d}T}{\mathrm{d}z},
\label{eq:fourier_z}
\end{equation}

where $k$ is the effective thermal conductivity, which depends on lithology and on the local pressure and temperature conditions. By conservation of energy in steady state, the change in heat flux with depth is exclusively due to the internal volumetric production $A(z)$, so that the one-dimensional conduction equation in \autoref{eq:heat_equation} takes the form:

\begin{equation}
\frac{\mathrm{d}}{\mathrm{d}z}\!\left[k(z,T,P)\,
\frac{\mathrm{d}T}{\mathrm{d}z}\right] + A(z) = 0.
\label{eq:conduccion}
\end{equation}
where $A(z)$ is the volumetric rate of radiogenic heat production defined in \autoref{eq:A_exponencial}. The remaining quantity to be specified is the effective thermal conductivity $k(z,T,P)$, which encodes the dependence of heat transport on mineralogy, 
temperature, and pressure. Its parametrization is described in the following subsection.

\subsubsection{Thermal conductivity}

The thermal conductivity $k$ of geological materials is not constant; rather, it depends strongly on temperature, pressure, and mineralogical composition (see e.g. \citealt{Hofmeister1999, Hasterok2011}). In crystalline solids such as the silicates that make up the crust and mantle, heat transport occurs through two main physical mechanisms: phonon conduction and radiative transfer (see e.g. \citealt{Hofmeister1999, Hofmeister2005}).

The dominant component at low and moderate temperatures is the lattice (phononic) conductivity $k_{\rm lat}$, associated with the propagation of vibrations in the crystal lattice. As temperature increases, phonon–phonon scattering reduces the efficiency of heat transport; conversely, increasing pressure compresses the crystal lattice, thereby enhancing conductivity. To capture both effects simultaneously, we adopt the parametrization of \citet{Hasterok2011} (hereafter HC2011):

\begin{equation}
k_{\rm lat}(T, P) = k_0 \left( \frac{298}{T} \right)^{n} 
\left( 1 + \frac{K'_T}{K_T}\,P \right),
\label{eq:k_lat}
\end{equation}
where $k_0$ is the mineral conductivity under standard conditions (298\,K, 1\,bar), $n$ is an empirical exponent that governs the temperature dependence—typically between 0.4 and 0.6 for silicates—, $K_T\equiv(\partial P/\partial V)_T$ is the isothermal incompressibility modulus and $K'_T\equiv \partial K_T/\partial P$ its derivative with respect to pressure. With this formulation, shallow (low-pressure, $P$) and hot (high $T$) regions reduce the value of the thermal conductivity, acting as relative thermal insulators and giving rise to high geothermal gradients in the upper crust. 

At sufficiently high temperatures, characteristic of the lithospheric mantle ($T \gtrsim 1200$\,K), infrared photons begin to transport energy significantly through translucent crystals. To model this radiative conductivity $k_{\rm rad}$, the unstable classical approximation ($k_{\rm rad} \propto T^3$) is avoided and a smoothed transition is adopted via the error function proposed by \citet{Hofmeister1999}:

\begin{equation}
k_{\rm rad}(T) = \frac{1}{2}\,k_{R,\rm max} 
\left[ 1 + \mathrm{erf}\!\left( \frac{T - T_R}{\omega} \right) \right],
\label{eq:k_rad}
\end{equation}
where $k_{R,\rm max}$ is the maximum asymptotic limit of the radiative conductivity, $T_R$ is the central temperature of the transition, and $\omega$ is the width of that transition. The total conductivity of an individual mineral is the sum of both contributions:

\begin{equation}
k_{\rm mineral} = k_{\rm lat} + k_{\rm rad}.
\label{eq:k_total}
\end{equation}

Since rocks are polycrystalline aggregates, the effective conductivity of each layer is computed from the volumetric fractions $f_i$ of its constituent minerals using the Voigt-Reuss-Hill (VRH) average (HC2011):

\begin{equation}
k_{\rm ef} = \frac{1}{2} \left( \sum_{i} f_i\,k_i + 
\left[ \sum_{i} \frac{f_i}{k_i} \right]^{-1} \right).
\label{eq:VRH}
\end{equation}

This average combines the Voigt upper bound (uniform strain) and the Reuss lower bound (uniform stress), providing a robust and physically consistent estimate for multicomponent mineralogical mixtures (see e.g. \citealt{Berryman1995, Turcotte2014}). In the temperature range of the upper and middle lithosphere, the radiative contribution is secondary, and phononic conductivity together with the VRH average constitute the dominant terms for the astrobiologically relevant regions considered in this work.

\bigskip

Estimating $k_{\rm mineral}$ from the properties of each phase requires knowing two fundamental ingredients: the volumetric fractions $f_i$ of each mineral present in the rock, and the intrinsic conductivity $k_i$ of each of those phases under the local temperature and pressure conditions. Determining these quantities is not trivial, since it depends on the chemical composition of the planet and its differentiation history. In the following sub-subsection we describe how the mineralogical composition of the relevant layers is parameterized and how the values of $f_i$ and $k_i$ used in this work are obtained.

\subsection{Parametric Mineralogical Model}
\label{subsec:parametric_mineralogy}

Earth’s continental crust—enriched in felsic minerals such as quartz and alkali feldspars—is the result of billions of years of magmatic distillation driven by plate tectonics, a process that likely represents an exception in the exoplanetary context (see e.g. \citealt{Foley2018, Putirka2024}). Most rocky exoplanets will operate under a stagnant-lid regime, where basaltic volcanism dominates crust formation (see e.g. \citealt{Noack2017, Foley2018}). This geological diversity has direct consequences for subsurface habitability: quartz-poor crusts exhibit higher bulk density, which accelerates lithostatic compaction and reduces the porous volume available to host liquid water at depth.

To capture this compositional spectrum, we have developed a continuous parametric model of mineralogical mixing defined by three geochemical variables:

\begin{itemize}
    \item \textbf{Felsic fraction ($f_{\mathrm{felsic}}$):} controls the bulk composition of the lithosphere via a continuous interpolation between a terrestrial felsic crust, $f_{\mathrm{felsic}} = 1$, dominated by quartz and sodic plagioclase, following \citealt{Rudnick2014} and a primary basaltic-ultramafic crust, $f_{\mathrm{felsic}} = 0$, dominated by anorthite and pyroxenes, after \citealt{Putirka2019}.

    \item \textbf{Iron–magnesium ratio (Fe/Mg):} governs solid solutions in ferromagnesian silicates, modulating the partitioning between the forsterite–fayalite end-members in olivines and enstatite–ferrosilite in pyroxenes 
    (see e.g. \citealt{Unterborn2023, Putirka2024}). Since iron-rich phases are significantly denser, this parameter alters the lithostatic pressure gradient $dP/dz$ and, consequently, the depth at which the macrofractures required to host fluids are sealed.

    \item \textbf{Degree of hydration ($h$):} models retrograde metamorphism across the amphibolite–greenschist facies transition, transforming primary anhydrous silicates into secondary hydrated phases—predominantly amphiboles and micas—\citep{Winter2014}. This mineralogical alteration reduces the effective rock density and modifies thermal diffusivity in the upper layers of the lithosphere.
\end{itemize}

The mineralogical fraction of any mineral $i$ in a given layer is obtained by linear interpolation between the two compositional end-members:
\begin{equation}
    M_i^{\rm layer} = f_{\rm felsic}\, M_i^{\rm fel,layer} 
    + (1 - f_{\rm felsic})\, M_i^{\rm maf},
    \label{eq:interpolacion}
\end{equation}
where $M_i^{\rm fel,layer}$ is the reference fraction of mineral $i$ derived from the CIPW norm applied to the oxide compositions of \citet{Rudnick2014}, and $M_i^{\rm maf}$ is its fraction in the mafic end-member according to \citet{Putirka2019}, both based on the stellar abundances in the Hypatia catalog \citep{Hinkel2014} for more than 4000 solar-type stars. The adopted numerical values are summarized in \autoref{tab:composiciones_ref}.

\begin{table*}[ht]
\centering
\caption{Reference mineralogical compositions (mass \%) for the felsic and mafic end-members of each lithospheric layer. \textit{Note: the felsic values are derived values}, obtained by applying the standard CIPW norm \citep{Cross1902} to the oxide compositions of \citet{Rudnick2014}. The mafic end-member follows the predictions of \citet{Putirka2019} for primary crusts of rocky exoplanets.}
\label{tab:composiciones_ref}
\begin{tabular}{lcccc}
\hline\hline
Mineral & Felsic UCC & Felsic MCC & Felsic LCC & Mafic end-member \\
        & ($f_{\rm felsic}=1$) & ($f_{\rm felsic}=1$) & ($f_{\rm felsic}=1$) 
        & ($f_{\rm felsic}=0$) \\
\hline
Quartz (Qz)            & 20.6 & 10.0 &  2.0 &  0.0 \\
Orthoclase (Or)        & 11.4 &  8.0 &  2.0 &  0.0 \\
Albite (Ab)            & 21.0 & 18.0 & 10.0 & 10.0 \\
Anorthite (An)         &  8.0 & 15.0 & 28.0 & 30.0 \\
Total pyroxenes (Px)   &  4.0 & 14.0 & 30.0 & 40.0 \\
Total olivine (Ol)     &  1.0 &  3.0 &  8.0 & 20.0 \\
\hline
\end{tabular}
\end{table*}

\textbf{Iron–magnesium partitioning in ferromagnesian silicates.}
The total fractions of olivine and pyroxene obtained from \autoref{eq:interpolacion} are subdivided between their magnesian and ferrous end-members using the ideal solid-solution model of \citet{Unterborn2023}:
\begin{equation}
    X_{\rm Fe} = \frac{{\rm Fe/Mg}}{1 + {\rm Fe/Mg}}, \qquad 
    X_{\rm Mg} = 1 - X_{\rm Fe}.
    \label{eq:XFeMg}
\end{equation}
Pyroxenes are distributed between orthopyroxenes and clinopyroxenes in proportions that vary with depth, reflecting the progressive increase of clinopyroxene with pressure documented in seismic profiles of the lithospheric mantle (see e.g. \citealt{Dziewonski1981, 
Stacey2008}); the adopted layer-by-layer ratios are listed in \autoref{tab:px_ratios}.

\begin{table}[ht]
\centering
\caption{Adopted orthopyroxene/clinopyroxene ratios by lithospheric layer.}
\label{tab:px_ratios}
\begin{tabular}{lcc}
\hline\hline
Layer & $f_{\rm Opx}$ & $f_{\rm Cpx}$ \\
\hline
Upper crust    & 0.70 & 0.30 \\
Middle crust   & 0.60 & 0.40 \\
Lower crust    & 0.50 & 0.50 \\
Upper mantle   & 0.40 & 0.60 \\
\hline
\end{tabular}
\end{table}

\textbf{Retrograde metamorphism and hydration.}
The parameter $h \in [0,1]$ transforms primary anhydrous silicates into hydrated phases following the petrological reactions of \citet{Winter2014}: pyroxenes and anorthite are converted into 
hornblende, and orthoclase into phlogopite. Defining the mass-transfer rate $\tau = 0.4\,h$, calibrated to reproduce the typical modal proportions of amphibolite-facies assemblages \citep{Yardley2009, Winter2014}, the transferred masses are:

\begin{align}
    \Delta_{\rm Px} &= M_{\rm Px,tot}\cdot\tau, 
    \nonumber \\
    \Delta_{\rm An} &= M_{\rm An}\cdot 0.3\,\tau, 
    \nonumber \\
    M_{\rm Hbl}     &= \Delta_{\rm Px} + \Delta_{\rm An}, 
    \label{eq:hidratacion} \\
    M_{\rm Phl}     &= M_{\rm Or}\cdot 0.5\,h, 
    \nonumber
\end{align}
where the coefficient $0.3$ on anorthite reflects the partial calcium contribution to hornblende stoichiometry \citep{Winter2014}, and the factor $0.5$ on orthoclase limits the conversion to phlogopite to the fraction compatible with the availability of Mg in typical crustal fluids \citep{Yardley2009}. 
The source mineral masses are reduced accordingly 
($M'_i = M_i - \Delta_i$), and this alteration reduces the effective density of the rock and modifies the thermal diffusivity in the upper layers of the lithosphere.

\textbf{Mantle composition and final normalization.}
The lithospheric mantle is modeled as a standard depleted peridotite: 60\,\% olivine, 35\,\% pyroxenes, and 5\,\% garnets, following \citet{McDonough1995} and consistent with lithospheric xenoliths compiled by \citet{Stacey2008}. This composition does not vary with $f_{\rm felsic}$ or $h$, since the lithospheric mantle does not participate in crustal magmatic distillation processes or in the circulation of meteoric fluids. Garnets are distributed between pyrope and almandine following the same Fe/Mg partitioning as in \autoref{eq:XFeMg}. 

\bigskip

At each stage, the resulting fractions are renormalized to $\sum_i M'_i = 1$, ensuring mass conservation before computing the effective thermophysical properties.

With the mineralogical composition of each layer defined and the thermophysical properties—thermal conductivity, density, and radiogenic heat production—fully parameterized, the next step is to integrate these ingredients into a self-consistent geotherm. In the next subsubsection we describe the numerical scheme used to obtain the temperature and heat-flow profiles throughout the planetary lithosphere.

\subsection{Integration of the geothermal profile}

Continental geotherms provide the fundamental reference framework for assessing subsurface habitability. To first order, assuming constant thermal conductivity and uniform heat production, ~\autoref{eq:conduccion} admits a parabolic analytical solution:

\begin{equation}
T(z) = T_0 + \frac{q_s}{k}\,z - \frac{A}{2k}\,z^2, \qquad 
q(z) = q_s - A\,z,
\label{eq:geotherm}
\end{equation}
where $T_0$ and $q_s$ are the temperature and heat flux at the surface, respectively. This solution elegantly illustrates the combined effect of basal heat flow and radiogenic heating, but it does not capture the inherent nonlinearities of a real lithosphere
with variable $k(z,T,P)$ and $A(z)$.

In practice, the geothermal profile is constructed by direct numerical integration of ~\autoref{eq:conduccion}, iteratively coupling the temperature, pressure, and density profiles along the lithospheric column, and updating $k(z,T,P)$ at each step using equations~\autoref{eq:k_lat}--\autoref{eq:VRH} and $A(z)$ using ~\autoref{eq:A_exponencial}, in accordance with the compositional stratigraphy defined by the parametric mineralogical model (~\autoref{subsec:parametric_mineralogy}). This procedure, detailed by HC2011, allows one to integrate the temperature and pressure profile step by step down to depths of hundreds of kilometers in a physically consistent way.

\bigskip

However, knowing where water is thermodynamically stable is not sufficient to assess the presence of macroscopic amounts of liquid water: it is also necessary to determine whether the rocky medium has enough pore space to host it. This leads us to consider the crustal porosity profile.

\subsection{Crustal porosity}\label{subsec:porosity}

The temperature and pressure profiles inside the crust determine where water can exist in the liquid phase, but not whether physical space is available to contain it. In the interior 
of a rocky planet, that space is controlled by the porosity of the crust: the fraction of the rock volume occupied by interconnected pores, fractures, and cavities that can host fluids. Porosity thus introduces a second, independent dimension into the assessment of 
subsurface aquability and habitability.

The shallow crust of rocky planets exhibits a high initial porosity, resulting from fracturing produced by meteoritic impacts, tectonic processes, and chemical weathering 
(see e.g. \citealt{Clifford1993, Han2014}). This porosity decreases rapidly with depth as a consequence of mechanical compaction induced by lithostatic pressure. Based on empirical 
compaction models and lunar gravimetric observations, \citet{Han2014} parametrizes the fractional porosity by an exponential law:

\begin{equation}
\phi(P) = \phi_0 \exp\left(-c\frac{P}{P_c}\right),
\label{eq:porosity_pressure}
\end{equation}
where $\phi_0$ is the surface porosity, $P_c$ is the characteristic closure pressure, and $c$ controls the efficiency of pore closure with depth \citep{Han2014}. In massive exoplanets, high surface gravity closes pores at shallow depths; in smaller bodies such as Mars or the Moon, aquifer networks can extend much deeper into the crust (see e.g. \citealt{Clifford1993, Han2014}).

The decay of porosity with pressure motivates a physically motivated distinction between two habitability-relevant regimes. For aquability, the relevant criterion is simply $\phi > 0$: any connected pore space where $T$ and $P$ fall within the liquid stability field of water qualifies, even at very low porosities, since thin films along grain boundaries can host liquid water at porosities well below $10^{-3}$ \citep{Parnell2016}. Habitability, 
however, imposes a stricter requirement: fluid must circulate to supply nutrients and sustain metabolic activity (see e.g. \citealt{Parnell2016, Magnabosco2018}). Studies of the deep continental subsurface in basaltic and granitic formations document microbial activity in rocks with porosities as low as ${\sim}0.1$--$0.2\,\%$ \citep{Orcutt2013, Magnabosco2018}, establishing an empirical lower bound on the hydraulic connectivity required for life. We therefore adopt a minimum habitability threshold $\phi_{\rm min}^{\rm hab} = 10^{-3}$, below which the rock is considered astrobiologically inert regardless of its thermodynamic state; the sensitivity of the results to this choice is explored in Section~\ref{sec:results}.

The porosity profile $\phi(z)$ thus becomes the third ingredient of the model, together with the temperature $T(z)$ and pressure $P(z)$ profiles, and enters differently into the aquability and habitability criteria. The self-consistent numerical integration of these three profiles constitutes the quantitative basis on which the effective aquable and eirenespheres of each planet are evaluated.

\section{Numerical implementation}

To solve the system of coupled equations described in the previous sections, an iterative numerical integration scheme was implemented, based on the \textit{bootstrap} method introduced by \citet{Chapman1986} and adopted by HC2011 for the calculation of continental geotherms.

This method starts from an initial temperature estimate—obtained from the parabolic analytical solution of ~\autoref{eq:geotherm}—and refines it iteratively by updating $k(z,T,P)$, $\rho(z,T,P)$ and $A(z)$ at each node, until strict convergence criteria in both temperature and pressure are simultaneously satisfied.

The integration proceeds over a discrete spatial mesh with step size $\Delta z$ from the surface ($z = 0$), where the surface temperature $T_0$—obtained from the radiative balance of ~\autoref{eq:Teq_transmitancia}—and the surface heat flow $q_s$ are imposed as boundary conditions. The latter is assigned as a free parameters, with the reference value being the present terrestrial heat flux $q_{s,\oplus} \approx 65\,\mathrm{mW\,m^{-2}}$  \citep{Davies2010}. At each depth node $z_i$, the algorithm sequentially updates the following properties:

\begin{enumerate}

    \item \textbf{Local porosity} $\phi(P_i)$, evaluated using \autoref{eq:porosity_pressure} at the local pressure. This quantity is updated first, since it controls both the effective density and the thermal conductivity. We adopt $\phi_0 = 0.15$, representative of a highly fractured crust \citep{Han2014}, a characteristic closing pressure $P_c = 200$ MPa, consistent with typical values for silicate rocks under lithostatic compression (see e.g. \citealt{Stacey2008, Turcotte2014}), and a dimensionless parameter $c = 4$ (see \autoref{eq:porosity_pressure}), within the empirical range derived from compaction models (see e.g. \citealt{Athy1930, Han2014}).

    \item \textbf{Effective density} $\rho_i$, obtained by combining the density of the solid mineral grain $\rho_{\rm grain}$—calculated at local conditions $(T_i, P_i)$ using the equations of state implemented in \texttt{BurnMan} (see e.g. \citealt{Cottaar2014, Myhill2023})—with the porosity correction:
    \begin{equation}
        \rho_i = \rho_{\rm grain}(T_i, P_i)\,(1 - \phi_i).
        \label{eq:rho_efectiva}
    \end{equation}

    \item \textbf{Enclosed mass} $M(r_i)$, calculated as the total mass of the planet minus the mass accumulated in the overlying layers that have already been integrated, that is:
    \begin{equation}
            M(<r_i) = M_{\rm tot} - M_{\rm above}
            \label{eq:closed_mass}
    \end{equation}
    
    The mass of each layer is evaluated as $\Delta M = 4\pi r_i^2 \Delta r_i \rho_i$, with $\rho_i$ the local effective density, which includes the effect of porosity.

    \item \textbf{Local gravity} $g_i$, calculated at each node from the enclosed mass $M(<r_i)$ accumulated from the surface, updating the gravitational profile in a self-consistent way with the density:
    \begin{equation}
        g_i = \frac{G\,M(<r_i)}{r_i^2}, \qquad 
        r_i = R_{\rm planet} - z_i.
        \label{eq:gravedad_local}
    \end{equation}

    \item \textbf{Lithostatic pressure} $P_{i+1}$, obtained by integrating the hydrostatic equilibrium (~\autoref{eq:hidrostatico}) with the updated local density and gravity:
    \begin{equation}
        P_{i+1} = P_i + \rho_i\,g_i\,\Delta z.
        \label{eq:P_update}
    \end{equation}
    The calculation of the effective density, the enclosed mass, and the local gravity in each layer is performed iteratively until pressure convergence is achieved ($|\Delta P / P| < \epsilon_P$), which guarantees a self-consistent coupling between density, gravity, and pressure.

    \item \textbf{Effective thermal conductivity} $k_{\rm ef}(z_i, T_i, P_i)$, calculated by means of the weighted geometric average between the thermal conductivity of the solid grain $k_{\rm grain}$ —obtained from the VRH average of Eqs.~\ref{eq:k_lat}--\ref{eq:VRH}— and the conductivity of the interstitial fluid $k_{\rm fluid}$:
    \begin{equation}
        k_{\rm ef} = k_{\rm grain}{(1-\phi)}\cdot k_{\rm fluid}{\phi},
        \label{eq:k_geometrico}
    \end{equation}
    where it is adopted $k_{\rm fluid} = 0.6\,\mathrm{W\,m^{-1}\,K^{-1}}$, 
     typical value for liquid water in the temperature range of interest (see, e.g., \citealt{Wagner2002}).
    
    \item \textbf{Effective radiogenic production} $A_{\rm bulk}(z_i)$, corrected by the porosity fraction, since only the solid fraction of the rock contributes to heat production:
    \begin{equation}
        A_{\rm bulk}(z_i) = A(z_i)\,(1 - \phi_i),
        \label{eq:A_bulk}
    \end{equation}
    where $A(z_i)$ is evaluated using ~\autoref{eq:A_exponencial}.

    \item \textbf{Temperature} $T_{i+1}$ \textbf{and heat flow} 
    $q_{i+1}$, updated by the equation \textit{bootstrap}:
    \begin{align}
        T_{i+1} &= T_i + \frac{q_i}{k_{\rm ef}}\,\Delta z 
        - \frac{A_{\rm bulk}}{2\,k_{\rm ef}}\,\Delta z^2,
        \label{eq:bootstrap_T}\\
        q_{i+1} &= q_i - A_{\rm bulk}\,\Delta z.
        \label{eq:bootstrap_q}
    \end{align}
   Steps (v)–(vii) are iterated until convergence in temperature is reached
    ($|\Delta T| < \epsilon_T$).

\end{enumerate}

The integration stops when the temperature exceeds $T_{\rm max} \approx 2150\,\mathrm{K}$, a limit above which \texttt{BurnMan} does not guarantee the validity of the equations of state 
for the minerals considered. In practice, this thermal criterion acts as the dominant stopping condition, since the lithosphere–asthenosphere 
rheological transition typically occurs around 
${\sim}1300\,\mathrm{K}$ (see e.g. \citealt{ Artemieva2006, Stacey2008}), 
so that the integration always covers the lithospheric region of astrobiological interest before approaching the validity limit of 
\texttt{BurnMan}.

As a geophysical reference, the lithosphere–asthenosphere rheological transition typically occurs around ${\sim}1300\,\mathrm{K}$ (see e.g. \citealt{Artemieva2006, Stacey2008}), so in practice the integration always covers the lithospheric region of astrobiological interest.

The reference thermophysical parameters at standard conditions ($k_0$, $n$, $K_T$, $K'_T$, $k_{R,\rm max}$, $T_R$, $\omega$) for each mineral in the assemblage are taken from the compilations of HC2011 and \citet{Hofmeister1999}, complemented with the data of \citet{Pertermann2006} for ferromagnesian minerals. The sensitivity of the model to the adopted values of thermal conductivity, basal heat flux, and radiogenic decay rate will be explored by means of parametric analyses in ~\autoref{sec:results}.

\bigskip
 

The numerical integration described in this section yields, for each modeled planet, a set of discretized profiles $\{T_i, P_i, \rho_i, \phi_i\}$ evaluated on a mesh of equally spaced depth nodes, where the index $i$ runs through the layers from the surface ($i=0$, $z=0$) down to the base of the lithosphere ($i=N$, $z=z_{\rm max}$). Each node represents an elemental volume of crust at depth $z_i$, with well-defined local thermodynamic conditions. The evaluation of subsurface habitability is thus reduced to classifying each node according to the physical and biological criteria defined in the following sections.

\section{Aquability and habitability criteria}\label{sec:criteria}

\subsection{Aquability condition}

Based on the water phase diagram presented in \autoref{sec:sshz}, we define the condition of Aquability at a node $(T_i, P_i)$ as:

\begin{equation}
T_{\rm m}(P_i) < T_i < \min\!\left[T_{\rm vap}(P_i),\, T_c\right],
\quad P_i > P_{\rm tp},
\label{eq:aguabilidad}
\end{equation}
where $T_{\rm m}(P)$ is the local melting temperature of ice and $T_{\rm vap}(P)$ is the boiling temperature—both are phase transition curves from the diagram shown in \autoref{fig:phase_water_diagram}—, $T_c = 647.1\,\mathrm{K}$ is the critical temperature, and $P_{\rm tp} = 611.7\,\mathrm{Pa}$ is the pressure of the triple point, below which liquid water cannot exist at any temperature. 

The curves $T_{\rm m}(P)$ and $T_{\rm vap}(P)$ are evaluated numerically from the IAPWS reference thermodynamic formulations (see e.g. \citealt{IAPWS1995, Wagner2002, IAPWSRelease2009}), incorporating the high-pressure ice polymorphs (Ih, III, V, VI, VII) that are relevant at increasing depths \citep{Choukroun2010}. We adopt the critical point as a conservative upper limit for “aquability”, since supercritical water exhibits radically different physicochemical properties—such as altered solubility of biomolecules—that introduce fundamental uncertainties regarding its ability to sustain known biological processes (see e.g. \citealt{SchulzeMakuch2008, Lingam2018}). 

A node that satisfies the conditions in \autoref{eq:aguabilidad} is termed \emph{aquable} and contributes to the effective eirenesphere or \textit{eirensphere} if it also meets the biological criteria defined in the following subsections.

\subsubsection{Aquability minimum and maximum depths}

Inside the planet, the temperature profile $T(z)$ increases monotonically with depth following the geotherm calculated in \autoref{sec:model}, while the pressure $P(z)$ follows the lithostatic profile. The intersection of these profiles with the phase boundaries of water defines two characteristic depths:

\begin{itemize}

    \item \textbf{Minimum depth of water stability} $z_{\rm min}$: the depth at which the pressure and temperature conditions required for the existence of liquid water are simultaneously satisfied. In practice, the pressure condition ($P > P_{\rm tp}$) is trivially reached in the lithosphere: even in the absence of an atmosphere, a rock column just a few meters thick is enough to exceed the $611.7\,\mathrm{Pa}$ of the triple point, given that the density of rocky materials ($\rho \sim 2700$--$3000\,\mathrm{kg\,m^{-3}}$) generates lithostatic pressures far greater than atmospheric pressure at very shallow depths. The determining factor for $z_{\rm min}$ is therefore temperature: on planets without an atmosphere or with very cold surfaces, the surface temperature may be lower than $T_{\rm m}(P)$, so that $z_{\rm min}$ is set by the depth at which the geotherm reaches the melting curve $T_{\rm m}(P)$, that is, the base of the cryosphere.

    \item \textbf{Maximum depth of water stability} $z_{\rm max}$: the depth at which the geothermal profile reaches the boiling curve $T_{\rm vap}(P)$ or the critical temperature $T_c$, whichever occurs first.

\end{itemize}

The difference $\Delta z_{\rm aq} = z_{\rm max} - z_{\rm min}$ defines the \textit{thickness of the aquable layer} (TAL). This thickness is not solely a surface property of the planet: it depends on the processes that occur beneath the surface, in particular on the basal heat flux $q_s$, the lithospheric thermal conductivity $k$—which controls the slope of the geotherm—and the surface gravity $g$, which sets the rate of increase of pressure with depth, through the scaling relations in \autoref{subsec:scaling}. 

On planets with higher surface gravity than Earth ($g > g_\oplus$), the lithostatic pressure increases more rapidly with depth, which on the one hand raises $T_{\rm vap}(P)$—widening the upper thermal margin of the aquable zone—but on the other hand compresses the thickness of the available porous crust, as discussed in \autoref{subsec:porosity}. This competition between thermal and mechanical effects implies that the aquable zone does not grow monotonically with planetary mass, but instead exhibits a non-trivial behavior that we explore in \autoref{sec:results}.

\bigskip

That a layer is aquable does not imply that it is habitable. Habitability requires additional conditions: temperature and pressure must fall within the ranges that living organisms can tolerate.  
On Earth, these limits are determined empirically by the most resistant known extremophile organisms (see e.g. \citealt{Rothschild2001, Merino2019}). Hereafter we will assume, without loss of generality, that these limits of biological tolerance are universal; that is, any form of life based on aqueous biochemistry—regardless of its planet of origin—operates within physical ranges similar to those of terrestrial extremophiles. This hypothesis, widely adopted in the astrobiological literature (see e.g. \citealt{Rothschild2001, Cockell2016, Merino2019}), constitutes the most conservative criterion available in the absence of evidence for extraterrestrial life. In the following subsections we formally define these limits.

\subsection{Habitability condition: biological limits}
\label{sec:habitable_limits}

\subsubsection{Temperature limits}

Temperature is one of the most decisive environmental factors for life: it regulates the stability of macromolecules (proteins, nucleic acids, lipids), the kinetics of metabolic reactions, and the structural integrity of cellular membranes (see e.g. \citealt{Daniel2000, Bains2015}).

In the terrestrial biosphere, experimental evidence places the range of biological activity from cryogenic environments—with microbial activity documented below $-20\,^\circ\mathrm{C}$—to hyperthermophilic communities that thrive in hydrothermal settings. The most robust records of growth and reproduction at high temperature come from hyperthermophilic archaea. For example, \emph{Pyrolobus fumarii} remains viable up to ${\sim}113\,^\circ\mathrm{C}$ \citep{Blochl1997}, 
while \emph{Geogemma barossii} (Strain 121) and cultures of \emph{Methanopyrus kandleri} under high pressure have shown proliferation up to $121$--$122\,^\circ\mathrm{C}$ \citep{Kashefi2003, Takai2008}. These limits underscore 
that pressure, by keeping water in the liquid phase, is an essential requirement for known biochemistry to function at such elevated temperatures.

However, the fact that some species can grow up to ${\sim}122\,^\circ\mathrm{C}$ does not imply that life can be extended indefinitely to higher temperatures. Analyses of the kinetics of hydrolysis and decomposition of essential metabolites suggest that many core metabolic molecules—peptide bonds, phosphodiester bonds in DNA/RNA, and the pyrophosphate bond in ATP—undergo progressively faster decomposition as temperature increases \citep{White1984}. From this perspective, \citet{Bains2015} estimated a conservative theoretical limit on the order of ${\sim}150$–$180\,^\circ\mathrm{C}$ for water-based terrestrial biochemistry, albeit with uncertainties associated with reaction conditions and the water activity of the medium. These estimates make explicit why pressure, although it widens the thermal window for liquid water, does not remove the fundamental chemical constraints on the stability of metabolites and membranes.

For subsurface habitability models, the empirically observed upper growth limit is usually adopted as a conservative operational criterion, 
$T_{\rm bio}^{\rm max} \approx 395\,\mathrm{K}$ (${\sim}122\,^\circ\mathrm{C}$), consistent with the most resistant known hyperthermophilic organism 
\citep{Kashefi2003, Takai2008}. This value provides a lower bound on the eirenesphere: any region that exceeds this thermal threshold is excluded from the calculation, regardless of whether water is thermodynamically stable. Note, however, that since the model adopts $T_{\rm bio}^{\rm max} = 423.15\,\mathrm{K}$ ($150\,^\circ\mathrm{C}$) following the theoretical analysis of \citet{Bains2015}, the calculated eirenesphere actually represents a more conservative upper bound than that derived solely from the empirical record.

\bigskip
While temperature imposes the best-studied upper limit for subsurface life, the increase in pressure with depth introduces an independent and equally fundamental constraint. In the next subsubsection we analyze the biological limits associated with pressure and the piezophilic extremophiles that define them.

\subsubsection{Pressure limits}

Pressure is a fundamental parameter in the assessment of subsurface habitability, particularly in deep environments where lithostatic or hydrostatic pressure can reach values on the order of hundreds of megapascals, equivalent to several thousand atmospheres. Studies of Earth’s deep biosphere have shown that life can persist under extreme pressure conditions in both marine and continental environments. In the Mariana Trench ($\approx 11\,\mathrm{km}$ depth), complex organisms and metabolically active microorganisms have been identified at pressures close to $110\,\mathrm{MPa}$ (see e.g. \citealt{Yayanos1995, Danovaro2010}), while microorganisms of the continental deep biosphere have been detected at depths where pressures exceed $100$–$130\,\mathrm{MPa}$ (see e.g. \citealt{Orcutt2013, Magnabosco2018}).

Among the most extreme piezophilic organisms known are the hyperthermophilic archaeon \textit{Thermococcus piezophilus}, capable of growing between $0.1$ and $125\,\mathrm{MPa}$ with an optimum near $50\,\mathrm{MPa}$ \citep{Dalmasso2016}, and the psychrophilic, obligate piezophilic bacterium \textit{Colwellia marinimaniae}, isolated from the Challenger Deep of the Mariana Trench, which grows between $80$ and $140\,\mathrm{MPa}$ with an optimum around $120\,\mathrm{MPa}$ \citep{Kusube2017}. The latter currently represents the highest pressure value experimentally documented for sustained microbial growth under natural conditions.

At the molecular level, pressure directly affects the structural stability of biological systems. Biophysical experiments show that many globular proteins undergo irreversible denaturation typically in the range $400$–$800\,\mathrm{MPa}$, while macromolecular complexes can dissociate at pressures as low as $200$–$300\,\mathrm{MPa}$ \citep{Meersman2006}. Similarly, lipid membranes undergo pressure-induced phase transitions to more ordered states around $200$–$300\,\mathrm{MPa}$, compromising their fluidity and functionality \citep{Winter2009}. Nevertheless, experiments with diamond anvil cells have shown that certain bacteria can maintain transient metabolic activity near $1\,\mathrm{GPa}$ ($\approx 10^4$ atm), while bacterial spores can withstand even higher pressure pulses \citep{Sharma2002}. These results suggest that the ultimate threshold for cellular survival could considerably exceed the limit associated with sustained active metabolism.

In the astrobiological context, these constraints become especially relevant for deep oceans and water-rich interior environments. For example, the subsurface oceans of icy moons such as \textit{Europa} could experience pressures on the order of $100$–$200\,\mathrm{MPa}$ at the ocean floor, within the range tolerated by the most extreme terrestrial piezophiles known to date \citep{Vance2018}. In contrast, ocean worlds and super-Earths with deep global hydrospheres could reach considerably higher pressures. However, so far no absolute upper limit has been empirically established for active metabolism under high pressure, provided that thermal and chemical conditions suitable for biological processes are maintained (see e.g. \citealt{SchulzeMakuch2008, Merino2019}).

Consequently, within the framework adopted in this work, pressure is treated primarily as a permissive factor rather than a strictly limiting one for subsurface habitability. Its influence is assumed to be predominantly indirect, acting through its effects on the thermodynamic stability of liquid water, the kinetics of biochemical reactions, fluid–mineral interactions, and the structural stability of biomolecules, rather than through the existence of a clearly defined biological pressure limit.

\bigskip

With the biological limits of temperature and pressure thus established, we can precisely formulate the habitability criterion that will be applied to the geothermal and pressure profiles calculated in \autoref{sec:model}.

\section{Formal definition of the eirenesphere}
\label{sec:eirenesphere}

An aquable node $(T_i, P_i)$ of the planetary interior is considered \emph{habitable}, ie. it belongs to the eirenesphere of the planet, if it simultaneously satisfies the biological limits established in the previous subsubsections and the hydraulic connectivity condition discussed in \autoref{subsec:porosity}:

\begin{equation}
\mathrm{max}[T_{\rm bio}^{\rm min}, T_m(P_i)] < T_i < T_{\rm bio}^{\rm max},
\qquad
P_i < P_{\rm bio}^{\rm max},
\qquad
\phi_i \geq \phi_{\rm min}^{\rm hab},
\label{eq:habitabilidad}
\end{equation}
where the values adopted in this work are:

\begin{itemize}
    \item $T_{\rm bio}^{\rm min} = 253.15\,\mathrm{K}$ ($-20\,^\circ\mathrm{C}$): lower temperature limit, set by the minimum temperature for metabolic activity documented in extreme psychrophiles in permafrost and subglacial brines (see e.g. \citealt{Rivkina2000, Merino2019}).

    \item $T_{\rm bio}^{\rm max} = 423.15\,\mathrm{K}$ ($150\,^\circ\mathrm{C}$): upper temperature limit, adopted from the analysis of kinetic stability of essential metabolites against hydrolysis in water by \citet{Bains2015}, which places the conservative theoretical threshold for Earth-like biochemistry at $\sim\!150$--$180\,^\circ\mathrm{C}$. This value is more conservative than the empirical record for growth at $\sim\!122\,^\circ\mathrm{C}$ (see e.g. \citealt{Kashefi2003, Takai2008}) and provides an upper bound for the potentially eirenesphere.

    \item $P_{\rm bio}^{\rm max} = 200\,\mathrm{MPa}$: upper pressure limit, conservatively adopted above the documented record for active growth ($140\,\mathrm{MPa}$, \citealt{Kusube2017}). This choice incorporates a margin with respect to the uncertainties inherent in the calculated pressure profiles and keeps the criterion within a range still compatible with the physiology of piezophilic extremophiles, in agreement with general reviews on the limits of life in extreme and subsurface environments (see e.g. \citealt{SchulzeMakuch2008, Merino2019}).

    \item $\phi_{\rm min}^{\rm hab} = 10^{-3}$: minimum porosity threshold for biologically relevant hydraulic connectivity, discussed in \autoref{subsec:porosity}.
\end{itemize}

The eirenesphere is therefore a strict subset of the aquable layer, bounded simultaneously by thermodynamic, mechanical, and biological criteria. \autoref{eq:habitabilidad} is evaluated node by node along the discretized geothermal profile, and the corresponding crustal volume is obtained by spherical integration. This volume constitutes the direct input for the \textit{Eirenesphere Volumetric Index} defined in the next section.

\section{The Eirenesphere Volumetric Index (EVI)}\label{sec:index}

In order to comprehensively quantify the subsurface habitability of rocky planets and to facilitate systematic comparison between exoplanets, we propose the \emph{Eirenesphere Volumetric Index}, EVI. Unlike classical approaches based purely on surface irradiation, this metric is constructed in two stages: first, by quantifying the actual physical volume available for biologically viable liquid water in the planet’s interior, ie. the volume of the eirenesphere; and second, by averaging this volume over the orbital domain of the stellar habitable zone.

\subsection{Volume of the eirenesphere}\label{subsec:v3d}

The first step is to determine the physical space effectively available for a subsurface biosphere. In a rocky medium, life does not occupy the total geometric volume of the crust, but only the interstitial space controlled by the local porosity $\phi(z)$, defined in \autoref{subsec:porosity}.

We define a binary local habitability indicator function $\mathcal{H}(T,P,\phi)$ such that $\mathcal{H} = 1$ if the node simultaneously satisfies the aquability condition (\autoref{eq:aguabilidad}), the biological limits of temperature and pressure (\autoref{eq:habitabilidad}), and the minimum threshold of hydraulic connectivity $\phi \geq \phi_{\rm min}^{\rm hab}$; and $\mathcal{H} = 0$ otherwise.

Under the assumption of planetary spherical symmetry, the volume of a layer located at depth $z$ and thickness $\mathrm{d}z$ is given by $\mathrm{d}V = 4\pi (R_p - z)^2 \mathrm{d}z$, however, only a fraction $\phi(z)$ of this volume corresponds to pore space that can potentially be filled with liquid water, and it will contribute to the eirenesphere volume only if the local conditions satisfy $\mathcal{H}=1$. Thus, the volume of the eirenesphere, $V_\mathrm{ES}$ is defined as:
\begin{equation}
V_\mathrm{ES} = 4\pi \int_{0}^{R_p} 
\mathcal{H}\!\left(T(z), P(z), \phi(z)\right)\, \phi(z)\, (R_p - z)^2 \, \mathrm{d}z,
\label{eq:V_eff}
\end{equation}
where $R_p$ is the planetary radius and $z$ is the depth measured from the surface.

In the numerical implementation, this integral is evaluated as a discrete sum over the nodes of the geothermal profile that satisfy $\mathcal{H}=1$:
\begin{equation}
V_\mathrm{ES} \approx \sum_{i \in \mathcal{H}=1} 
\phi_i \, 4\pi r_i^2 \Delta z,
\label{eq:V_eff_discreta}
\end{equation}
where $r_i = R_p - z_i$ corresponds to the geocentric radius of node $i$ and $\Delta z$ is the spatial integration step.

It is important to distinguish this quantity from the total volume of rock that satisfies the thermodynamic and biological conditions, regardless of the available porosity. Mathematically, this quantity would correspond to the integral in \autoref{eq:V_eff} without the factor $\phi(z)$. The volume of the eirenesphere is always less than or equal to the total habitable volume and represents the physically accessible fraction of the subsurface that can host fluids and, potentially, microbial ecosystems.

This distinction is especially relevant on massive planets, where the higher surface gravity increases lithostatic compaction and rapidly reduces porosity with depth. In such cases, a planet may possess an extensive thermodynamically aquable region but a relatively small eirenesphere due to the mechanical closure of pore spaces.

\bigskip

The eirenesphere thus defined quantifies the ability of a specific planet to host deep aquatic ecosystems under a given thermal and orbital state. However, in order to build a comparative index between planets of different masses and thermal stories, it is necessary to average this quantity over the range of distances compatible with the subsurface stellar habitable zone (SSHZ, \citealt{Escudero2023}). In the following subsection, we formally define this average and the resulting index as the \textit{Eirenesphere Volume Index}, EVI.

\subsection{Definition of the EVI}

The eirenesphere volume $V_\mathrm{ES}$ defined in \autoref{eq:V_eff} is not a static property of the planet, but rather depends on time and on the orbital distance $a$. The dependence on time comes from the secular evolution of heat flux resulting from the thermal evolution of the planet and the decay of the isotopes sustaining the radiogenic heat. On the other hand, surface temperature $T_{\rm surf}(a)$—the upper boundary condition of the geothermal profile—, which depends on the average distance to the host star $a$, determines the position and thickness of the aquable region within the crust.

Since one of the aims of our work—shared with many studies in this area—is to quantify planetary habitability beyond the mere question of detectability, our objective here is to estimate the total volume of the Galaxy that might host life. The goal is not only to infer how many planets of a certain type could exist, but also to assess the Galaxy’s overall quantitative capacity to sustain habitable niches. To this end, one approach is to assume a specific heat production and then, averaging over all possible orbital distances around a star, compute the cumulative volume of the einespheres that could arise on planets.

When the planet is close to the star, the increase in surface temperature shifts the beginning of the region compatible with liquid water to greater depths; depending on whether the surface remains within the aquability range or not, this warming can compress or expand the aquable zone from the upper layers. Conversely, at large orbital distances, surface cooling can bring the crustal temperature below the melting point of water, progressively reducing the extent of the aquable region or even eliminating it entirely. At both orbital extremes, the eirenesphere may tend toward zero.

We define the \textit{Eirenesphere Volume Index}, EVI of a planet with a given set of bulk ${\cal B}:\{M_p, R_p, q_s(t)\}$ and mineralogic ${\cal M}:\{f_{\mathrm{felsic}},X_\mathrm{Fe}, h\}$ properties  according to:

\begin{equation}
\mathrm{EVI}({\cal B,M}) = \frac{1}{\Delta a} \int_{a_{\mathrm{in}}}^{a_{\mathrm{out}}} 
V_\mathrm{ES}(a|{\cal B,M}) \, \mathrm{d}a,
\label{eq:I3D}
\end{equation}
where $a_{\mathrm{in}}$ and $a_{\mathrm{out}}$ correspond to the inner and outer edges of the subsurface stellar habitable zone and $\Delta a = a_{\mathrm{out}} - a_{\mathrm{in}}$ is the total width of this region.

In the numerical implementation, the integral is evaluated using the trapezoidal rule over a grid of $N$ orbital distances uniformly distributed in the interval $[a_{\mathrm{in}}, a_{\mathrm{out}}]$, computing $V_\mathrm{ES}(a)$ at each point via the procedure described in \autoref{eq:V_eff_discreta}.

The index $\mathrm{EVI}$ has a direct physical interpretation: it represents the average eirenesphere volume that the planet would maintain if it orbited at any location within the SSHZ. Hereafter, this volume will be expressed both in absolute units ($\mathrm{km}^3$) and normalized to the total volume of Earth’s oceans (1 TO $\approx 1.332 \times 10^9\,\mathrm{km}^3$, \citealt{Charette2010}), allowing a direct comparison of planets with different masses, compositions, and thermal regimes on a common scale.

A high value of $\mathrm{EVI}$ could indicate that the planet has an internal structure capable of sustaining porous, biologically viable aquifers under a wide range of insolation conditions, making it a robust astrobiological target even in the presence of orbital uncertainties. Conversely, zero or very low values may reflect: (i) lithostatic compaction intense enough to eliminate porosity before thermodynamically favorable conditions are reached, (ii) an insufficient heat flux to maintain liquid water in the subsurface, or (iii) surface temperatures incompatible with the stability of liquid water throughout the entire stellar habitable zone.

This diagnostic capability makes the $\mathrm{EVI}$ index a complementary—though not substitutive—tool to the classical criteria of surface habitability.

\section{Results}\label{sec:results}

\subsection{Thermal structure and internal profiles}\label{subsec:thermal_profiles}

\begin{figure*}
    \centering
    \includegraphics[width=\textwidth]{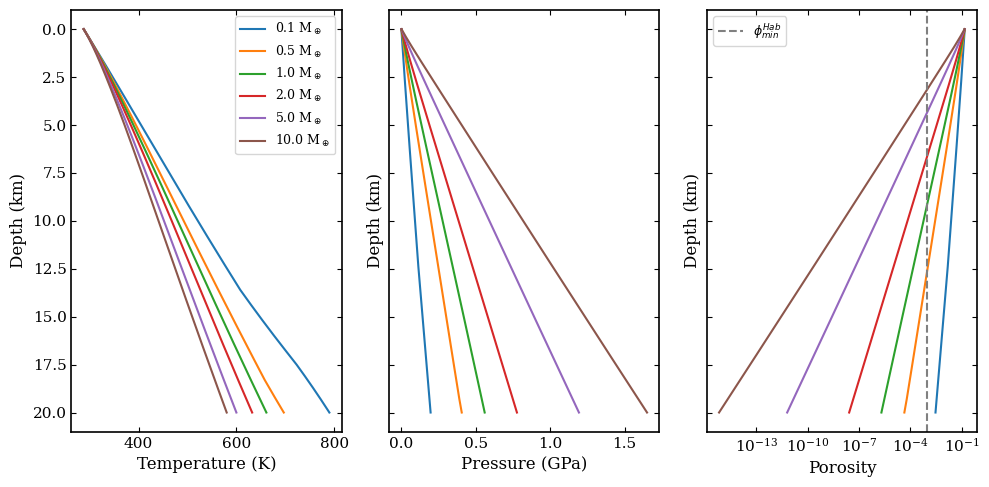}
    \caption{Internal profiles for rocky planets with masses from $0.1$ to $10\,M_\oplus$, assuming fixed terrestrial-like composition (HC2011) and surface heat flow $q_s = 65\,\mathrm{mW\,m^{-2}}$. Panels show temperature, pressure, and porosity versus depth. The vertical dashed line in the porosity panel marks the adopted minimum porosity for effective habitability ($\phi_{\rm min}^{\rm hab} = 10^{-3}$).}
    \label{fig:mass_profiles}
\end{figure*}

In order to isolate the effect of planetary gravity on thermal structure and crustal compaction, one-dimensional profiles of temperature, pressure, and porosity were calculated for rocky planets with masses between $0.1$ and $10\,M_\oplus$. In all cases, the same reference mineralogical composition, similar to that of Earth and based on the compilation of HC2011, was assumed. This composition will be used as a reference in most of the experiments and constitutes the basis of the results presented here. Likewise, a fixed surface heat flow of $q_s = 65\,\mathrm{mW\,m^{-2}}$, representative of mature continental provinces, was imposed in most of our test planets. Finally, the planetary radii corresponding to each mass were determined using the scaling relations described in \autoref{sec:model}. \autoref{fig:mass_profiles} shows the evolution of the internal profiles for the different planetary masses considered. 

As mass increases, the pressure gradient rises significantly due to the increase in surface gravity and in the average internal density. While a $0.1\,M_\oplus$ planet reaches pressures of only ${\sim}0.2\,\mathrm{GPa}$ at a depth of $20\,\mathrm{km}$, a $10\,M_\oplus$ planet exceeds ${\sim}1.5\,\mathrm{GPa}$ at the same depth. The increase in pressure produces a much more efficient compaction of crustal porosity. In low-mass planets, porosity remains relatively high even at depths of several kilometers, preserving a significant fraction of the porous volume that is potentially accessible to fluids. In contrast, more massive planets exhibit an extremely rapid collapse of porosity, reaching values close to zero within the first few kilometers of depth. This behavior directly reflects the exponential dependence of compaction on lithostatic pressure discussed in \autoref{subsec:porosity}.

Thermal differences between models are more moderate than those observed in pressure and porosity. Although all models were calculated with the same surface heat flow, variations in gravity and effective conductivity partially modify the geothermal gradient. Less massive planets develop slightly higher thermal gradients, associated with more persistent porosity that reduces the effective conductivity of the medium, whereas more massive planets exhibit relatively more moderate thermal profiles despite their higher internal pressures.

Taken together, these results show a significant decoupling between the thermal stability of liquid water and the preservation of porous space in the crust. Increasing mass promotes higher pressure and enlarges regions that are thermodynamically compatible with liquid water, but simultaneously accelerates the destruction of the available porous volume. This trade-off between internal heating and gravitational compaction is one of the fundamental mechanisms controlling subsurface habitability in rocky planets.

\begin{figure*}
    \centering
    \includegraphics[width=0.8\textwidth]{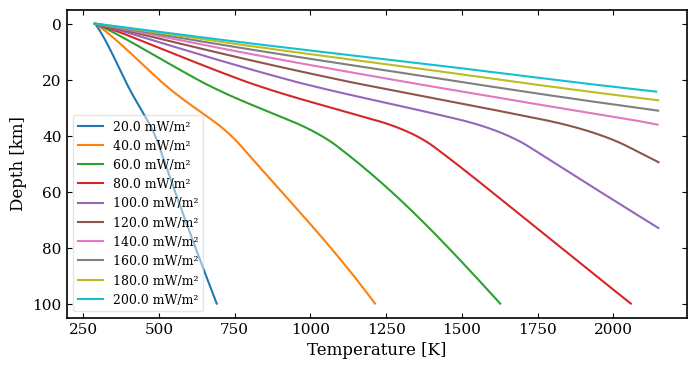}
    \caption{Geotherms for an Earth-like planet with fixed composition, mass, and radius were calculated while varying surface heat flow from $20$ to $200\,\mathrm{mW\,m^{-2}}$.}
    \label{fig:qs_profiles}
\end{figure*}

In order to explore the thermal sensitivity of the model, geotherms were calculated for a reference Earth-like planet, keeping the composition, mass, and planetary radius fixed, but varying the surface heat flow between $20$ and $200\,\mathrm{mW\,m^{-2}}$. This parameter directly controls the geothermal gradient and, consequently, the depth at which conditions compatible with liquid water can be reached. The results are shown in \autoref{fig:qs_profiles}.

Small variations in $q_s$ produce significant changes in the thermal structure of the crust. Models with low heat flow develop relatively cold geotherms, where temperatures compatible with liquid water are only reached at great depths. In contrast, high $q_s$ values generate much steeper thermal gradients, shifting the relevant isotherms toward shallower regions of the crust. For sufficiently high heat flows, the maximum temperatures compatible with habitability are reached at relatively small depths, considerably restricting the thickness of the potentially eirenesphere. Surface heat flow therefore acts as one of the most sensitive control parameters of the subsurface thermal structure: whereas planetary mass mainly regulates gravitational compaction and the preservation of porosity, $q_s$ determines the location and thickness of the regions that are thermally compatible with liquid water. 

It is precisely the interaction between these two effects—compaction and thermal regime—that defines the extent and geometry of the aquable and eirenespheres, whose analysis is presented in the following subsection.

\subsection{Aquability layer and eirenesphere}\label{layer_thickness}

Using the thermal and pressure profiles obtained in the previous subsection, we evaluated the thermodynamic stability of liquid water as a function of depth, employing the official IAPWS formulations (see e.g. \citealt{IAPWS1995, IAPWSRelease2009}). The \emph{aquable layer} is defined as the region of the lithosphere where the local pressure and temperature conditions allow the existence of liquid water. In turn, the \emph{eirenesphere} or eirenesphere corresponds to the subset of that region that also simultaneously satisfies the biological limits of temperature and pressure and the minimum porosity criterion adopted in this work (Equations~\ref{eq:aguabilidad} and \ref{eq:habitabilidad}).

\begin{figure*}
    \centering
    \includegraphics[width=0.95\textwidth]{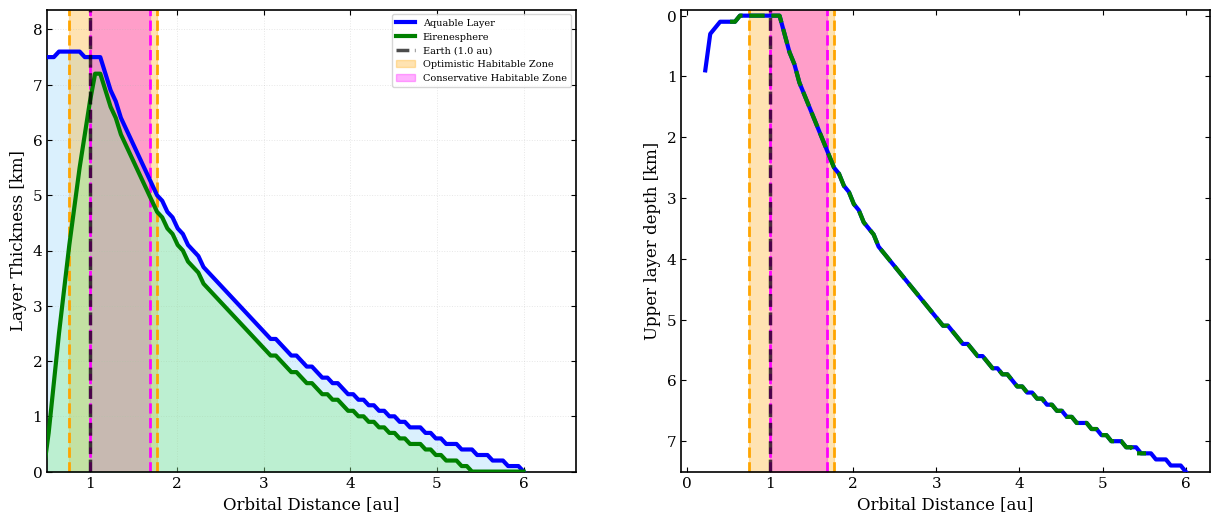}
    \caption{Thickness and depth of subsurface regions compatible with liquid water and effective habitability (eirenesphere) for an Earth-analog planet with Earth-like composition, based on HC2011 and a fixed surface heat flux of $q_s = 65\,\mathrm{mW\,m^{-2}}$. \textbf{Left:} thickness of the aquable layer (blue) and eirenesphere (green) versus orbital distance. Shaded areas mark the fraction of the aquable layer that meets both the thermal and porosity habitability criteria. \textbf{Right:} upper depth of the effective eirenesphere. Vertical dotted lines show the optimistic and conservative circumstellar habitable zone limits from \citet{Kopparapu2013, Kopparapu2014}.}
    \label{fig:habitable_fraction}
\end{figure*}

To explore the effect of surface conditions on the stability of subsurface liquid water, models were computed for an Earth-analog planet ($1\,M_\oplus$, $1\,R_\oplus$) using the reference composition of HC2011 and a fixed internal heat flux of 
$q_s = 65\,\mathrm{mW\,m^{-2}}$, while varying the orbital distance. \autoref{fig:habitable_fraction} shows the evolution of both the thickness of the aquable layer and the effective eirenesphere as a function of orbital distance around a Sun-like star.

The results show that the presence of subsurface liquid water persists over an extremely broad orbital range, extending from ${\sim}0.28\,\mathrm{au}$, where the surface temperature reaches ${\sim}547\,\mathrm{K}$ ($273.85\,^\circ\mathrm{C}$), out to ${\sim}5.94\,\mathrm{au}$, corresponding to equilibrium temperatures close to $118\,\mathrm{K}$ ($-155.15\,^\circ\mathrm{C}$). This interval far exceeds the classical limits of circumstellar habitability defined by \citet{Kasting1993} and refined by \citet{Kopparapu2013, Kopparapu2014}, highlighting the important thermal buffering effect of the planetary subsurface.

\autoref{fig:habitable_fraction}(a) shows that the thickness of the aquable layer reaches a maximum near ${\sim}1\,\mathrm{au}$ and progressively decreases toward both inner and outer orbits. It is noteworthy that this maximum coincides with Earth’s current orbital distance, suggesting that our planet is in a particularly favorable configuration for subsurface habitability under present geothermal conditions. In regions close to the star, the increase in surface temperature shifts the upper limit of liquid water stability to greater depths, significantly reducing the total available thickness. Conversely, in outer orbits, the low surface temperatures force the isotherms compatible with liquid water to be reached only at great depths, again restricting the aquable thickness.

The eirenesphere follows a similar trend, although it is systematically thinner due to the additional constraint imposed by the minimum porosity required for fluid circulation and storage. It is particularly noteworthy that the maximum eirenesphere thickness does not occur exactly at the current Earth orbit, but slightly beyond it, at around ${\sim}1.1\,\mathrm{au}$. Under these conditions, the combination of moderate surface temperatures and internal thermal gradients favors the maximum extent of potentially habitable subsurface environments.

The right panel of \autoref{fig:habitable_fraction} depicts how the upper boundary of the eirenesphere varies with orbital distance. At small orbital radii, elevated surface temperatures permit liquid water to exist essentially from the surface downward, whereas farther out the eirenesphere shifts progressively to deeper layers. Nonetheless, even beyond the classical habitable zone, subsurface domains capable of sustaining liquid water remain, maintained by internal geothermal heat. These findings indicate that subsurface habitable environments could endure on planets that are superficially inhospitable, whether extremely hot or globally glaciated.

\begin{figure*}
    \centering
    \includegraphics[width=0.6\textwidth]{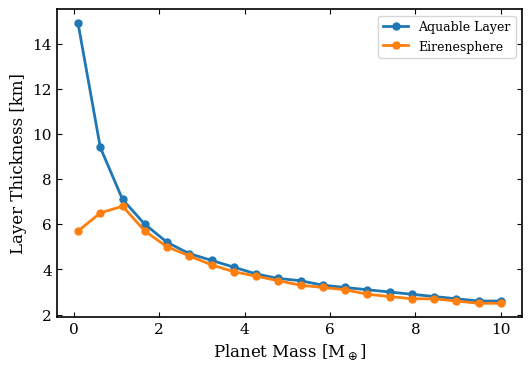}
    \caption{Thickness of the Aquable layer (blue) and eirenesphere (orange) versus planetary mass. All models assume an Earth-like composition (HC2011) and a fixed surface heat flow of $q_s = 65\,\mathrm{mW\,m^{-2}}$.}
    \label{fig:mass_habitable}
\end{figure*}

In addition to orbital dependence, we explored how planetary mass simultaneously modifies the thermal stability of liquid water and the preservation of cortical porous space. To this end, models between $0.1$ and $10\,M_\oplus$ were computed while keeping the reference mineralogical composition and the surface heat flow ($q_s = 65\,\mathrm{mW\,m^{-2}}$) fixed, whereas the radius, surface pressure, surface gravity, and radiogenic decay scale were scaled using the relationships described in \autoref{subsec:scaling}. The results are shown in \autoref{fig:mass_habitable}.

Both the thickness of the Aquable layer and that of the eirenesphere display a non-monotonic behavior with planetary mass, with a maximum around ${\sim}1.1\,M_\oplus$ controlled by different physical mechanisms on either side of that maximum.

On low-mass planets, gravitational compaction is weak and lithostatic pressure increases slowly with depth, so porosity is preserved to great depths and does not constitute the limiting factor. What constrains the eirenesphere in this regime is the steep geothermal gradient: because gravity is lower, the lithostatic pressure per unit depth is smaller, but the heat flux per unit area remains comparable to Earth’s, generating sharper thermal gradients that raise the temperature above the biological limit $T_{\rm bio}^{\rm max} = 423\,\mathrm{K}$ at relatively shallow depths. Thus, for example, for a planet of $0.1\,M_\oplus$ the aquable layer extends down to ${\sim}14.9\,\mathrm{km}$, but the eirenesphere is confined to only ${\sim}5.7\,\mathrm{km}$—just $38\,\%$ of the aquable zone—precisely because the biological thermal limit is reached much earlier than the pressure limit.

In contrast, on massive planets temperature is not the limiting factor: the geothermal gradient is more moderate and the aquable, eirenesphere extends to comparable depths. However, the high lithostatic pressure causes an accelerated collapse of porosity, eliminating the interstitial space required for fluid circulation before thermal conditions become unfavorable. In this regime it is porosity, and not temperature, that sets the limit of the deep biosphere.

The maximum eirenesphere thickness appears on planets close to Earth’s mass, where a balance between both mechanisms is achieved: the geothermal gradient is sufficiently gentle to keep temperatures below $T_{\rm bio}^{\rm max}$ down to several tens of kilometers in depth, and the lithostatic pressure is not yet high enough to completely collapse the porous network. It is noteworthy that this maximum roughly coincides with the mass of our own planet, suggesting that the Earth lies in a particularly favorable mass configuration for the development of extensive subsurface biospheres.

\bigskip
The results presented so far correspond to a static thermal state of the planet. However, the planetary interior is not stationary: secular cooling of the mantle progressively reduces the surface heat flux over geological time, modifying the position and thickness of the aquable and eirenespheres. The following subsection analyzes how this temporal evolution affects the extent of subsurface reservoirs over the lifetime of the planet.

\subsection{Temporal evolution of subsurface habitability}\label{subsec:temp_hab}

\begin{figure*}
    \centering
    \includegraphics[width=0.95\textwidth]{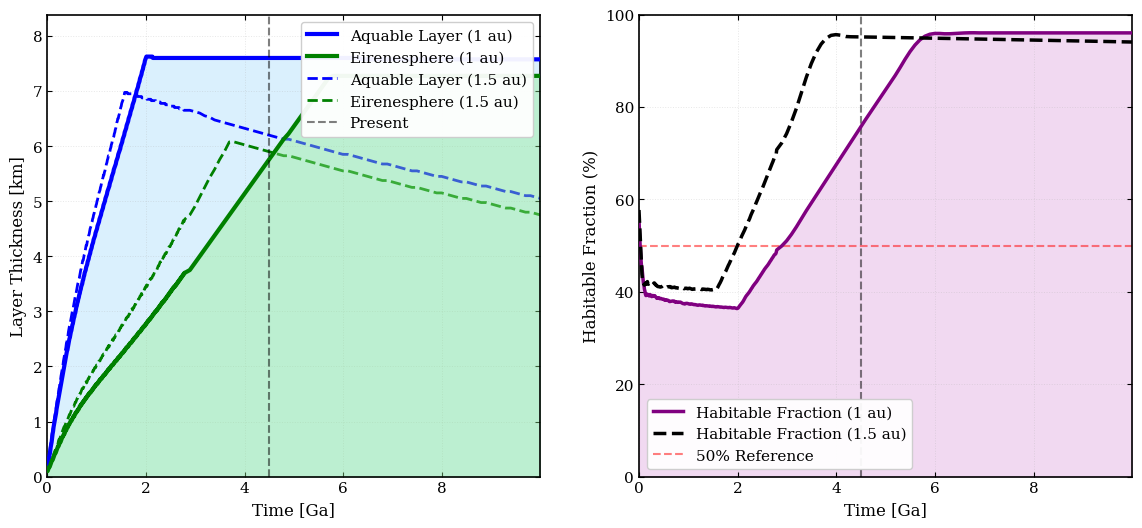}
    \caption{Temporal evolution of subsurface layers for an Earth-analog planet with HC2011-based mineralogy at two orbital distances: $1\,\mathrm{au}$ (solid) and $1.5\,\mathrm{au}$ (dashed). \textbf{Left:} thickness of the aquable layer (blue) and eirenesphere (green). \textbf{Right:} evolution of the eirenesphere, i.e. the portion of the aquable layer that meets the adopted thermal and porosity criteria. The horizontal red dashed line marks the $50\,\%$ threshold; the vertical dotted line marks Earth’s current age (${\sim}4.5\,\mathrm{Ga}$).}
    \label{fig:time_evolution}
\end{figure*}

In addition to the dependence on planetary mass and orbital distance, the extent of the eirenesphere depends strongly on the planet’s secular thermal evolution. In order to explore this temporal behavior, we computed the geothermal evolution of an Earth-analog planet ($1\,M_\oplus$, $1\,R_\oplus$) with a reference mineralogical composition, located at two representative orbital distances: $1\,\mathrm{au}$, corresponding to Earth’s current position, and $1.5\,\mathrm{au}$, within the conservative circumstellar habitable zone but with a significantly lower surface temperature. The temporal evolution of the internal heat flow was obtained from thermal evolution tracks derived from grids of terrestrial planets presented by \citet{Zuluaga2013}. The results are shown in \autoref{fig:time_evolution}.

In the earliest stages, the high geothermal heat flux generates extremely steep thermal gradients, causing the critical temperatures for the stability of liquid water and for biological habitability to be reached at very shallow depths. As a consequence, in both scenarios, the initial thickness of the aquable layer is identical (${\sim}0.17\,\mathrm{km}$), while that of the eirenesphere is only ${\sim}0.10\,\mathrm{km}$, with an initial habitable fraction of ${\sim}57\,\%$. These values reflect a thermal regime dominated by high primordial radiogenic heat production, whose effect on the geothermal gradient is independent of orbital distance.

As the planet evolves and the internal heat flux decreases, the geothermal gradient progressively relaxes and both layers expand. However, the evolutionary path differs significantly between the two orbital scenarios.

For the $1\,\mathrm{au}$ case, the aquable layer reaches a maximum thickness of ${\sim}7.62\,\mathrm{km}$ at around $2.0\,\mathrm{Ga}$, subsequently stabilizing near ${\sim}7.58\,\mathrm{km}$ at the end of the evolution, with a total change of $+7.40\,\mathrm{km}$ relative to the initial state. The eirenesphere continues to expand over a longer timespan, reaching its maximum of ${\sim}7.28\,\mathrm{km}$ near $5.8\,\mathrm{Ga}$ and remaining at that value until the end of the calculation. The fraction of the aquable layer represented by the eirenesphere increases steadily from the initial $57\,\%$ to ${\sim}96\,\%$ at late stages, indicating that in thermally mature planets the biological temperature limit ceases to be the dominant constraint and virtually the entire aquable layer becomes potentially habitable.

The scenario at $1.5\,\mathrm{au}$ follows a qualitatively similar but quantitatively different evolution. The aquable layer reaches a maximum of ${\sim}6.97\,\mathrm{km}$ at ${\sim}1.57\,\mathrm{Ga}$—earlier and with a smaller peak thickness than at 
$1\,\mathrm{au}$—and then decreases until stabilizing at ${\sim}5.05\,\mathrm{km}$ at the end of the evolution, with a total change of only $+4.88\,\mathrm{km}$. This reduction in the final thickness compared to the $1\,\mathrm{au}$ case reflects the effect of the lower surface temperature: because the upper crust is colder, the aquable zone extends from shallower depths, but the gentler geothermal gradient causes the lower boundary of the zone—where the temperature exceeds the boiling point of water at the local pressure—to be reached earlier. The eirenesphere shows analogous behavior, with a maximum of ${\sim}6.08\,\mathrm{km}$ at ${\sim}3.69\,\mathrm{Ga}$ and a final value of ${\sim}4.75\,\mathrm{km}$. The habitable fraction converges toward ${\sim}94\,\%$, slightly lower than in the $1\,\mathrm{au}$ case, consistent with a aquable layer that at late stages extends to depths where porosity has been partially compacted by lithostatic pressure.

The vertical dashed line in \autoref{fig:time_evolution} marks the current age of Earth (${\sim}4.5\,\mathrm{Ga}$). At this time, the planet at $1\,\mathrm{au}$ has already undergone most of its subsurface expansion and is close 
to its state of habitable saturation, whereas the $1.5\,\mathrm{au}$ scenario is still in an active expansion phase of the eirenesphere.

Taken together, these results show that internal habitability is not a static property, but an evolutionary feature controlled by the secular cooling 
of the planetary interior. Young planets exhibit subsurfaces that are excessively hot and restricted, whereas thermally mature worlds develop subsurface reservoirs that are substantially more extensive and biologically accessible. The comparison between the two orbital distances further illustrates that planets farther from their star do not necessarily develop more extensive subsurface biospheres: the lower surface temperature at $1.5\,\mathrm{au}$ does not offset the reduction in the final thickness of the aquable layer relative to the terrestrial case, so that the optimal orbital position for subsurface habitability depends on the balance among surface temperature, geothermal gradient, and the temporal evolution of the internal heat flux.

\bigskip
The following subsection explores how compositional and mineralogical variations modify the thermophysical properties of the lithosphere and, consequently, the extent of subsurface environments compatible with liquid water and effective habitability.

\begin{figure*}
    \centering
    \includegraphics[width=0.95\textwidth]{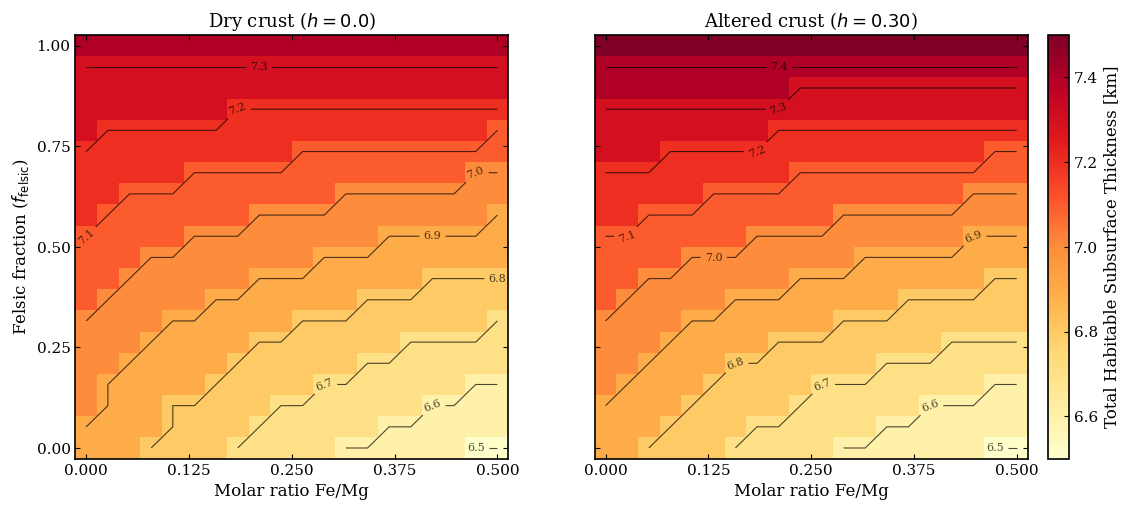}
    \caption{Thickness of the eirenesphere as a function of felsic fraction ($f_{\mathrm{felsic}}$) and Fe/Mg molar ratio for an Earth-analog planet ($1\,M_\oplus$, $1\, R_\oplus$, $q_s = 65\,\mathrm{mW\,m^{-2}}$, $1\,\mathrm{au}$), computed with the parametric mineralogical model in \autoref{subsec:parametric_mineralogy}. The left panel shows a dry crust ($h = 0.0$) and the right panel a hydrated, altered crust ($h = 0.30$).}
    \label{fig:minerology}
\end{figure*}

\subsection{Impact of mineralogical diversity on the extent of the deep biosphere}\label{div_minero}

With the aim of assessing how the chemical composition of the lithosphere modifies the extent of habitable subsurface environments, we explored the parameter space defined by the mineralogical model presented in \autoref{subsec:parametric_mineralogy}: the felsic fraction ($f_{\mathrm{felsic}}$), the Fe/Mg molar ratio, and the degree of aqueous alteration of the crust ($h$). For this analysis, the bulk properties of the planet were kept constant ($1\,M_\oplus$, $1\,R_\oplus$, $q_s = 65\,\mathrm{mW\,m^{-2}}$, $1\,\mathrm{au}$), varying only the effective mineralogical composition of the crust and upper mantle. The results are presented in \autoref{fig:minerology}.

The dominant trend observed in both panels of ~\autoref{fig:minerology} corresponds to the strong dependence of eirenesphere thickness on the felsic fraction of the crust. The contours show a predominantly horizontal stratification, indicating that $f_{\mathrm{felsic}}$ exerts the main control on the vertical extent of the deep biosphere. Compositions enriched in felsic minerals reach maximum eirenesphere thicknesses of ${\sim}7.3$–$7.5\,\mathrm{km}$, whereas purely mafic or ultramafic crusts exhibit considerably smaller thicknesses, on the order of ${\sim}6.5$–$6.7\,\mathrm{km}$.

This difference arises primarily from variations in the effective thermophysical properties of the mineral matrix. Mafic and ultramafic crusts, dominated by olivines and pyroxenes, have higher effective thermal conductivities than felsic crusts (see e.g. \citealt{Hofmeister1999, Hasterok2011, Hasterok2022}), promoting more efficient transport of internal heat. As a consequence, for the same surface geothermal flux, the geothermal gradient becomes steeper and the critical isotherms associated with the biological limit ($T_{\mathrm{bio}}^{\mathrm{max}} \simeq 423\,\mathrm{K}$) are reached at shallower depths. In contrast, crusts enriched in felsic minerals—particularly quartz and feldspars—act as more thermally insulating media, shifting the habitable thermal boundary to deeper regions and allowing the development of more extensive subsurface biospheres (see e.g. \citealt{Hofmeister1999, Hasterok2011}).

The effect of the Fe/Mg ratio is comparatively secondary with respect to that of $f_{\mathrm{felsic}}$, although not negligible. As can be seen from the diagonal tilt of the contours in \autoref{fig:minerology}, an increase in iron content introduces a systematic reduction in eirenesphere thickness over the entire range of $f_{\mathrm{felsic}}$ explored. This effect is consistent with the increase in effective density of ferromagnesian silicates and with the higher thermal conductivity of iron-rich phases compared to their magnesium-rich analogues (see e.g. \citealt{Hofmeister1999, Unterborn2023}). Iron-enriched compositions therefore tend to produce slightly steeper geothermal gradients and a somewhat more efficient compression of the crustal structure, moderately reducing the thickness of the eirenesphere.

Crustal hydration introduces an additional effect that becomes apparent when comparing both panels of \autoref{fig:minerology}. The altered scenario ($h = 0.30$) exhibits slightly greater eirenesphere thicknesses relative to the dry case ($h = 0.0$), particularly for intermediate felsic compositions. The incorporation of hydrated minerals — hornblende and phlogopite, formed at the expense of dry pyroxenes and feldspars according to the retrograde metamorphic reactions described in \autoref{subsec:parametric_mineralogy} — locally reduces the effective thermal conductivity of the lithosphere (see e.g. \citealt{Yardley2009, Winter2014}), promoting less abrupt thermal gradients and shifting the critical biological isotherm to greater depths. However, this effect remains subordinate to the dominant control exerted by the primary silicate composition.

Taken together, these results show that mineralogy is a first-order parameter in determining subsurface habitability. The initial chemistry of the planet not only controls the structure and density of the crust, but also the efficiency with which internal heat is transported to the surface. Consequently, worlds with more felsic and thermally insulating compositions could sustain more extensive deep biospheres than planets dominated by mafic or ultramafic lithologies, even under similar orbital and geothermal conditions.

\bigskip

The following subsection explores the global sensitivity of subsurface habitability to simultaneous variations in orbital conditions and internal heat flux, allowing us to identify the regions of parameter space where deep biospheres reach their maximum extent.

\subsection{Parametric sensitivity of aquability and subsurface habitability}

\begin{figure}
    \centering
    \includegraphics[width=0.6\textwidth]{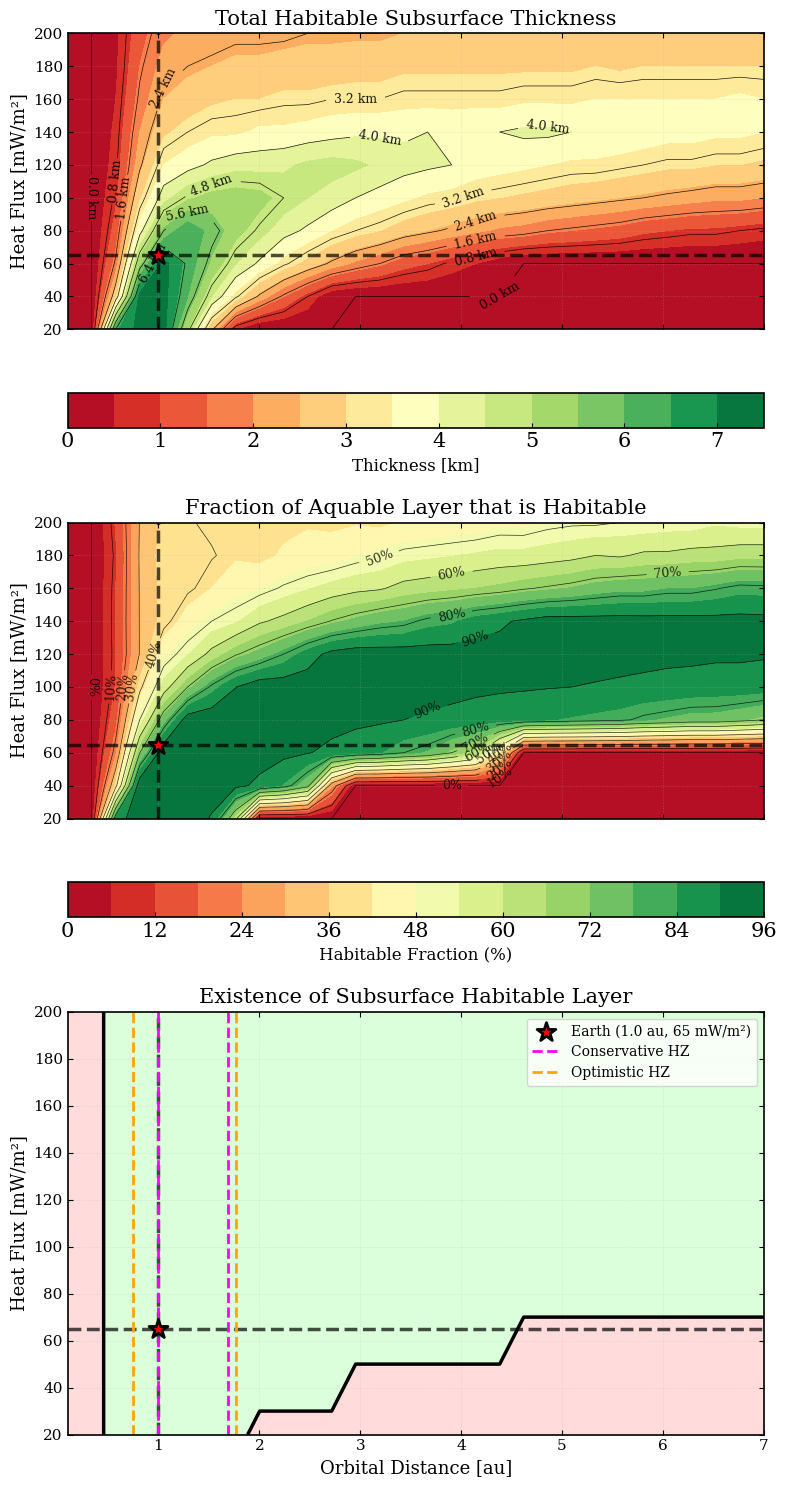}
    \caption{Parametric sensitivity of subsurface habitability versus orbital distance and surface heat flux for an Earth analog ($1\,M_\oplus$, $1\,R_\oplus$) with Earth-like composition from HC2011. (\emph{Left}) Total habitable subsurface thickness (km). (\emph{Center}) Fraction of the aquifer remaining within biological temperature and pressure limits. (\emph{Right}) Parameter regions with (green) and without (pink) a subsurface biosphere. Vertical dashed lines show circumstellar habitable zone limits from \citet{Kopparapu2013, Kopparapu2014}: yellow for the optimistic limit (\textit{Recent Venus -- Early Mars}) and magenta for the conservative limit (\textit{Moist Greenhouse -- Maximum Greenhouse}). The red star marks present Earth conditions ($d = 1.0\,\mathrm{au}$, $q_s = 65\,\mathrm{mW\,m^{-2}}$).}
    \label{fig:parametric_analysis}
\end{figure}

To evaluate the robustness of aquability and internal habitability against joint variations in the planet’s astronomical and geophysical conditions, a parametric analysis was carried out, simultaneously exploring the phase space defined by the orbital average distance $a$ and the surface heat flux $q_s$. For each combination of parameters, the total thickness of the eirenesphere, the biologically viable fraction of the deep aquifer reservoir, and the existence or absence of a stable subsurface biosphere were determined. \autoref{fig:parametric_analysis} summarizes this analysis by means of three complementary two‑dimensional maps. The left panel shows the total thickness of the subsurface eirenesphere; the central panel represents the fraction of the aquable layer that simultaneously satisfies the biological limits of temperature and pressure (\autoref{eq:habitabilidad}); finally, the right panel delineates the regions of parameter space where an eirenesphere does or does not exist. For comparison, the limits of the classical circumstellar habitable zone from \citet{Kopparapu2013, Kopparapu2014} and the current terrestrial scenario are included as references.

The results show that subsurface habitability constitutes a geophysically robust phenomenon across a wide range of planetary conditions. The maximum thickness of the eirenesphere reaches values of ${\sim}7.6\,\mathrm{km}$ in scenarios characterized by moderate to low heat fluxes ($q_s \lesssim 40\,\mathrm{mW\,m^{-2}}$), where the geothermal gradient is sufficiently gentle to keep temperatures below the biological sterilization threshold ($T_{\rm bio}^{\rm max} \simeq 423\,\mathrm{K}$) down to greater depths. In contrast, as the heat flux increases, the biologically critical isotherms rise toward the surface, progressively reducing the thickness available for the deep biosphere.

The topology of the left panel further reveals that the dependence on orbital distance is highly nonlinear. In regions close to the star ($d \lesssim 1.5\,\mathrm{au}$), the increase in surface temperature initially favors the presence of subsurface liquid water and expands the vertical extent of the biosphere. However, at extremely small distances, the surface thermal contribution can shift the system toward partially sterilizing regimes, locally decreasing the effective habitable fraction.

The central panel shows that the biological fraction of the aquable reservoir depends mainly on the balance between surface and internal heating. For moderate heat fluxes ($q_s \sim 50$–$120\,\mathrm{mW\,m^{-2}}$), a large proportion of the aquable layer remains within biologically permissible conditions, reaching habitable fractions above $90\,\%$. In contrast, scenarios dominated by low geothermal heating exhibit an abrupt reduction in this fraction, especially at large orbital distances where the cold surface forces the liquid layer to exist only at greater depths.

Particularly noteworthy is the result shown in the right panel: the existence of a subsurface biosphere is not strictly confined to the classical circumstellar habitable zone \citep{Kasting1993, Kopparapu2013}. Even beyond the outer edge of the optimistic habitable zone, extensive regions persist where it is possible to maintain habitable environments beneath the surface thanks to endogenous heating. This behavior implies that a planet’s internal energy can partly decouple habitability from the surface radiative balance, in line with what has been proposed in previous work on subsurface habitability in frozen worlds (see e.g. \citealt{McMahon2013, Lingam2018, Escudero2023}).

In this regime, heat generated by radiogenic processes dominates the thermodynamics of the lithosphere, allowing the formation of deep aquifers beneath completely frozen surfaces. As a consequence, cryogenic worlds that appear inhospitable from an atmospheric perspective may harbor biologically viable niches several kilometers below the surface. This result significantly extends the potentially eirenesphere of the galaxy and reinforces the idea that planetary habitability must be understood as a three-dimensional phenomenon simultaneously controlled by internal thermal structure and surface dynamics.

\bigskip

The strong dependence of subsurface habitability on the thermophysical properties of the crust suggests that rocky exoplanets cannot be classified, in terms of their potential to host life-compatible conditions, solely on the basis of their orbital position with respect to the host star. As shown in the previous subsection, mineralogy controls the efficiency of heat transport and, consequently, the depth at which the requirements of liquid water, open porosity, and biological stability are simultaneously met. However, these internal properties interact nonlinearly with the external astronomical conditions and with the planet’s endogenous heat flux. Motivated by this interdependence, the following subsection constructs a map of physical regimes of subsurface habitability, simultaneously exploring orbital distance and surface heat flux as control parameters.

\begin{figure*}
    \centering
    \includegraphics[width=0.7\textwidth]{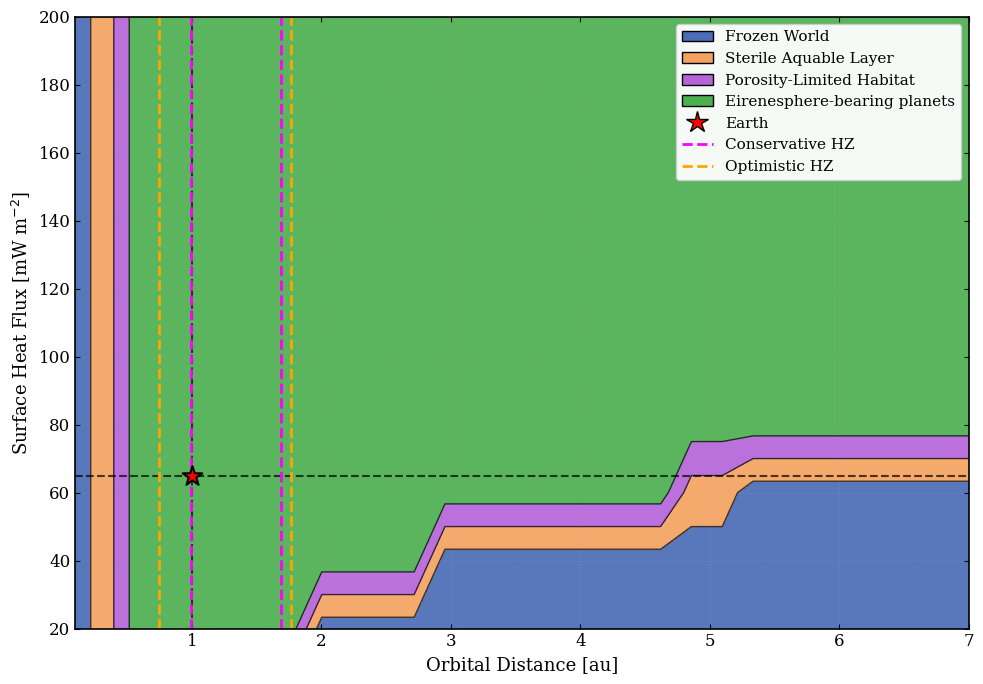}
    \caption{Map of subsurface habitability regimes as a function of orbital distance and surface heat flux for a terrestrial planet with HC2011 crustal composition. Colored regions mark four lithospheric thermodynamic states: completely frozen worlds (\textit{Frozen World}, blue), overheated sterile aquifers (\textit{Sterile Aquable Layer}, orange), habitats limited by porosity collapse (\textit{Porosity-Limited Habitat}, purple), and eirenesphere-bearing planets (\textit{Habitable Subsurface}, green). Dashed lines show circumstellar habitable zone limits from \citet{Kopparapu2013, Kopparapu2014}: the yellow line is the optimistic limit (\textit{Recent Venus -- Early Mars}) and the magenta line the conservative limit (\textit{Moist Greenhouse -- Maximum Greenhouse}). The red star marks present Earth conditions ($d = 1.0\,\mathrm{au}$, $q_s = 65\,\mathrm{mW\,m^{-2}}$).}
    \label{fig:habitability_regimes}
\end{figure*}

\subsection{Physical regimes of subsurface habitability}

With the aim of identifying the different thermodynamic states that can emerge in rocky exoplanets, a physical classification was carried out of the parameter space defined by the orbital distance $d$ and the surface heat flux $q_s$. Unlike the continuous thickness and depth maps discussed previously, this analysis makes it possible to directly distinguish between different \emph{astrobiological regimes}, defined by the simultaneous presence or absence of liquid water, biologically viable temperatures, and permeable porosity. Below we present the regimes that will be used:

\begin{itemize}
    \item \textbf{Frozen World:} completely frozen worlds where the temperature remains below the melting point of water throughout the entire crustal column, preventing the existence of subsurface liquid water.

    \item \textbf{Sterile Aquable Layer:} scenarios in which thermodynamically stable liquid water exists, but temperatures exceed the adopted biological limit ($T_{\rm bio}^{\rm max} \simeq 423\,\mathrm{K}$, \citealt{Kashefi2003, Bains2015}), preventing the persistence of complex biomolecules and active metabolism.

    \item \textbf{Porosity-Limited Habitat:} environments where liquid water and life-compatible temperatures coexist, but the lithostatic pressure causes the collapse of the porous network below the threshold $\phi_{\rm min}^{\rm hab} = 10^{-3}$, preventing the fluid circulation necessary to sustain a subterranean biosphere (see e.g. \citealt{Parnell2016, Magnabosco2018}).

    \item \textbf{Eirenesphere-bearing planets:} regimes in which liquid water, biologically tolerable temperatures, and sufficient porosity to sustain fluid circulation and microbial habitats all converge simultaneously.
\end{itemize}

\autoref{fig:habitability_regimes} summarizes this classification for a planet with Earth-like mass and radius, and a reference crustal composition. Each region in the diagram corresponds to a different physical state of the subsurface lithosphere:

The most notable result from \autoref{fig:habitability_regimes} is that the \textit{eirenesphere-bearing} regime dominates most of the explored parameter space, extending well beyond the limits of the classical circumstellar habitable zone. For heat fluxes above ${\sim}30\,\mathrm{mW\,m^{-2}}$, subsurface habitability persists out to orbital distances of ${\sim}5$--$7\,\mathrm{au}$, far beyond the optimistic outer edge of \citet{Kopparapu2013}. This result quantitatively reinforces the conclusion that planetary internal heat can decouple subsurface habitability from the surface radiative balance (see e.g. \citealt{McMahon2013, Lingam2018, Escudero2023}).

The transition between regimes reveals the hierarchy of limiting mechanisms. At the inner edge of the system ($d \lesssim 1\,\mathrm{au}$), intense stellar irradiation raises crustal temperatures to the point where the aquable layer becomes thermally sterile, establishing the \textit{Sterile Aquable Layer} regime. At large orbital distances and low heat fluxes ($q_s \lesssim 40\,\mathrm{mW\,m^{-2}}$), the drop in surface temperature combined with insufficient internal heating prevents the cryosphere from melting, generating the \textit{Frozen World} regime. Between these two extremes, the \textit{Porosity-Limited Habitat} regime appears as a narrow transition band, indicating that porosity loss due to lithostatic compaction acts as a secondary mechanism that only becomes dominant under specific conditions of intermediate heat flux and moderate-to-large orbital distance.

Notably, Earth’s current position lies firmly within the \textit{Eirenesphere-bearing} regime, consistent with the documented existence of a deep continental biosphere extending down to several kilometers (see e.g. \citealt{Orcutt2013, Magnabosco2018, lloyd2025intraterrestrials}). This result provides a qualitative validation of the model, as it correctly reproduces the astrobiological state of the only planet known to host life.

\bigskip
The regime map qualitatively establishes the conditions under which subsurface habitability is possible. The next step is to quantify this habitability through a single metric that simultaneously integrates the eirenesphere and the orbital position of the planet. In the following subsection we present the results of the $\mathrm{EVI}$ calculated for the full set of models explored.

\subsection{Volumes of the aquable layer and eirenespheres}

\begin{figure*}
    \centering
    \includegraphics[width=0.6\textwidth]{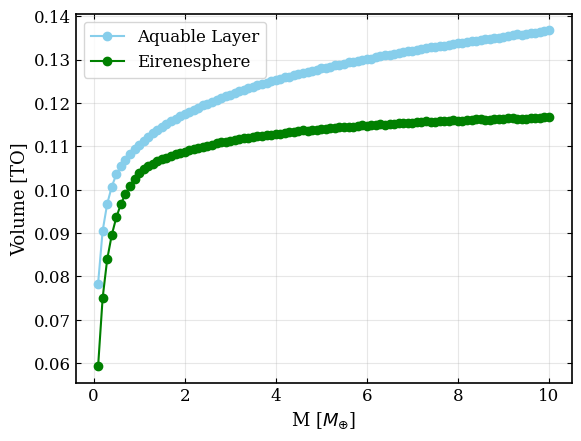}
    \caption{Effective aquable (light blue) and habitable (dark green) volumes versus planetary mass for $d = 1.0\,\mathrm{au}$, $q_s = 65\,\mathrm{mW\,m^{-2}}$, and the HC2011 reference composition.}
    \label{fig:volume_mass}
\end{figure*}

Having described the aquable and eirenespheres in terms of their depth, we next quantify the three-dimensional extent of these internal reservoirs to evaluate the planet’s biological carrying capacity. In contrast to surface habitability models—which are constrained to a two-dimensional framework—our method incorporates both the geometric volume $V_{\rm geom}$ and the effective volume $V_{\rm eff}$, the latter being weighted by the crustal porosity profile specified in \autoref{subsec:porosity}. To cast these volumes in astrobiologically relevant terms, we normalize them by the total volume of Earth’s modern surface oceans, or terrestrial oceans, TO~$= 1.332\times10^9\,\mathrm{km^3}$ \citep{Charette2010}.

For a terrestrial analog ($d = 1.0\,\mathrm{au}$, 
$q_s = 65\,\mathrm{mW\,m^{-2}}$) with the reference-model mineralogy, the model predicts a water-bearing layer with a thickness of ${\sim}7.5\,\mathrm{km}$, of which ${\sim}6.8\,\mathrm{km}$ is compatible with the thermal biological limits defined in \autoref{sec:criteria}. When these thicknesses are integrated globally under spherical symmetry, the total geometric volume of the water-bearing zone is $V_{\rm aq} \approx 2.88\,\mathrm{TO}$, whereas the volume of the eirenesphere is $V_{\rm hab} \approx 2.61\,\mathrm{TO}$. These values highlight that the physical space theoretically available for life in the subsurface far exceeds Earth’s surface water inventory.

Nevertheless, the biologically relevant metric is the water that is actually accessible within the porous medium. When incorporating the degradation of porosity with pressure (~\autoref{eq:porosity_pressure}), the effective volumes are reduced by approximately a factor of 26: 
$V_\mathrm{SO} \approx 0.11\,\mathrm{TO}$ and 
$V_\mathrm{ES} \approx 0.10\,\mathrm{TO}$. Despite representing only ${\sim}10\,\%$ of a surface ocean, ${\sim}91\,\%$ of the effective subsurface fluid lies within the eirenesphere. This result indicates that, under the thermal conditions of modern Earth, porosity—rather than temperature—is the main limiting factor for the extent of the deep biosphere, in agreement with estimates for Earth’s continental deep biosphere (see e.g. \citealt{Magnabosco2018, lloyd2025intraterrestrials}).

In order to assess how these quantities vary with planetary mass, we compute the aquable and eirenesphere volumes for rocky planets with masses between $0.1$ and $10\,M_\oplus$, keeping fixed the orbital distance ($1\,\mathrm{au}$), the surface heat flux ($q_s = 65\,\mathrm{mW\,m^{-2}}$), and the reference mineralogical composition. The results are shown in \autoref{fig:volume_mass}. We see that volumes exhibit a steep increase for masses below ${\sim}1\,M_\oplus$ and gradually approach saturation values at higher masses. However, the gap between the aquable and eirenespheres widens progressively from ${\sim}1\,M_\oplus$ onward: while the aquable volume continues to grow slowly toward ${\sim}0.135\,\mathrm{TO}$ at $10\,M_\oplus$, the eirenesphere levels off at around ${\sim}0.115\,\mathrm{TO}$. This divergence shows that, in massive planets, gravitational compaction reduces the available pore space more rapidly than the thermodynamically habitable space, so that an increasingly large fraction of the aquable layer lacks sufficient porosity to sustain fluid circulation and active metabolism. In this regime, porosity becomes the dominant limiting factor for the deep biosphere, surpassing thermal criteria.

Taken together, these results show that volumetric subsurface habitability does not scale linearly with planetary mass: more massive planets may display larger eirenespheres in absolute terms, but the reduction in effective porosity progressively limits the growth of the biologically accessible volume. This trade-off between geometric volume and effective pore space is one of the main conclusions of the model and has direct implications for the prioritization of astrobiological targets among rocky exoplanets of different masses.

\subsection{The EVI for rocky planets of different masses}
\label{sec:evi_results}

Once the eirenesphere has been measured for an Earth-analog planet, the next step is to explore how this metric varies across planetary parameter space. Following the formulation introduced in \autoref{sec:index}, we calculate the EVI on a two-dimensional grid defined by planetary mass $M$ and surface heat flux $q_s$, keeping fixed the structural and thermodynamic parameterizations described in \autoref{subsec:scaling} and the reference mineralogy. The index was calculated by integrating the eirenesphere across the conservative circumstellar habitable zone for a Sun-like star, as defined by \citet{Kopparapu2013, Kopparapu2014} via \autoref{eq:I3D}.

\begin{figure*}
\centering
\includegraphics[width=0.82\textwidth]{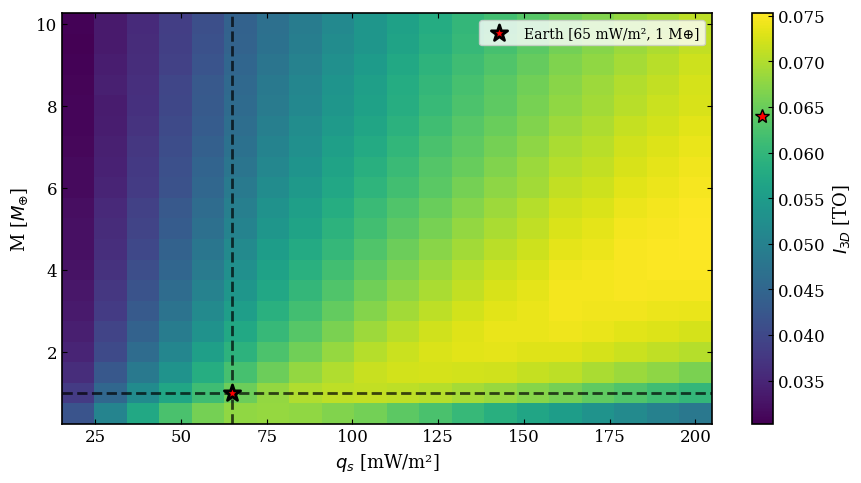}
\caption{Eirenesphere Volumetric Index ($\mathrm{EVI}$) as a function of planetary mass $M$ and surface heat flux $q_s$, given in units of the total volume of Earth’s present‑day oceans (\textit{TO}). The crustal mineralogy follows HC2011, and the index is integrated over the conservative circumstellar habitable zone of \citet{Kopparapu2013, Kopparapu2014}. The star marks Earth’s present conditions 
($M = 1\,M_\oplus$, $q_s = 65\,\mathrm{mW\,m^{-2}}$), which correspond to moderate internal habitability.}
\label{fig:hab3d_mass_qs}
\end{figure*}

\autoref{fig:hab3d_mass_qs} shows the distribution of $\mathrm{EVI}$ in the $(M, q_s)$ space. The most immediate result is that the index increases steadily with both parameters over the entire explored range, with no evidence of saturation within the limits of the calculation. This behavior reflects two independent effects that act in the same direction.

On the one hand, increasing planetary mass enlarges the total geometric volume of the lithosphere and, with it, the available space for deep reservoirs. Although gravitational compaction reduces the effective porosity in massive planets, this effect does not become dominant within the explored mass range ($0.1$–$10\,M_\oplus$), so the increase in geometric volume outweighs the porosity loss and the index continues to grow. This implies that, within this range, moderate super‑Earths host potentially more extensive subsurface biospheres than Earth‑analog planets, a result consistent with previous studies on the habitability of massive worlds (see e.g. \citealt{Noack2017}).

On the other hand, the surface heat flux $q_s$ directly controls the thickness of the water‑bearing zone: a higher geothermal flux generates steeper thermal gradients that shift the isotherms of water stability toward more accessible depths, expanding the aquable layer and, with it, the eirenesphere. Note that, although very high fluxes could in principle exceed the biological limit $T_{\rm bio}^{\rm max}$ at shallow depths, the explored range ($20$–$200\,\mathrm{mW\,m^{-2}}$) does not systematically reach that regime, which explains the increasing trend observed across the entire map.

Earth’s current position ($1\,M_\oplus$, $q_s = 65\,\mathrm{mW\,m^{-2}}$), marked with a star in \autoref{fig:hab3d_mass_qs}, lies in the region of intermediate index values, with $\mathrm{EVI}\approx 0.064$~TO. This result indicates that, under its present conditions, Earth may not provide the most optimal, but nonetheless a highly favorable scenario for subsurface habitability, while more massive planets or those with greater geothermal activity could host substantially more extensive deep biospheres. However, Earth’s position on this map also shows that our planet resides in a regime of moderate but robust internal habitability, consistent with the documented existence of an active deep continental biosphere (see e.g. \citealt{Orcutt2013, Magnabosco2018, lloyd2025intraterrestrials}).

Taken together, these results indicate that subsurface habitability is not restricted to Earth‑analog planets, but can be significantly greater in more massive or more geothermally active worlds. The $\mathrm{EVI}$ index thus makes it possible to extend the traditional concept of habitability toward a volumetric description of the planetary interior, incorporating the thermal and porous structure of the lithosphere as fundamental factors in assessing the biological potential of rocky exoplanets.

\subsection{The Galactic Volumetric Capacity for Subsurface Biospheres}
\label{sec:galactic_volume}

To fully grasp the astrobiological implications of subsurface habitability, it is instructive to extrapolate our volumetric findings to a galactic scale. Traditional surface-based habitability models estimate the number of potentially habitable worlds using the occurrence rate of rocky planets in the circumstellar habitable zone, $\eta_\oplus$. However, our results demonstrate that the Subsurface Stellar Habitable Zone (SSHZ) extends significantly further---out to $\sim 5-7$ au for planets with moderate geothermal heat fluxes \citep[see also][]{Lingam2018}. 

We can estimate the total volume of biologically viable subsurface environments in the Milky Way, $V_{\text{gal}}$, around solar-type (G-type) stars alone using the following relation:
\begin{equation}
    V_{\text{gal}} \approx N_{\text{G}} \cdot \eta_{\text{SSHZ}} \cdot \langle \mathrm{EVI} \rangle
\end{equation}
where $N_{\text{G}}$ is the total number of G-type dwarf stars in the Galaxy, $\eta_{\text{SSHZ}}$ is the occurrence rate of rocky planets (Earths and super-Earths) within the extended subsurface habitable zone, and $\langle EVI \rangle$ is the average Eirenesphere Volumetric Index for this planetary population.

Current estimates place the total number of stars in the Milky Way at roughly $100$ to $400$ billion, with G-type dwarfs comprising approximately $7\%$ of this population, yielding $N_{\text{G}} \approx 2 \times 10^{10}$ stars. Based on demographic data from the Kepler mission, conservative estimates for $\eta_\oplus$ in the classical habitable zone range from $0.3$ to $0.8$ for solar-type stars \citep{Bryson_2021}. Because the SSHZ spans a much wider orbital parameter space than the classical surface HZ, the occurrence rate of rocky planets within this extended range is significantly higher. For this order-of-magnitude estimate, we conservatively assume $\eta_{\text{SSHZ}} \approx 1.0$, implying at least one rocky planet per G-type star resides within the $\sim 0.5 - 7.0$ au range.

As calculated in \autoref{sec:evi_results}, Earth under its present conditions possesses an $EVI \approx 0.064$ TO. Given that mature super-Earths with higher geothermal activity can sustain eirenespheres several times larger than Earth's, we adopt a conservative average $\langle \mathrm{EVI} \rangle \approx 0.07$ TO across the rocky exoplanet population.

Evaluating our expression yields:
\begin{equation}
    V_{\text{gal}} \approx (2 \times 10^{10}) \cdot (1.0) \cdot (0.07 \text{ TO}) \approx 1.4 \times 10^{9} \text{ TO}
\end{equation}
This indicates that the Milky Way harbours approximately two billion times the volume of Earth's modern surface oceans in the form of deep, biologically viable interstitial fluids, purely around solar-type stars. If M-dwarf systems were included---which dominate the stellar population and where surface habitability is often jeopardised by intense stellar flaring \citep{Shields_2016}---this volumetric capacity would increase by at least an order of magnitude.

Ultimately, this galactic extrapolation reinforces a central thesis of this work: from a volumetric perspective, Earth's surface ocean does not represent the standard for habitability, but rather a thin boundary layer atop a vast, three-dimensional cosmic biosphere. Subsurface eirenespheres likely constitute the dominant habitat for life in the universe.

\section{Discussion}\label{sec:discusion}

\subsection{The astrobiological importance of interior habitability}

The main motivation for studying subsurface habitability arises from a fundamental fact observed on Earth: the planetary interior constitutes one of the most biologically stable and persistent environments known. Unlike the surface, where conditions can vary drastically due to climate changes, stellar radiation, or external impacts, the subsurface provides a sheltered environment that is thermally and chemically buffered, as demonstrated by Earth’s deep biosphere, in which life persists under extreme conditions of pressure, isolation, and total absence of sunlight, sustained by geochemical gradients and energy sources derived from water–rock interactions (see e.g. \citealt{Orcutt2013, Magnabosco2018, lloyd2025intraterrestrials}). This observation has direct implications for exoplanetary astrobiology, since the traditional search for life—biased toward surface conditions similar to those on Earth under the paradigm that habitability simultaneously requires surface liquid water, a stable atmosphere, and moderate temperatures—automatically excludes a vast population of worlds that could host internally habitable reservoirs even when their surfaces are completely hostile, including: planets that have partially or totally lost their atmospheres; worlds that are frozen at the surface but maintain persistent geothermal activity; super-Earths with surface pressures incompatible with liquid water; bodies subjected to intense irradiation from M-type stars; and icy moons with internal oceans decoupled from the surface environment (see e.g. \citealt{Chyba2000, Hand2009, Lingam2018}).

Our results quantify this scenario in concrete terms. For an Earth-analog planet, the model predicts an eirenesphere of ${\sim}0.10\,\mathrm{TO}$, with ${\sim}94\,\%$ of the water-bearing reservoir lying within biological limits. The $\mathrm{EVI}$ calculated for this case is ${\sim}0.06\,\mathrm{TO}$, a value that increases systematically in super-Earths with higher geothermal activity, reaching values several times greater than Earth’s for masses of ${\sim}3$--$10\,M_\oplus$ and heat fluxes of 
${\sim}150$--$200\,\mathrm{mW\,m^{-2}}$. This result suggests that, from a volumetric perspective, Earth does not represent the optimal scenario for subsurface habitability, but rather a moderate case within a much broader spectrum.

\subsection{Metabolic constraints and energy availability}

Although the presence of liquid water and physical conditions compatible with life are necessary requirements, the actual habitability of a subsurface environment also depends on the availability of free energy capable of sustaining active metabolism. In the absence of photosynthesis, the deep biosphere must draw on geochemical sources associated with the planet’s internal processes (see e.g. \citealt{Chyba2000, Hand2009, lloyd2025intraterrestrials}).

Among the most relevant mechanisms are redox reactions produced by water–rock interactions. Serpentinization, for example, generates molecular hydrogen and methane capable of sustaining chemolithotrophic ecosystems over geological timescales \citep{Orcutt2013}. Nevertheless, there is a minimum energy flux threshold below which biochemical reactions cannot sustain cellular repair or active reproduction, a threshold that remains a subject of debate (see e.g. \citealt{Merino2019, lloyd2025intraterrestrials}).

The model developed in this work explicitly addresses thermodynamic and porosity criteria, but it does not incorporate a detailed description of metabolic energy balances or geochemical energy production rates. Therefore, our results should be interpreted as an estimate of the \emph{physical potential for habitability}, rather than as a direct prediction of biological activity. The incorporation of geochemical models of energy production represents a natural extension of this work that would allow refinement of the estimate of the \emph{metabolically active} volume within the calculated habitable zone.

\subsection{Model validation and comparison with previous work}

An indirect way to validate the model is to verify that the results for Earth are consistent with observations of the deep continental biosphere. The maximum documented depth of microbial activity in the Earth's crust is ${\sim}5$--$7\,\mathrm{km}$ in crystalline formations (see e.g. \citealt{Magnabosco2018, lloyd2025intraterrestrials}), a value that reasonably agrees with the thickness of the eirenesphere predicted by the model, namely ${\sim}6.8\,\mathrm{km}$ for $1\,M_\oplus$, $q_s = 65\,\mathrm{mW\,m^{-2}}$, $d = 1\,\mathrm{au}$. Although this agreement does not constitute a rigorous quantitative validation, it suggests that the model correctly captures the order of magnitude of the relevant parameters.

Regarding comparisons with previous models, \citet{McMahon2013} estimated that Earth’s habitable subsurface biosphere extends to depths compatible with the predictions of this work. For their part, \citet{Noack2017} showed that the internal habitability of super-Earths depends critically on tectonic regime and heat flux, in line with the dependence on $q_s$ that we observe in the $\mathrm{EVI}$ map. However, none of these studies simultaneously incorporates crustal porosity structure and mineralogical parametrization as first-order variables, which sets the approach presented here apart from those existing in the literature.

\subsection{Model limitations}

The model presented contains several simplifications that must be taken into account when interpreting the results.

The thermal structure was treated using one-dimensional steady-state geothermal models, ignoring dynamic processes such as mantle convection, active tectonics, localized volcanism, or three-dimensional hydrothermal circulation. On planets with active plate tectonic regimes, these processes could both locally increase the heat flow—expanding or decreasing the eirenesphere—as well as redistribute the internal energy in a heterogeneous way. On planets with a stagnant-lid regime, the absence of lithospheric recycling could reduce the long-term heat flow more rapidly than predicted by the model of \citet{Turcotte2014}, affecting the temporal evolution calculated in \autoref{subsec:temp_hab}.

The mineralogical composition was parameterized with terrestrial analogues based on HC2011 and \citet{Rudnick2014}. Although the parametric model developed in \autoref{subsec:parametric_mineralogy} explores variations in $f_{\rm felsic}$, Fe/Mg, and hydration, the actual properties of rocky exoplanets could differ substantially depending on the stellar abundances of the host system (see e.g. \citealt{Putirka2019, Unterborn2023}). In particular, planets with high carbon content or sulfur-bearing compounds could exhibit thermal conductivity and porosity properties significantly different from those considered here.

The treatment of porosity depends solely on lithostatic pressure, without including tectonic fracturing, hydrothermal alteration, mineral recrystallization, or secondary porosity generation. These mechanisms can drastically modify
the effective permeability in real environments, especially in volcanically active crusts. The minimum porosity threshold $\phi_{\rm min}^{\rm hab} = 10^{-3}$ was conservatively adopted based on observations of the deep terrestrial biosphere, but extrapolating it to exoplanets with different lithologies introduces an additional uncertainty that warrants systematic exploration.

Finally, the adopted biological criteria ($T_{\rm bio}^{\rm max} = 423\,\mathrm{K}$, $P_{\rm bio}^{\rm max} = 200\,\mathrm{MPa}$) correspond to empirical limits based on known terrestrial life. Extremophile organisms with biochemistries different from those studied could tolerate more extreme conditions, which would increase the calculated eirenesphere. The sensitivity of the results to these parameters was partially explored in \autoref{sec:results}, where it was shown that the temperature threshold is the dominant limiting factor on low-mass planets, whereas pressure and porosity dominate on massive planets.

\subsection{Implications for the search for life on exoplanets}

The results obtained in this work significantly expand the traditional geographic limits of observational astrobiology. As demonstrated in the regime map of \autoref{fig:habitability_regimes}, the \textit{Habitable Subsurface} domain persists at orbital distances of up to ${\sim}5$–$7\,\mathrm{au}$ for endogenous heat fluxes greater than ${\sim}30\,\mathrm{mW\,m^{-2}}$. This thermal resilience implies that a wide set of exoplanets currently cataloged as uninhabitable or as “icy worlds” under surface-habitable-zone criteria possess the volumetric potential to host extensive internal biospheres. However, turning this theoretical framework into a remote-detection strategy requires redefining the nature of planetary biosignatures. Unlike a surface photosynthetic biosphere, which directly and massively alters atmospheric composition, a subsurface biosphere interacts with the exterior indirectly, mediated by volcanic degassing and diffusion through the crust \citep{Schwieterman2018, Meadows2018}.

Despite lithospheric confinement, this chemical coupling between the subsurface and the atmosphere is not negligible. Anaerobic chemolithotrophic metabolisms, characteristic of Earth’s deep biosphere, release volatile byproducts such as methane (CH$_4$), molecular hydrogen (H$_2$), hydrogen sulfide (H$_2$S), and nitrous oxide (N$_2$O) \citep{Seager2013, Schwieterman2018, Hu2022}. More than the detection of a single gas, the most robust signature of a deep biosphere lies in the simultaneous thermodynamic disequilibrium of complementary species, notably the CH$_4$–CO$_2$ pair \citep{Krissansen-Totton2018}. Since the long-term coexistence of these gases requires a continuous flux of reductants that abiotic geological processes rarely sustain on their own, this disequilibrium constitutes a critical indicator of metabolically active subsurface environments.

However, interpreting these signals requires extreme caution due to abiotic false positives. Non-biological processes such as serpentinization and radiolysis in confined water—mechanisms that paradoxically provide energy to internal life—produce large quantities of H$_2$ and CH$_4$ of purely geochemical origin \citep{Sherwood2013, Tarnas2018}. Therefore, the unambiguous identification of a deep biosphere will depend on characterizing complex sets of gases in chemical disequilibrium within their geophysical context \citep{Schwieterman2018}. Such an analytical challenge will exceed the capabilities of current observatories, demanding the transit spectroscopy sensitivity of next-generation telescopes such as the \textit{Extremely Large Telescope} (ELT) or concepts for space missions dedicated to the atmospheric characterization of rocky exoplanets \citep{Meadows2018, Quanz2022}.

Beyond direct spectroscopic characterization, the index $\mathrm{EVI}$ formulated in this work offers a fundamental predictive tool for the statistical prioritization of targets prior to atmospheric analysis. Since the key variables of the index—planetary mass, theoretical heat flux, and inferences of mineralogical composition—can be derived from internal structure models and transit or radial-velocity observations, $\mathrm{EVI}$ makes it possible to quantitatively classify the internal habitability potential of known, real exoplanets. Compact and well-characterized stellar systems such as TRAPPIST-1 \citep{Agol2021}, where stringent constraints on the densities and core fractions of multiple rocky worlds are already available, constitute the immediate natural laboratory for a future, operational application of this index.

The high habitability efficiency calculated by our model (where $94.08\%$ of the effective aquable volume is found under conditions compatible with life) introduces a crucial statistical implication. If the emergence of life is linked to the long-term persistence of liquid water in porous media, intermediate-mass planets with moderate heat fluxes maximize not only the instantaneous eirenesphere, but also its stability on geological timescales, as discussed in \autoref{subsec:temp_hab}. In this way, three-dimensional habitability analysis overturns the temporal bias of surface habitability, positioning thermally mature worlds with deep biospheres as the most stable, resilient, and volumetrically significant biological containers in the local universe.

Finally, the staggering volumetric capacity of the Galaxy to host subsurface biospheres---estimated here at billions of terrestrial oceans even for a conservative population of solar-type stars---fundamentally reshapes our statistical expectations for the search for life. If the vast majority of the Milky Way's biologically viable real estate is subterranean, observational astrobiology must urgently prioritize the development and refinement of indirect biosignature detection frameworks, such as tracing tectonic degassing and volatile disequilibrium \citep{Schwieterman2018}. Surface biospheres may represent the exception rather than the rule, implying that our most robust chance of detecting extraterrestrial life lies in deciphering the subtle atmospheric signatures exhaled by these volumetrically dominant, deep planetary environments.

\section{Conclusions}\label{sec:conclusions}

The results obtained in this work show that planetary habitability can extend far beyond the classical concept based exclusively on surface conditions. The thermal and structural analysis carried out demonstrates that the stability of liquid water and the existence of biologically viable subsurface niches depend strongly on internal geophysical properties—heat flow, thermal conductivity, crustal structure, and the evolution of porosity with depth—that are independent of the planet’s orbital position. In this context, habitability ceases to be solely a radiative problem and becomes a three-dimensional property governed by the interaction between internal geodynamics, the thermodynamics of water, and the lithospheric structure. This conceptual transition has profound implications for astrobiology, as it substantially broadens the range of potentially habitable planetary bodies.

In this work, we have developed an integrated geophysical model to quantify the subsurface habitability of rocky planets, consistently incorporating the thermal structure, mineralogical composition, porosity evolution with pressure, and the thermodynamic and biological criteria required for the existence of liquid water and microbial life in the planetary interior. Building on this framework, we have proposed the Eigenesphere Volumetric Index ($\mathrm{EVI}$) as a quantitative metric for comparing the potential for internal habitability among rocky planets of different mass, composition, and thermal regime. The main conclusions of this work are as follows.

\begin{enumerate}

\item \textbf{Subsurface habitability is a geophysically robust phenomenon that extends well beyond the classical habitable zone.} Parametric analysis in the space of (orbital distance, heat flux) shows that the eirenesphere region dominates most of the explored parameter space, persisting out to orbital distances of ${\sim}5$–$7\,\mathrm{au}$ for heat fluxes above ${\sim}30\,\mathrm{mW\,m^{-2}}$. This result demonstrates that a planet’s internal heat can decouple subsurface habitability from the surface radiative balance, turning superficially hostile worlds—frozen, atmosphere-less, or subjected to intense stellar irradiation—into viable astrobiological candidates from an interior perspective.

\item \textbf{There is an optimal range of planetary mass for subsurface habitability, around ${\sim}1\,M_\oplus$.} The thickness of the effective eirenesphere does not increase monotonically with mass; instead, it exhibits a maximum near Earth’s mass, as a result of the balance between two opposing mechanisms: in low-mass planets, the steep geothermal gradient causes the temperature to exceed the biological limit $T_{\rm bio}^{\rm max}$ at shallow depths, confining the deep biosphere; in massive planets, the high lithostatic pressure collapses crustal porosity before thermal conditions become unfavorable. It is noteworthy that this maximum approximately coincides with Earth’s mass, suggesting that our planet lies in a particularly favorable mass configuration for the development of subsurface biospheres.

\item \textbf{The distinction between aquability and habitability is physically relevant and quantitatively significant.} In thermally mature planets, ${\sim}94\,\%$ of the aquable layer is biologically accessible, with porosity — rather than temperature — being the dominant limiting factor. In young planets, by contrast, only ${\sim}57\,\%$ of the aquable layer satisfies biological criteria, and it is the thermal limit that constrains habitability. This distinction implies that the mere detection of subsurface liquid water is not a sufficient condition to infer habitability, and that the thermal history and porous structure of the lithosphere must be explicitly considered in any astrobiological assessment.

\item \textbf{Subsurface habitability is an evolutionary, not a static, property.} The analysis of temporal evolution shows that, for an Earth-analog planet located at $1\,\mathrm{au}$, the aquable layer grows from ${\sim}0.17\,\mathrm{km}$ at the initial instant to ${\sim}7.6\,\mathrm{km}$ after ${\sim}2\,\mathrm{Ga}$ of secular cooling, stabilizing at that value during late evolution. The eirenesphere follows a similar but more prolonged trajectory, reaching its maximum of ${\sim}7.28\,\mathrm{km}$ near $5.8\,\mathrm{Ga}$. These results indicate that young planets host confined subsurface biospheres and that internal habitability is maximized in thermally mature worlds, with direct implications for prioritizing exoplanetary systems of different ages.

\item \textbf{$\mathrm{EVI}$ provides a quantitative metric for the systematic comparison of internal habitability among exoplanets.} The index, defined as the average eirenesphere integrated along the circumstellar habitable zone, increases steadily with planetary mass and surface heat flux within the explored range ($0.1$--$10\,M_\oplus$, $20$–$200\,\mathrm{mW\,m^{-2}}$). The Earth, with $EVI \approx 0.06\,\mathrm{TO}$, lies in a region of moderate internal habitability, whereas super-Earths with higher geothermal activity could reach values several times larger. Since the input parameters of the index are in principle observationally accessible or inferable through internal structure models, $\mathrm{EVI}$ can be applied to real exoplanetary systems as their geophysical characterization improves, such as the TRAPPIST-1 system \citep{Agol2021}.

\item \textbf{Crustal mineralogy is a first-order parameter for subsurface habitability.} The parametric model developed here shows that felsic crusts, dominated by quartz and feldspars with low thermal conductivity, favor more extensive deep biospheres than mafic or ultramafic crusts under the same geothermal conditions, with differences in eirenesphere thickness of up to ${\sim}0.9\,\mathrm{km}$ between the compositional endmembers explored. The Fe/Mg ratio and the degree of crustal hydration introduce secondary but non-negligible effects, consistent with their impact on the effective density and thermal conductivity of the mineralogical assemblage. This result implies that the stellar composition of the host system — which to first order determines the planet’s mineralogy — is a relevant factor in assessing the subsurface habitability potential of rocky exoplanets (see e.g. \citealt{Putirka2019, Unterborn2023}).

\item \textbf{The galactic capacity for subsurface life is volumetrically staggering}. Extrapolating the EVI to the Milky Way's population of rocky planets around solar-type stars reveals a potential habitable reservoir on the order of billions of terrestrial oceans. This vast capacity underscores a profound astrobiological paradigm shift: Earth's surface ocean, long considered the standard for habitability, is likely dwarfed by the cumulative volume of sheltered, deep planetary eirenespheres. Consequently, these subsurface environments may constitute the most abundant, stable, and resilient biological niches in the universe.

\end{enumerate}

Taken together, these results reinforce the idea that planetary habitability must be understood as a three-dimensional and evolutionary phenomenon, not reducible to purely surface-based or radiative criteria. The $\mathrm{EVI}$ introduced in this work provides a concrete tool for quantifying this interior dimension of habitability, complementing classical approaches based on the circumstellar habitable zone. Future extensions of the model that incorporate geochemical models of energy production, alternative tectonic regimes, and non-terrestrial biochemistries will make it possible to refine these estimates and broaden their applicability to the diversity of rocky worlds.

\section*{ACKNOWLEDGEMENTS}

The authors gratefully acknowledge the community of the \textit{Semillero de Investigación en Astronomía} of the Institute of Physics at Universidad de Antioquia, where some of the foundational ideas for this work first emerged. In particular, S.O. wishes to thank Sebastian Rodríguez-Numpaque and Camilo Ospinal for their insightful early discussions on this topic.

This work has been possible due to the availability of open-source, general-purpose {\tt Python} packages, including 
\  {\tt Matplotlib} \citep{Matplotlib2007}, {\tt Numpy} \citep{Numpy2020}, {\tt Scipy} \citep{virtanenSciPy10Fundamental2020}, {\tt pandas} \citep{mckinney_proc_scipy_2010}, {\tt BurnMaN} \citep{Cottaar2014, Myhill2023, myhill_2024_14238360}, {\tt IAPWS} \citep{IAPWS1995, IAPWSRelease2009, IAPWSSeawater2010}.

\section*{Data Availability}

All the codes used for generating  the figures in this paper are available at the \texttt{GitHub} repository \url{https://github.com/Santiago-Orjuela/Eirenesphere}.





\appendix

\section{Circumstellar Habitable Zone}

The classical criterion for surface habitability is based on the radiative equilibrium between the incident energy from the host star and the thermal energy emitted by the planet. Under the assumption of global radiative equilibrium and considering an idealized single-layer atmosphere with effective infrared transmittance $\tau$ (where $\tau=0$ implies an atmosphere that absorbs all the longwave radiation emitted by the planet, while $\tau=1$ corresponds to an atmosphere that is completely transparent to infrared), the surface temperature can be parameterized as (see Appendix~\ref{app:balance_radiativo}):

\begin{equation}
T_{\rm surf} = 331\,\mathrm{K}\,
\left[
\frac{L_\star(1-A)}{2\tau\,d^2}
\right]^{1/4},
\label{eq:Teq_transmitancia}
\end{equation}
where $L_\star$ is the stellar luminosity in solar units, $A$ the Bond albedo, and $d$ the orbital distance in astronomical units. This formulation compactly summarizes atmospheric thermal enhancement and is particularly useful for parametrically exploring the impact of the greenhouse effect on the stability of surface liquid water. For Earth, the empirical value $\tau = 0.6$ reasonably reproduces the observed global mean temperature. 

~\autoref{eq:Teq_transmitancia} generalizes the purely radiative, atmosphere-free case, which is recovered in the limit $\tau \rightarrow 1$, and provides a simple physical basis for the operational definition of the circumstellar habitable zone.

In its most elementary form, the habitable zone (HZ) is defined as the range of orbital distances for which, given $L_\star$, $A$, and $\tau$, the surface temperature remains within the interval compatible with the existence of liquid water under an atmospheric pressure similar to that of Earth, that is, $273\,\mathrm{K} \leq T_{\rm surf} \leq 373\,\mathrm{K}$. Solving for $d$ from equation~\eqref{eq:Teq_transmitancia}:

\begin{equation}
d = \left[\frac{L_\star(1-A)}{\tau}\right]^{1/2}
\!\left(\frac{331\,\mathrm{K}}{T_{\rm surf}}\right)^{2}.
\label{eq:d_hz}
\end{equation}

For a solar-type star ($L_\star = 1\,L_\odot$), a Bond albedo $A = 0.3$, and $\tau = 0.6$, the inner and outer limits of the HZ are:

\begin{align}
d_{\rm in}  &\approx 0.85\,\mathrm{au} \quad (T_{\rm surf} = 373\,\mathrm{K}),\\
d_{\rm out} &\approx 1.59\,\mathrm{au} \quad (T_{\rm surf} = 273\,\mathrm{K}).
\end{align}

The Earth, at $d = 1\,\mathrm{au}$, lies within this zone, qualitatively validating the model.

However, this zeroth-order model has important limitations arising from its static nature: it does not capture the response of the atmosphere to changes in insolation, nor the interdependence between temperature and atmospheric composition. In particular, as insolation increases, a planet with surface water will experience a rise in water-vapour concentration, which reduces the effective transmittance $\tau$ and amplifies the warming through a positive feedback, a phenomenon known as a runaway greenhouse effect (\textit{runaway greenhouse}) (see e.g. \citealt{Ingersoll1969, Goldblatt2012}). In the opposite direction, a decrease in temperature can lead to the condensation or freezing of greenhouse gases, increasing $\tau$ and accelerating the cooling of the planet toward a \textit{snowball Earth} state \citep{Pierrehumbert2004}. These processes introduce fundamental nonlinearities that a constant $\tau$ parameter cannot capture.

\begin{figure}
    \centering
    \includegraphics[width=0.9\linewidth]{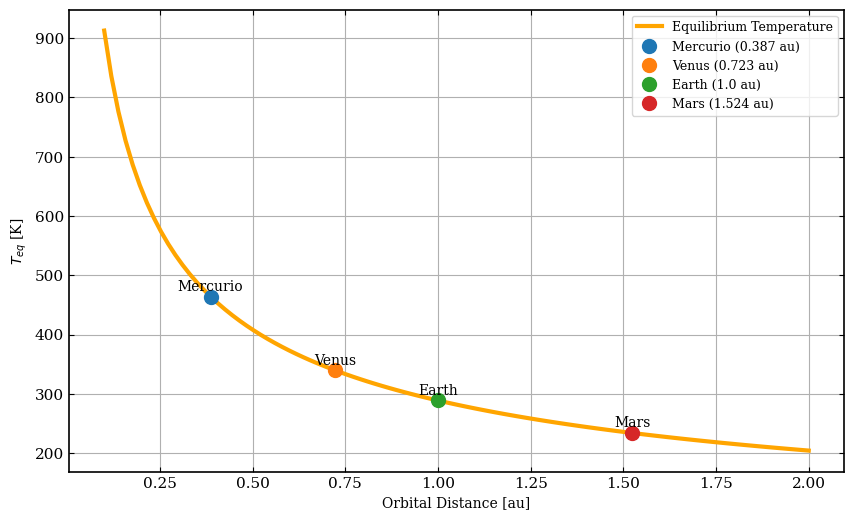}
    \caption{Equilibrium surface temperature as a function of orbital distance for a solar-type star ($L_\star = 1\,L_\odot$), computed from ~\autoref{eq:Teq_transmitancia} with Bond albedo $A = 0.3$ and infrared transmittance $\tau = 0.6$. The four terrestrial planets of the Solar System are shown for reference, confirming that Earth lies within the 
    temperature range compatible with liquid water at its surface.}
    \label{fig:Temperature_eq}
\end{figure}

To incorporate these effects in a rigorous way, it is necessary to use one-dimensional (1-D) climate models that explicitly solve radiative transport and atmospheric photochemistry as a function of temperature and composition. Seminal works such as those of \citet{Kasting1993} and their subsequent refinements by \citet{Kopparapu2013, Kopparapu2014} have established the limits of the HZ on more robust physical grounds, incorporating feedback from water vapor, ice albedo, and phase transitions of greenhouse gasses. In ~\autoref{subsubsec:zh_limits}, we will use the parameterization presented in ~\autoref{eq:Teq_transmitancia} as the initial condition for the surface temperature of the test exoplanets, from which we will derive the temperature and pressure profiles in the planetary interior.

\subsection{Limits of the Habitable Zone}\label{subsubsec:zh_limits}

In order to refine the inner and outer boundaries of the habitable zone, \cite{Kopparapu2013,Kopparapu2014} introduced an empirical parametrization based on one-dimensional radiative–convective climate models. Within this framework, the characteristic orbital distance associated with a given HZ boundary is expressed as
\begin{equation}
d = \left( \frac{L_\star}{S_{\rm eff}} \right)^{1/2},
\end{equation}
where $S_{\rm eff}$ is the effective stellar flux normalized to the solar value. This flux depends on the effective temperature of the star according to
\begin{equation}
S_{\rm eff} =
S_{\rm eff,\odot} + aT' + bT'^2 + cT'^3 + dT'^4,
\end{equation}
with $T' = T_{\rm eff} - 5780\,\mathrm{K}$ and coefficients determined for different climate criteria, such as the runaway greenhouse effect, the maximum greenhouse, among others. In Table 1 of \cite{Kopparapu2014} the values of the coefficients are presented; we will use them in ~\autoref{layer_thickness} to compare these limits with those obtained here when interior habitability conditions are included.

\section{Radiative balance in the single-layer atmosphere model}\label{app:balance_radiativo}

\subsection{Model description}

We consider a simplified model of a planetary atmosphere consisting of a single homogeneous layer that interacts with the infrared radiation emitted by the surface, as illustrated schematically in~\autoref{fig:radiative_balance}. 
The following assumptions are adopted:

\begin{enumerate}
    \item The atmosphere is a grey body with emissivity $\varepsilon = 1$ (blackbody approximation).

    \item The atmosphere transmits a fraction $\tau_{\rm layer} \in [0,1]$ of the surface infrared radiation directly to space, and absorbs the remaining fraction $(1 - \tau_{\rm layer})$.

    \item There is no specular reflection in the atmosphere ($r = 0$).

    \item The planet has a Bond albedo $A$. The mean irradiance absorbed per unit surface area is $(1-A)\,L_\star/(16\pi d^2)$, where the factor $1/4$ reflects the ratio between the planet's cross-sectional 
    area and its total surface area.

    \item The system is in stationary radiative equilibrium.
\end{enumerate}

\subsection{Radiative balance equations}

\subsubsection*{Balance of the atmospheric layer}

As shown by the flux vectors in~\autoref{fig:radiative_balance}, the atmosphere 
absorbs the fraction $(1 - \tau_{\rm layer})$ of the radiation emitted by the surface, $\sigma T_{\rm surf}^4$. To preserve thermal equilibrium, it re-emits that energy equally upward to space and downward back to the surface. The resulting energy balance of the atmospheric layer is therefore:

\begin{equation}
2\,\sigma T_{\rm atm}^4 = (1 - \tau_{\rm layer})\,\sigma T_{\rm surf}^4,
\label{eq:app_atm_balance}
\end{equation}

from which the relation between the atmospheric and surface temperatures is obtained:

\begin{equation}
T_{\rm atm}^4 = \frac{1 - \tau_{\rm layer}}{2}\,T_{\rm surf}^4.
\label{eq:app_Tatm_Tsurf}
\end{equation}

\subsubsection*{Surface balance}

\begin{figure*}
    \centering
    \includegraphics[width=0.85\textwidth]{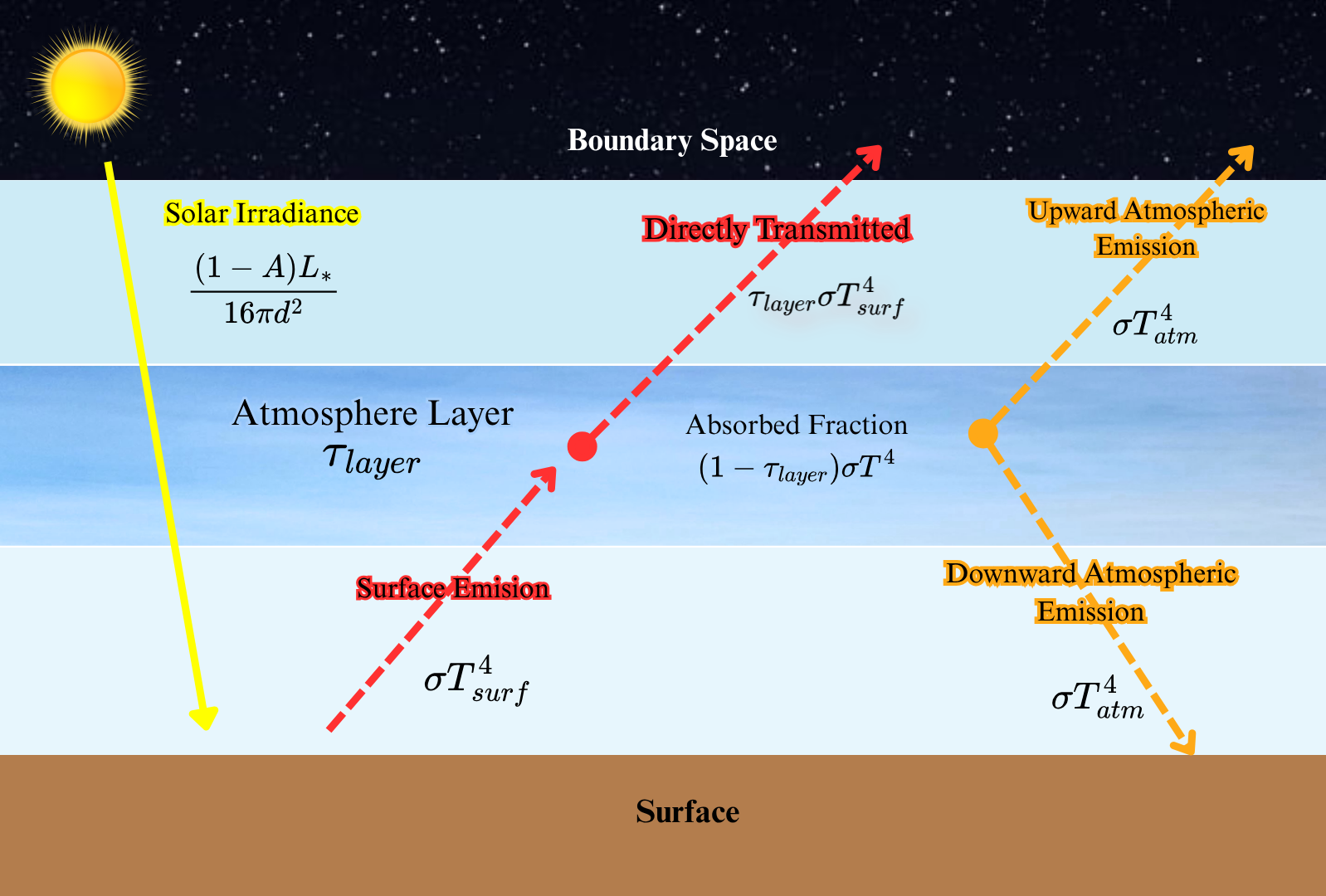}
    \caption{Schematic of the 1-layer atmospheric radiative balance model. Net shortwave solar irradiance, $F_{\rm solar} = (1-A)L_\star / (16\pi d^2)$, passes through the atmosphere and is fully absorbed by the surface. The surface at temperature $T_{\rm surf}$ emits blackbody longwave radiation $\sigma T_{\rm surf}^4$. A fraction $\tau_{\rm layer}$ of this emission escapes directly to space; the remainder $(1 - \tau_{\rm layer})$ is absorbed by the atmospheric layer. In thermodynamic equilibrium, the atmosphere at temperature $T_{\rm atm}$ re-emits this energy isotropically, with $\sigma T_{\rm atm}^4$ radiated upward to space and $\sigma T_{\rm atm}^4$ radiated downward to the surface (greenhouse back-radiation).}
    \label{fig:radiative_balance}
\end{figure*}

Tracing the downward arrows targeting the lower boundary in~\autoref{fig:radiative_balance}, the surface receives the mean absorbed shortwave solar irradiance $(1-A)\,L_\star/(16\pi d^2)$ plus the downward longwave irradiance from the atmosphere $\sigma T_{\rm atm}^4$. 
In equilibrium, this total incoming energy must equal the blackbody thermal emission at temperature $T_{\rm surf}$:

\begin{equation}
\sigma T_{\rm surf}^4 = \frac{(1-A)\,L_\star}{16\pi d^2}
+ \sigma T_{\rm atm}^4.
\label{eq:app_ground_balance}
\end{equation}

\subsection{Solution of the system}

Substituting~\autoref{eq:app_Tatm_Tsurf} into~\autoref{eq:app_ground_balance}:

\begin{equation}
\sigma T_{\rm surf}^4 = \frac{(1-A)\,L_\star}{16\pi d^2}
+ \frac{1 - \tau_{\rm layer}}{2}\,\sigma T_{\rm surf}^4.
\end{equation}

Grouping the $T_{\rm surf}^4$ terms on the left-hand side:

\begin{equation}
\sigma T_{\rm surf}^4
\left[1 - \frac{1 - \tau_{\rm layer}}{2}\right]
= \frac{(1-A)\,L_\star}{16\pi d^2}.
\end{equation}

The bracketed factor simplifies as:

\begin{equation}
1 - \frac{1 - \tau_{\rm layer}}{2}
= \frac{2 - (1 - \tau_{\rm layer})}{2}
= \frac{1 + \tau_{\rm layer}}{2}.
\end{equation}

Substituting back:

\begin{equation}
\sigma T_{\rm surf}^4 \cdot \frac{1 + \tau_{\rm layer}}{2}
= \frac{(1-A)\,L_\star}{16\pi d^2},
\end{equation}

and solving for $T_{\rm surf}^4$:

\begin{equation}
T_{\rm surf}^4
= \frac{(1-A)\,L_\star}{8\pi\,\sigma\,(1 + \tau_{\rm layer})\,d^2}.
\label{eq:app_Tsurf4_Lstar}
\end{equation}

\subsection{Effective transmittance and notational convention}

The factor $(1 + \tau_{\rm layer})/2$ that appears naturally in the solution has a clear physical interpretation: it represents the effective 
fraction of surface radiation that escapes to space, accounting for both the direct transmission $\tau_{\rm layer}$ through the layer and the upward re-emission of the absorbed fraction $(1 - \tau_{\rm layer})$. For notational simplicity, we define the \emph{effective transmittance}:

\begin{equation}
\tau \equiv \frac{1 + \tau_{\rm layer}}{2},
\qquad \tau \in \left[\tfrac{1}{2},\,1\right].
\label{eq:app_tau_def}
\end{equation}

Under this convention, $\tau = 1$ when $\tau_{\rm layer} = 1$ (fully transparent atmosphere, no greenhouse effect) and 
$\tau = 1/2$ when $\tau_{\rm layer} = 0$ (fully opaque atmosphere, maximum greenhouse effect). For Earth, the empirical value $\tau = 0.6$ corresponds to $\tau_{\rm layer} = 0.2$, meaning that the atmosphere 
directly transmits 20\,\% of surface infrared radiation and absorbs the remaining 80\,\%. Substituting $(1 + \tau_{\rm layer}) = 2\tau$ into~\autoref{eq:app_Tsurf4_Lstar}:

\begin{equation}
T_{\rm surf}^4
= \frac{(1-A)\,L_\star}{16\pi\,\sigma\,\tau\,d^2}.
\label{eq:app_Tsurf4_tau}
\end{equation}

\subsection{Dimensionless form and the 331\,K constant}

Expressing $L_\star$ in units of solar luminosity $L_\odot$ and $d$ in astronomical units (AU), ~\autoref{eq:app_Tsurf4_tau} takes the form:

\begin{equation}
T_{\rm surf}
= \left[\frac{L_\odot}{16\pi\,\sigma\,\mathrm{au}^2}\right]^{1/4}
\left[\frac{(1-A)\,L_\star}{\tau\,d^2}\right]^{1/4}.
\end{equation}

The dimensional prefactor is evaluated numerically with
$L_\odot = 3.828\times10^{26}$\,W,
$\sigma = 5.670\times10^{-8}$\,W\,m$^{-2}$\,K$^{-4}$, and
$1\,\mathrm{au} = 1.496\times10^{11}$\,m:

\begin{align}
\left[\frac{L_\odot}{16\pi\,\sigma\,\mathrm{au}^2}\right]^{1/4}
&= \left[
\frac{3.828\times10^{26}}
{16\pi \times 5.670\times10^{-8}
\times (1.496\times10^{11})^2}
\right]^{1/4} \nonumber \\
&= \left[5.998\times10^{9}\,\mathrm{K}^4\right]^{1/4}
\approx 278\,\mathrm{K}.
\label{eq:app_const278}
\end{align}

Absorbing the factor $2^{1/4}$ inside the bracket ---that is, dividing its argument by 2--- and using the identity 
$278\,\mathrm{K} \approx 331\,\mathrm{K}/2^{1/4}$ 
(verified as $331/2^{1/4} \approx 278.3\,\mathrm{K}$), the equation can be rewritten in the equivalent and more compact form:

\begin{equation}
\boxed{
T_{\rm surf}
= 331\,\mathrm{K}
\left[\frac{L_\star\,(1-A)}{2\,\tau\,d^2}\right]^{1/4},
}
\label{eq:app_Tsurf_final}
\end{equation}
where $L_\star$ is in units of $L_\odot$ and $d$ in AU. This is the expression used throughout the main text 
(~\autoref{eq:Teq_transmitancia}), with $\tau$ denoting the 
effective infrared transmittance defined in 
~\autoref{eq:app_tau_def}. In the limit $\tau \rightarrow 1$ ($\tau_{\rm layer} \rightarrow 1$, no greenhouse effect), the standard atmosphere-free equilibrium temperature is recovered.

\label{lastpage}

\begin{thebibliography}{111}
\providecommand{\natexlab}[1]{#1}
\providecommand{\url}[1]{\texttt{#1}}
\expandafter\ifx\csname urlstyle\endcsname\relax
  \providecommand{\doi}[1]{doi: #1}\else
  \providecommand{\doi}{doi: \begingroup \urlstyle{rm}\Url}\fi

\bibitem[IAP(2010)]{IAPWSSeawater2010}
Guideline on the thermophysical properties of seawater.
\newblock Technical report, IAPWS, 2010.

\bibitem[Agol et~al.(2021)Agol, Dorn, Grimm, Turbet, Ducrot, Delrez, Gillon, Demory, Burdanov, Barkaoui, et~al.]{Agol2021}
E.~Agol, C.~Dorn, S.~L. Grimm, M.~Turbet, E.~Ducrot, L.~Delrez, M.~Gillon, B.-O. Demory, A.~Burdanov, K.~Barkaoui, et~al.
\newblock Refining the transit-timing and photometric analysis of trappist-1: masses, radii, densities, dynamics, and ephemerides.
\newblock \emph{The planetary science journal}, 2\penalty0 (1):\penalty0 1, 2021.

\bibitem[Artemieva(2006)]{Artemieva2006}
I.~M. Artemieva.
\newblock Global 1°×1° thermal model tc1 for the continental lithosphere: Implications for lithosphere secular evolution.
\newblock \emph{Tectonophysics}, 416\penalty0 (1):\penalty0 245--277, 2006.
\newblock ISSN 0040-1951.
\newblock \doi{https://doi.org/10.1016/j.tecto.2005.11.022}.
\newblock URL \url{https://www.sciencedirect.com/science/article/pii/S0040195105006256}.
\newblock The Heterogeneous Mantle.

\bibitem[Athy(1930)]{Athy1930}
L.~F. Athy.
\newblock Density, porosity, and compaction of sedimentary rocks.
\newblock \emph{Aapg Bulletin}, 14\penalty0 (1):\penalty0 1--24, 1930.

\bibitem[Atri(2025)]{Atri2025}
D.~Atri.
\newblock New research suggests life could survive beneath the surface of mars and other planets using high energy particles from space.
\newblock \emph{Astrobiology}, 2025.
\newblock URL \url{https://astrobiology.com/2025/07/new-research-suggests-life-could-survive-beneath-the-surface-of-mars-and-other-planets-using-high-energy-particles-from-space.html}.

\bibitem[Bains(2004)]{Bains2004}
W.~Bains.
\newblock Many chemistries could be used to build living systems.
\newblock \emph{Astrobiology}, 4\penalty0 (2):\penalty0 137--167, 2004.
\newblock \doi{10.1089/153110704323175124}.

\bibitem[Bains et~al.(2015)Bains, Xiao, and Yu]{Bains2015}
W.~Bains, Y.~Xiao, and C.~Yu.
\newblock Prediction of the maximum temperature for life based on the stability of metabolites to decomposition in water.
\newblock \emph{Life}, 5\penalty0 (2):\penalty0 1054--1100, 2015.

\bibitem[Batalha(2014)]{Batalha2014}
N.~M. Batalha.
\newblock Exploring exoplanet populations with nasa’s kepler mission.
\newblock \emph{Proceedings of the National Academy of Sciences}, 111\penalty0 (35):\penalty0 12647--12654, 2014.
\newblock \doi{10.1073/pnas.1304196111}.
\newblock URL \url{https://www.pnas.org/doi/abs/10.1073/pnas.1304196111}.

\bibitem[Berryman(1995)]{Berryman1995}
J.~G. Berryman.
\newblock Mixture theories for rock properties.
\newblock \emph{Rock Physics and Phase Relations: A Handbook of Physical Constants}, 3:\penalty0 205--228, 1995.
\newblock \doi{10.1029/RF003p0205}.

\bibitem[Bl{\"o}chl et~al.(1997)Bl{\"o}chl, Rachel, Burggraf, Hafenbradl, Jannasch, and Stetter]{Blochl1997}
E.~Bl{\"o}chl, R.~Rachel, S.~Burggraf, D.~Hafenbradl, H.~W. Jannasch, and K.~O. Stetter.
\newblock \textit{Pyrolobus fumarii}, gen. and sp. nov., represents a novel group of archaea, extending the upper temperature limit for life to 113 degrees {C}.
\newblock \emph{Extremophiles}, 1\penalty0 (1):\penalty0 14--21, 1997.
\newblock \doi{10.1007/s007920050010}.

\bibitem[Bryson et~al.(2021)Bryson, Kunimoto, Kopparapu, Coughlin, Borucki, Koch, Aguirre, Allen, Barentsen, Batalha, et~al.]{Bryson_2021}
S.~Bryson, M.~Kunimoto, R.~K. Kopparapu, J.~L. Coughlin, W.~J. Borucki, D.~Koch, V.~S. Aguirre, C.~Allen, G.~Barentsen, N.~M. Batalha, et~al.
\newblock The occurrence of rocky habitable-zone planets around solar-like stars from kepler data.
\newblock \emph{The Astronomical Journal}, 161\penalty0 (1):\penalty0 36, 2021.
\newblock \doi{10.3847/1538-3881/abc418}.

\bibitem[Chapman(1986)]{Chapman1986}
D.~S. Chapman.
\newblock Thermal gradients in the continental crust.
\newblock \emph{Geological Society Special Publications}, 24\penalty0 (1):\penalty0 63--70, 1986.
\newblock \doi{10.1144/GSL.SP.1986.024.01.07}.

\bibitem[Charette and Smith(2010)]{Charette2010}
M.~A. Charette and W.~H.~F. Smith.
\newblock The volume of {Earth}'s ocean.
\newblock \emph{Oceanography}, 23\penalty0 (2):\penalty0 112--114, 2010.
\newblock \doi{10.5670/oceanog.2010.51}.

\bibitem[Choukroun and Grasset(2010)]{Choukroun2010}
M.~Choukroun and O.~Grasset.
\newblock Thermodynamic data and modeling of the water and ammonia-water phase diagrams up to 2.2 gpa for planetary geophysics.
\newblock \emph{The Journal of chemical physics}, 133\penalty0 (14), 2010.

\bibitem[Chyba(2000)]{Chyba2000}
C.~F. Chyba.
\newblock Energy for microbial life on europa.
\newblock \emph{Nature}, 403\penalty0 (6768):\penalty0 381--382, 2000.
\newblock \doi{https://doi.org/10.1038/35000281}.

\bibitem[Clifford(1993)]{Clifford1993}
S.~M. Clifford.
\newblock A model for the hydrologic and climatic behavior of water on {Mars}.
\newblock \emph{Journal of Geophysical Research: Planets}, 98\penalty0 (E6):\penalty0 10973--11016, 1993.
\newblock \doi{10.1029/93JE00225}.

\bibitem[Cockell(2014)]{Cockell2014}
C.~Cockell.
\newblock \emph{The subsurface habitability of terrestrial rocky planets: Mars}, pages 225--260.
\newblock 05 2014.
\newblock ISBN 978-3-11-030013-0.
\newblock \doi{10.1515/9783110300130.225}.

\bibitem[Cockell et~al.(2016)Cockell, Bush, Bryce, Direito, Fox-Powell, Harrison, Lammer, Landenmark, Martin-Torres, Nicholson, et~al.]{Cockell2016}
C.~S. Cockell, T.~Bush, C.~Bryce, S.~Direito, M.~Fox-Powell, J.~P. Harrison, H.~Lammer, H.~Landenmark, J.~Martin-Torres, N.~Nicholson, et~al.
\newblock Habitability: a review.
\newblock \emph{Astrobiology}, 16\penalty0 (1):\penalty0 89--117, 2016.

\bibitem[Cottaar et~al.(2014)Cottaar, Heister, Rose, and Unterborn]{Cottaar2014}
S.~Cottaar, T.~Heister, I.~Rose, and C.~Unterborn.
\newblock Burnman: A lower mantle mineral physics toolkit.
\newblock \emph{Geochemistry, Geophysics, Geosystems}, 15\penalty0 (4):\penalty0 1164--1179, 2014.
\newblock \doi{https://doi.org/10.1002/2013GC005122}.
\newblock URL \url{https://agupubs.onlinelibrary.wiley.com/doi/abs/10.1002/2013GC005122}.

\bibitem[Cross et~al.(1902)Cross, Iddings, Pirsson, and Washington]{Cross1902}
W.~Cross, J.~P. Iddings, L.~V. Pirsson, and H.~S. Washington.
\newblock A quantitative chemico-mineralogical classification and nomenclature of igneous rocks.
\newblock \emph{The Journal of Geology}, 10\penalty0 (6):\penalty0 555--690, 1902.
\newblock \doi{10.1086/621030}.

\bibitem[Dalmasso et~al.(2016)Dalmasso, Oger, Selva, Courtine, L’haridon, Garlaschelli, Roussel, Miyazaki, Reveillaud, Jebbar, et~al.]{Dalmasso2016}
C.~Dalmasso, P.~Oger, G.~Selva, D.~Courtine, S.~L’haridon, A.~Garlaschelli, E.~Roussel, J.~Miyazaki, J.~Reveillaud, M.~Jebbar, et~al.
\newblock Thermococcus piezophilus sp. nov., a novel hyperthermophilic and piezophilic archaeon with a broad pressure range for growth, isolated from a deepest hydrothermal vent at the mid-cayman rise.
\newblock \emph{Systematic and Applied Microbiology}, 39\penalty0 (7):\penalty0 440--444, 2016.

\bibitem[Daniel* and Cowan(2000)]{Daniel2000}
R.~M. Daniel* and D.~A. Cowan.
\newblock Biomolecular stability and life at high temperatures.
\newblock \emph{Cellular and Molecular Life Sciences CMLS}, 57\penalty0 (2):\penalty0 250--264, 2000.

\bibitem[Danovaro et~al.(2010)Danovaro, Dell'Anno, Pusceddu, Gambi, Heiner, and Kristensen]{Danovaro2010}
R.~Danovaro, A.~Dell'Anno, A.~Pusceddu, C.~Gambi, I.~Heiner, and R.~Kristensen.
\newblock The first metazoa living in permanently anoxic conditons.
\newblock \emph{BMC biology}, 8:\penalty0 30, 04 2010.
\newblock \doi{10.1186/1741-7007-8-30}.

\bibitem[Davies and Davies(2010)]{Davies2010}
J.~H. Davies and D.~R. Davies.
\newblock Earth's surface heat flux.
\newblock \emph{Solid Earth}, 1\penalty0 (1):\penalty0 5--24, 2010.
\newblock \doi{10.5194/se-1-5-2010}.

\bibitem[Dorn et~al.(2015)Dorn, Khan, Heng, Connolly, Alibert, Benz, and Tackley]{Dorn2015}
C.~Dorn, A.~Khan, K.~Heng, J.~A.~D. Connolly, Y.~Alibert, W.~Benz, and P.~Tackley.
\newblock Can we constrain the interior structure of rocky exoplanets from mass and radius measurements?
\newblock \emph{Astronomy \& Astrophysics}, 577:\penalty0 A83, 2015.
\newblock \doi{10.1051/0004-6361/201424915}.

\bibitem[Dziewonski and Anderson(1981)]{Dziewonski1981}
A.~M. Dziewonski and D.~L. Anderson.
\newblock Preliminary reference {Earth} model.
\newblock \emph{Physics of the Earth and Planetary Interiors}, 25\penalty0 (4):\penalty0 297--356, 1981.
\newblock \doi{10.1016/0031-9201(81)90046-7}.

\bibitem[Escudero and Amils(2023)]{Escudero2023}
C.~Escudero and R.~Amils.
\newblock Hard rock dark biosphere and habitability.
\newblock \emph{Frontiers in Astronomy and Space Sciences}, Volume 10 - 2023, 2023.
\newblock ISSN 2296-987X.
\newblock \doi{10.3389/fspas.2023.1203845}.
\newblock URL \url{https://www.frontiersin.org/journals/astronomy-and-space-sciences/articles/10.3389/fspas.2023.1203845}.

\bibitem[Foley and Smye(2018)]{Foley2018}
B.~J. Foley and A.~J. Smye.
\newblock Carbon cycling and habitability of earth-sized stagnant lid planets.
\newblock \emph{Astrobiology}, 18\penalty0 (7):\penalty0 873--896, 2018.
\newblock \doi{10.1089/ast.2017.1695}.

\bibitem[Goldblatt and Watson(2012)]{Goldblatt2012}
C.~Goldblatt and A.~J. Watson.
\newblock The runaway greenhouse: implications for future climate change, geoengineering and planetary atmospheres.
\newblock \emph{Philosophical Transactions of the Royal Society A: Mathematical, Physical and Engineering Sciences}, 370\penalty0 (1974):\penalty0 4197--4216, 2012.

\bibitem[{Grimm, Simon L.} et~al.(2018){Grimm, Simon L.}, {Demory, Brice-Olivier}, {Gillon, Michaël}, {Dorn, Caroline}, {Agol, Eric}, {Burdanov, Artem}, {Delrez, Laetitia}, {Sestovic, Marko}, {Triaud, Amaury H. M. J.}, {Turbet, Martin}, {Bolmont, Émeline}, {Caldas, Anthony}, {de Wit, Julien}, {Jehin, Emmanuël}, {Leconte, Jérémy}, {Raymond, Sean N.}, {Van Grootel, Valérie}, {Burgasser, Adam J.}, {Carey, Sean}, {Fabrycky, Daniel}, {Heng, Kevin}, {Hernandez, David M.}, {Ingalls, James G.}, {Lederer, Susan}, {Selsis, Franck}, and {Queloz, Didier}]{Grimm2018}
{Grimm, Simon L.}, {Demory, Brice-Olivier}, {Gillon, Michaël}, {Dorn, Caroline}, {Agol, Eric}, {Burdanov, Artem}, {Delrez, Laetitia}, {Sestovic, Marko}, {Triaud, Amaury H. M. J.}, {Turbet, Martin}, {Bolmont, Émeline}, {Caldas, Anthony}, {de Wit, Julien}, {Jehin, Emmanuël}, {Leconte, Jérémy}, {Raymond, Sean N.}, {Van Grootel, Valérie}, {Burgasser, Adam J.}, {Carey, Sean}, {Fabrycky, Daniel}, {Heng, Kevin}, {Hernandez, David M.}, {Ingalls, James G.}, {Lederer, Susan}, {Selsis, Franck}, and {Queloz, Didier}.
\newblock The nature of the trappist-1 exoplanets.
\newblock \emph{A\&A}, 613:\penalty0 A68, 2018.
\newblock \doi{10.1051/0004-6361/201732233}.
\newblock URL \url{https://doi.org/10.1051/0004-6361/201732233}.

\bibitem[Han et~al.(2014)Han, Schmerr, Neumann, and Holmes]{Han2014}
S.-C. Han, N.~Schmerr, G.~Neumann, and S.~Holmes.
\newblock Global characteristics of porosity and density stratification within the lunar crust from grail gravity and lunar orbiter laser altimeter topography data.
\newblock \emph{Geophysical Research Letters}, 41\penalty0 (6):\penalty0 1882--1889, 2014.

\bibitem[Hand et~al.(2007)Hand, Carlson, and Chyba]{Hand2009}
K.~P. Hand, R.~W. Carlson, and C.~F. Chyba.
\newblock Energy, chemical disequilibrium, and geological constraints on europa.
\newblock \emph{Astrobiology}, 7\penalty0 (6):\penalty0 1006--1022, 2007.
\newblock \doi{10.1089/ast.2007.0156}.
\newblock URL \url{https://doi.org/10.1089/ast.2007.0156}.
\newblock PMID: 18163875.

\bibitem[{Harris} et~al.(2020){Harris}, {Millman}, {van der Walt}, et~al.]{Numpy2020}
C.~R. {Harris}, K.~J. {Millman}, S.~J. {van der Walt}, et~al.
\newblock {Array programming with NumPy}.
\newblock \emph{Nature}, 585\penalty0 (7825):\penalty0 357--362, Sept. 2020.
\newblock \doi{10.1038/s41586-020-2649-2}.

\bibitem[Hasterok and Chapman(2011)]{Hasterok2011}
D.~Hasterok and D.~Chapman.
\newblock Heat production and geotherms for the continental lithosphere.
\newblock \emph{Earth and Planetary Science Letters}, 307\penalty0 (1):\penalty0 59--70, 2011.
\newblock ISSN 0012-821X.
\newblock \doi{https://doi.org/10.1016/j.epsl.2011.04.034}.
\newblock URL \url{https://www.sciencedirect.com/science/article/pii/S0012821X11002500}.

\bibitem[Hasterok et~al.(2022)Hasterok, Halpin, Collins, Hand, Kreemer, Gard, and Glorie]{Hasterok2022}
D.~Hasterok, J.~A. Halpin, A.~S. Collins, M.~Hand, C.~Kreemer, M.~G. Gard, and S.~Glorie.
\newblock New maps of global geological provinces and tectonic plates.
\newblock \emph{Earth-Science Reviews}, 231:\penalty0 104069, 2022.

\bibitem[Hinkel et~al.(2014)Hinkel, Timmes, Young, Pagano, and Turnbull]{Hinkel2014}
N.~R. Hinkel, F.~Timmes, P.~A. Young, M.~D. Pagano, and M.~C. Turnbull.
\newblock Stellar abundances in the solar neighborhood: the hypatia catalog.
\newblock \emph{The Astronomical Journal}, 148\penalty0 (3):\penalty0 54, 2014.

\bibitem[Hofmeister(2005)]{Hofmeister2005}
A.~Hofmeister.
\newblock Dependence of diffusive radiative transfer on grain-size, temperature, and fe-content: Implications for mantle processes.
\newblock \emph{Journal of Geodynamics}, 40\penalty0 (1):\penalty0 51--72, 2005.
\newblock ISSN 0264-3707.
\newblock \doi{https://doi.org/10.1016/j.jog.2005.06.001}.
\newblock URL \url{https://www.sciencedirect.com/science/article/pii/S0264370705000384}.

\bibitem[Hofmeister(1999)]{Hofmeister1999}
A.~M. Hofmeister.
\newblock Mantle values of thermal conductivity and the geotherm from phonon lifetimes.
\newblock \emph{Science}, 283\penalty0 (5408):\penalty0 1699--1706, 1999.
\newblock \doi{10.1126/science.283.5408.1699}.
\newblock URL \url{https://www.science.org/doi/abs/10.1126/science.283.5408.1699}.

\bibitem[Hu et~al.(2021)Hu, Bello-Arufe, Damiano, Roy, Ito, Scheucher, and Wunderlich]{Hu2022}
R.~Hu, A.~Bello-Arufe, M.~Damiano, P.-A. Roy, Y.~Ito, M.~Scheucher, and F.~Wunderlich.
\newblock Unveiling shrouded oceans on temperate exoplanets.
\newblock \emph{The Astrophysical Journal Letters}, 921\penalty0 (1):\penalty0 L8, 2021.
\newblock \doi{10.3847/2041-8213/ac1f92}.

\bibitem[{Hunter}(2007)]{Matplotlib2007}
J.~D. {Hunter}.
\newblock {Matplotlib: A 2D Graphics Environment}.
\newblock \emph{Computing in Science and Engineering}, 9\penalty0 (3):\penalty0 90--95, Jan. 2007.
\newblock \doi{10.1109/MCSE.2007.55}.

\bibitem[{IAPWS}(1995)]{IAPWS1995}
{IAPWS}.
\newblock Release on the iapws formulation 1995 for the thermodynamic properties of ordinary water substance for general and scientific use.
\newblock Technical report, International Association for the Properties of Water and Steam, 1995.

\bibitem[{IAPWS}(2009)]{IAPWSRelease2009}
{IAPWS}.
\newblock Revised release on the iapws industrial formulation 1997 for the thermodynamic properties of water and steam.
\newblock Technical report, International Association for the Properties of Water and Steam, 2009.

\bibitem[Ingersoll(1969)]{Ingersoll1969}
A.~P. Ingersoll.
\newblock The runaway greenhouse: A history of water on venus.
\newblock \emph{Journal of Atmospheric Sciences}, 26\penalty0 (6):\penalty0 1191--1198, 1969.

\bibitem[Jaupart et~al.(2007)Jaupart, Labrosse, Lucazeau, and Mareschal]{Jaupart2007}
C.~Jaupart, S.~Labrosse, F.~Lucazeau, and J.~Mareschal.
\newblock 7.06-temperatures, heat and energy in the mantle of the earth.
\newblock \emph{Treatise on geophysics}, 7:\penalty0 223--270, 2007.

\bibitem[Kashefi and Lovley(2003)]{Kashefi2003}
K.~Kashefi and D.~R. Lovley.
\newblock Extending the upper temperature limit for life.
\newblock \emph{Science}, 301\penalty0 (5635):\penalty0 934--934, 2003.
\newblock \doi{10.1126/science.1086823}.
\newblock URL \url{https://www.science.org/doi/abs/10.1126/science.1086823}.

\bibitem[Kasting et~al.(1993)Kasting, Whitmire, and Reynolds]{Kasting1993}
J.~F. Kasting, D.~P. Whitmire, and R.~T. Reynolds.
\newblock Habitable zones around main sequence stars.
\newblock \emph{Icarus}, 101\penalty0 (1):\penalty0 108--128, 1993.
\newblock \doi{https://doi.org/10.1006/icar.1993.1010}.

\bibitem[Kivelson et~al.(2002)Kivelson, Khurana, and Volwerk]{Kivelson2000}
M.~Kivelson, K.~Khurana, and M.~Volwerk.
\newblock The permanent and inductive magnetic moments of ganymede.
\newblock \emph{Icarus}, 157\penalty0 (2):\penalty0 507--522, 2002.
\newblock ISSN 0019-1035.
\newblock \doi{https://doi.org/10.1006/icar.2002.6834}.
\newblock URL \url{https://www.sciencedirect.com/science/article/pii/S001910350296834X}.

\bibitem[Kopparapu et~al.(2013)Kopparapu, Ramirez, Kasting, Eymet, Robinson, Mahadevan, Terrien, Domagal-Goldman, Meadows, and Deshpande]{Kopparapu2013}
R.~K. Kopparapu, R.~Ramirez, J.~F. Kasting, V.~Eymet, T.~D. Robinson, S.~Mahadevan, R.~C. Terrien, S.~Domagal-Goldman, V.~Meadows, and R.~Deshpande.
\newblock Erratum: “habitable zones around main-sequence stars: New estimates” (2013, apj, 765, 131).
\newblock \emph{The Astrophysical Journal}, 770\penalty0 (1):\penalty0 82, may 2013.
\newblock \doi{10.1088/0004-637X/770/1/82}.
\newblock URL \url{https://doi.org/10.1088/0004-637X/770/1/82}.

\bibitem[Kopparapu et~al.(2014)Kopparapu, Ramirez, SchottelKotte, Kasting, Domagal-Goldman, and Eymet]{Kopparapu2014}
R.~K. Kopparapu, R.~M. Ramirez, J.~SchottelKotte, J.~F. Kasting, S.~Domagal-Goldman, and V.~Eymet.
\newblock Habitable zones around main-sequence stars: Dependence on planetary mass.
\newblock \emph{The Astrophysical Journal Letters}, 787\penalty0 (2):\penalty0 L29, may 2014.
\newblock \doi{10.1088/2041-8205/787/2/L29}.
\newblock URL \url{https://doi.org/10.1088/2041-8205/787/2/L29}.

\bibitem[Krissansen-Totton et~al.(2018)Krissansen-Totton, Olson, and Catling]{Krissansen-Totton2018}
J.~Krissansen-Totton, S.~Olson, and D.~C. Catling.
\newblock Disequilibrium biosignatures over {Earth} history and implications for detecting exoplanet life.
\newblock \emph{Science Advances}, 4\penalty0 (1):\penalty0 eaao5747, 2018.
\newblock \doi{10.1126/sciadv.aao5747}.

\bibitem[Kusube et~al.(2017)Kusube, Kyaw, Tanikawa, Chastain, Hardy, Cameron, and Bartlett]{Kusube2017}
M.~Kusube, T.~S. Kyaw, K.~Tanikawa, R.~A. Chastain, K.~M. Hardy, J.~Cameron, and D.~H. Bartlett.
\newblock \textit{Colwellia marinimaniae} sp. nov., a hyperpiezophilic species isolated from an amphipod within the {Challenger Deep}, {Mariana Trench}.
\newblock \emph{International Journal of Systematic and Evolutionary Microbiology}, 67\penalty0 (4):\penalty0 824--831, 2017.
\newblock \doi{10.1099/ijsem.0.001671}.

\bibitem[Lingam and Loeb(2018)]{Lingam2018}
M.~Lingam and A.~Loeb.
\newblock Subsurface exolife.
\newblock \emph{International Journal of Astrobiology}, 18\penalty0 (2):\penalty0 112–141, Apr. 2018.
\newblock ISSN 1475-3006.
\newblock \doi{10.1017/s1473550418000083}.
\newblock URL \url{http://dx.doi.org/10.1017/S1473550418000083}.

\bibitem[Lloyd(2025)]{lloyd2025intraterrestrials}
K.~Lloyd.
\newblock \emph{Intraterrestrials: Discovering the Strangest Life on Earth}.
\newblock Princeton University Press, 2025.
\newblock ISBN 9780691236124.
\newblock URL \url{https://books.google.com.co/books?id=okExEQAAQBAJ}.

\bibitem[Magnabosco et~al.(2018)Magnabosco, Lin, Dong, Bomberg, Ghiorse, Stan-Lotter, Pedersen, Kieft, van Heerden, and Onstott]{Magnabosco2018}
C.~Magnabosco, L.-H. Lin, H.~Dong, M.~Bomberg, W.~Ghiorse, H.~Stan-Lotter, K.~Pedersen, T.~L. Kieft, E.~van Heerden, and T.~C. Onstott.
\newblock The biomass and biodiversity of the continental subsurface.
\newblock \emph{Nature Geoscience}, 11\penalty0 (10):\penalty0 707--717, 2018.

\bibitem[McDonough and Sun(1995)]{McDonough1995}
W.~F. McDonough and S.-S. Sun.
\newblock The composition of the {Earth}.
\newblock \emph{Chemical Geology}, 120\penalty0 (3--4):\penalty0 223--253, 1995.
\newblock \doi{10.1016/0009-2541(94)00140-4}.

\bibitem[McDonough et~al.(2020)McDonough, {\v{S}}r{\'a}mek, and Wipperfurth]{McDonough2020}
W.~F. McDonough, O.~{\v{S}}r{\'a}mek, and S.~A. Wipperfurth.
\newblock Radiogenic power and geoneutrino luminosity of the earth and other terrestrial bodies through time.
\newblock \emph{Geochemistry, Geophysics, Geosystems}, 21\penalty0 (7):\penalty0 e2019GC008865, 2020.

\bibitem[{McKinney}(2010)]{mckinney_proc_scipy_2010}
W.~{McKinney}.
\newblock Data structures for statistical computing in python.
\newblock In \emph{Proceedings of the 9th Python in Science Conference}, pages 56--61, 2010.
\newblock \doi{10.25080/Majora-92bf1922-00a}.

\bibitem[McMahon et~al.(2013)McMahon, O’Malley-James, and Parnell]{McMahon2013}
S.~McMahon, J.~O’Malley-James, and J.~Parnell.
\newblock Circumstellar habitable zones for deep terrestrial biospheres.
\newblock \emph{Planetary and Space Science}, 85:\penalty0 312--318, 2013.
\newblock ISSN 0032-0633.
\newblock \doi{https://doi.org/10.1016/j.pss.2013.07.002}.
\newblock URL \url{https://www.sciencedirect.com/science/article/pii/S0032063313001815}.

\bibitem[Meadows et~al.(2018)Meadows, Reinhard, Arney, Parenteau, Schwieterman, Domagal-Goldman, Lincowski, Stapelfeldt, Rauer, DasSarma, et~al.]{Meadows2018}
V.~S. Meadows, C.~T. Reinhard, G.~N. Arney, M.~N. Parenteau, E.~W. Schwieterman, S.~D. Domagal-Goldman, A.~P. Lincowski, K.~R. Stapelfeldt, H.~Rauer, S.~DasSarma, et~al.
\newblock Exoplanet biosignatures: understanding oxygen as a biosignature in the context of its environment.
\newblock \emph{Astrobiology}, 18\penalty0 (6):\penalty0 630--662, 2018.

\bibitem[Meersman et~al.(2006)Meersman, Smeller, and Heremans]{Meersman2006}
F.~Meersman, L.~Smeller, and K.~Heremans.
\newblock Protein stability and dynamics in the pressure–temperature plane.
\newblock \emph{Biochimica et Biophysica Acta (BBA) - Proteins and Proteomics}, 1764\penalty0 (3):\penalty0 346--354, 2006.
\newblock ISSN 1570-9639.
\newblock \doi{https://doi.org/10.1016/j.bbapap.2005.11.019}.
\newblock URL \url{https://www.sciencedirect.com/science/article/pii/S1570963905004322}.
\newblock Proteins Under High Pressure.

\bibitem[M{\'e}ndez(2001)]{mendez2001planetary}
A.~M{\'e}ndez.
\newblock Planetary habitable zones: The spatial distribution of life on planetary bodies.
\newblock In J.~Chela-Flores, T.~Owen, and F.~Raulin, editors, \emph{First Steps in the Origin of Life in the Universe}, pages 211--214. Kluwer Academic Publishers, 2001.

\bibitem[Merino et~al.(2019)Merino, Aronson, Bojanova, Feyhl-Buska, Wong, Zhang, and Giovannelli]{Merino2019}
N.~Merino, H.~S. Aronson, D.~P. Bojanova, J.~Feyhl-Buska, M.~L. Wong, S.~Zhang, and D.~Giovannelli.
\newblock Living at the extremes: Extremophiles and the limits of life in a planetary context.
\newblock \emph{Frontiers in Microbiology}, 10:\penalty0 780, 2019.
\newblock \doi{10.3389/fmicb.2019.00780}.

\bibitem[Metzger(2022)]{Metzger2022}
P.~T. Metzger.
\newblock The reclassification of asteroids from planets to non-planets.
\newblock \emph{Icarus}, 374:\penalty0 114768, 2022.
\newblock \doi{10.1016/j.icarus.2021.114768}.

\bibitem[Myhill et~al.(2023)Myhill, Cottaar, Heister, Rose, Unterborn, Dannberg, and Gassmoeller]{Myhill2023}
R.~Myhill, S.~Cottaar, T.~Heister, I.~Rose, C.~Unterborn, J.~Dannberg, and R.~Gassmoeller.
\newblock Burnman -- a python toolkit for planetary geophysics, geochemistry and thermodynamics.
\newblock \emph{Journal of Open Source Software}, 8\penalty0 (87):\penalty0 5389, 2023.
\newblock \doi{10.21105/joss.05389}.
\newblock URL \url{https://doi.org/10.21105/joss.05389}.

\bibitem[Myhill et~al.(2024)Myhill, Cottaar, Heister, Rose, Unterborn, Dannberg, Gassmoeller, and Farla]{myhill_2024_14238360}
R.~Myhill, S.~Cottaar, T.~Heister, I.~Rose, C.~Unterborn, J.~Dannberg, R.~Gassmoeller, and R.~Farla.
\newblock Burnman – a python toolkit for planetary geophysics, geochemistry and thermodynamics, Nov. 2024.
\newblock URL \url{https://doi.org/10.5281/zenodo.14238360}.

\bibitem[Nimmo and Manga(2009)]{Nimmo2009}
F.~Nimmo and M.~Manga.
\newblock Geodynamics of europa’s icy shell.
\newblock \emph{Europa}, pages 381--404, 2009.

\bibitem[Nimmo and Pappalardo(2016)]{Nimmo2016}
F.~Nimmo and R.~T. Pappalardo.
\newblock Ocean worlds in the outer solar system.
\newblock \emph{Journal of Geophysical Research: Planets}, 121\penalty0 (8):\penalty0 1378--1399, 2016.

\bibitem[Noack et~al.(2017)Noack, Rivoldini, and {Van Hoolst}]{Noack2017}
L.~Noack, A.~Rivoldini, and T.~{Van Hoolst}.
\newblock Volcanism and outgassing of stagnant-lid planets: Implications for the habitable zone.
\newblock \emph{Physics of the Earth and Planetary Interiors}, 269:\penalty0 40--57, 2017.
\newblock ISSN 0031-9201.
\newblock \doi{https://doi.org/10.1016/j.pepi.2017.05.010}.
\newblock URL \url{https://www.sciencedirect.com/science/article/pii/S0031920116301509}.

\bibitem[Orcutt et~al.(2013)Orcutt, LaRowe, Biddle, Colwell, Glazer, Reese, Kirkpatrick, Lapham, Mills, Sylvan, Wankel, and Wheat]{Orcutt2013}
B.~N. Orcutt, D.~E. LaRowe, J.~F. Biddle, F.~S. Colwell, B.~T. Glazer, B.~K. Reese, J.~B. Kirkpatrick, L.~L. Lapham, H.~J. Mills, J.~B. Sylvan, S.~D. Wankel, and C.~G. Wheat.
\newblock Microbial activity in the marine deep biosphere: progress and prospects.
\newblock \emph{Frontiers in Microbiology}, Volume 4 - 2013, 2013.
\newblock ISSN 1664-302X.
\newblock \doi{10.3389/fmicb.2013.00189}.
\newblock URL \url{https://www.frontiersin.org/journals/microbiology/articles/10.3389/fmicb.2013.00189}.

\bibitem[Parnell and McMahon(2016)]{Parnell2016}
J.~Parnell and S.~McMahon.
\newblock Physical and chemical controls on habitats for life in the deep subsurface beneath continents and ice.
\newblock \emph{Philosophical Transactions of the Royal Society A: Mathematical, Physical and Engineering Sciences}, 374\penalty0 (2059):\penalty0 20140293, 01 2016.
\newblock ISSN 1364-503X.
\newblock \doi{10.1098/rsta.2014.0293}.
\newblock URL \url{https://doi.org/10.1098/rsta.2014.0293}.

\bibitem[Pertermann and Hofmeister(2006)]{Pertermann2006}
M.~Pertermann and A.~M. Hofmeister.
\newblock Thermal diffusivity of olivine-group minerals at high temperature.
\newblock \emph{American Mineralogist}, 91\penalty0 (11--12):\penalty0 1747--1760, 2006.
\newblock \doi{10.2138/am.2006.2105}.

\bibitem[Pierrehumbert(2004)]{Pierrehumbert2004}
R.~T. Pierrehumbert.
\newblock High levels of atmospheric carbon dioxide necessary for the termination of global glaciation.
\newblock \emph{Nature}, 429\penalty0 (6992):\penalty0 646--649, 2004.

\bibitem[Plesa et~al.(2018)Plesa, Breuer, and Stamenkovic]{Plesa2018}
A.-C. Plesa, D.~Breuer, and V.~Stamenkovic.
\newblock The depth of subsurface liquid water on mars as predicted from 3d thermal evolution models.
\newblock In \emph{2nd GeoPlaNet Thematic School - Fluid-Rock Interactions in the Solar System}, 2018.
\newblock URL \url{https://elib.dlr.de/125273/}.

\bibitem[Porco et~al.(2006)Porco, Helfenstein, Thomas, Ingersoll, Wisdom, West, Neukum, Denk, Wagner, Roatsch, Kieffer, Turtle, McEwen, Johnson, Rathbun, Veverka, Wilson, Perry, Spitale, Brahic, Burns, DelGenio, Dones, Murray, and Squyres]{Porco2006}
C.~C. Porco, P.~Helfenstein, P.~C. Thomas, A.~P. Ingersoll, J.~Wisdom, R.~West, G.~Neukum, T.~Denk, R.~Wagner, T.~Roatsch, S.~Kieffer, E.~Turtle, A.~McEwen, T.~V. Johnson, J.~Rathbun, J.~Veverka, D.~Wilson, J.~Perry, J.~Spitale, A.~Brahic, J.~A. Burns, A.~D. DelGenio, L.~Dones, C.~D. Murray, and S.~Squyres.
\newblock Cassini observes the active south pole of enceladus.
\newblock \emph{Science}, 311\penalty0 (5766):\penalty0 1393--1401, 2006.
\newblock \doi{10.1126/science.1123013}.
\newblock URL \url{https://www.science.org/doi/abs/10.1126/science.1123013}.

\bibitem[Putirka(2024)]{Putirka2024}
K.~D. Putirka.
\newblock Exoplanet mineralogy, 2024.
\newblock URL \url{https://arxiv.org/abs/2404.15426}.

\bibitem[Putirka and Rarick(2019)]{Putirka2019}
K.~D. Putirka and J.~C. Rarick.
\newblock The composition and mineralogy of rocky exoplanets: A survey of >4000 stars from the hypatia catalog.
\newblock \emph{American Mineralogist}, 104\penalty0 (6):\penalty0 817--829, 2019.
\newblock \doi{10.2138/am-2019-6740}.

\bibitem[Quanz et~al.(2022)Quanz, Ottiger, Fontanet, Kammerer, Cantalloube, Kreidberg, Mollière, Nasedkin, Shulyak, and Wolf]{Quanz2022}
S.~P. Quanz, M.~Ottiger, E.~Fontanet, J.~Kammerer, F.~Cantalloube, L.~Kreidberg, P.~Mollière, E.~Nasedkin, D.~Shulyak, and S.~Wolf.
\newblock Large interferometer for exoplanets ({LIFE}): {I}. detecting rocky exoplanets in the habitable zones of sun-like stars.
\newblock \emph{Astronomy \& Astrophysics}, 664:\penalty0 A21, 2022.
\newblock \doi{10.1051/0004-6361/202140366}.

\bibitem[{Riu} et~al.(2022){Riu}, {Carter}, and {Poulet}]{Riu2022}
L.~{Riu}, J.~{Carter}, and F.~{Poulet}.
\newblock {Estimation of H2O content (in wt\%) stored in hydrated silicates at Mars}.
\newblock In \emph{EGU General Assembly Conference Abstracts}, EGU General Assembly Conference Abstracts, pages EGU22--9874, May 2022.
\newblock \doi{10.5194/egusphere-egu22-9874}.

\bibitem[Rivkina et~al.(2000)Rivkina, Friedmann, McKay, and Gilichinsky]{Rivkina2000}
E.~M. Rivkina, E.~I. Friedmann, C.~P. McKay, and D.~A. Gilichinsky.
\newblock Metabolic activity of permafrost bacteria below the freezing point.
\newblock \emph{Applied and Environmental Microbiology}, 66\penalty0 (8):\penalty0 3230--3233, 2000.
\newblock \doi{10.1128/AEM.66.8.3230-3233.2000}.

\bibitem[Rothschild and Mancinelli(2001)]{Rothschild2001}
L.~J. Rothschild and R.~L. Mancinelli.
\newblock Life in extreme environments.
\newblock \emph{Nature}, 409\penalty0 (6823):\penalty0 1092--1101, 2001.
\newblock \doi{10.1038/35059215}.

\bibitem[Rudnick and Gao(2014)]{Rudnick2014}
R.~L. Rudnick and S.~Gao.
\newblock Composition of the continental crust.
\newblock In H.~D. Holland and K.~K. Turekian, editors, \emph{Treatise on Geochemistry (Second Edition)}, pages 1--51. Elsevier, 2014.
\newblock \doi{10.1016/B978-0-08-095975-7.00301-6}.

\bibitem[Ruedas(2017)]{Ruedas2017}
T.~Ruedas.
\newblock Radioactive heat production of six geologically important nuclides.
\newblock \emph{Geochemistry, Geophysics, Geosystems}, 18\penalty0 (9):\penalty0 3530--3541, 2017.

\bibitem[Schulze-Makuch and Irwin(2008)]{SchulzeMakuch2008}
D.~Schulze-Makuch and L.~Irwin.
\newblock \emph{Life in the Universe: Expectations and Constraints}.
\newblock Advances in Astrobiology and Biogeophysics. Springer Berlin Heidelberg, 2008.
\newblock ISBN 9783540768166.
\newblock URL \url{https://books.google.com.co/books?id=qdjoTGAvrkkC}.

\bibitem[Schwieterman et~al.(2018)Schwieterman, Kiang, Parenteau, Harman, DasSarma, Fisher, Arney, Hartnett, Reinhard, Olson, et~al.]{Schwieterman2018}
E.~W. Schwieterman, N.~Y. Kiang, M.~N. Parenteau, C.~E. Harman, S.~DasSarma, T.~M. Fisher, G.~N. Arney, H.~E. Hartnett, C.~T. Reinhard, S.~L. Olson, et~al.
\newblock Exoplanet biosignatures: a review of remotely detectable signs of life.
\newblock \emph{Astrobiology}, 18\penalty0 (6):\penalty0 663--708, 2018.

\bibitem[Seager et~al.(2007)Seager, Kuchner, Hier-Majumder, and Militzer]{Seager2007}
S.~Seager, M.~Kuchner, C.~A. Hier-Majumder, and B.~Militzer.
\newblock Mass-radius relationships for solid exoplanets.
\newblock \emph{The Astrophysical Journal}, 669\penalty0 (2):\penalty0 1279--1297, 2007.
\newblock \doi{10.1086/521346}.

\bibitem[Seager et~al.(2013)Seager, Bains, and Hu]{Seager2013}
S.~Seager, W.~Bains, and R.~Hu.
\newblock Biosignature gases in h2-dominated atmospheres on rocky exoplanets.
\newblock \emph{The Astrophysical Journal}, 777\penalty0 (2):\penalty0 95, oct 2013.
\newblock \doi{10.1088/0004-637X/777/2/95}.
\newblock URL \url{https://doi.org/10.1088/0004-637X/777/2/95}.

\bibitem[Sharma et~al.(2002)Sharma, Scott, Cody, Fogel, Hazen, Hemley, and Huntress]{Sharma2002}
A.~Sharma, J.~H. Scott, G.~D. Cody, M.~L. Fogel, R.~M. Hazen, R.~J. Hemley, and W.~T. Huntress.
\newblock Microbial activity at gigapascal pressures.
\newblock \emph{Science}, 295\penalty0 (5559):\penalty0 1514--1516, 2002.

\bibitem[Sherwood~Lollar et~al.(2002)Sherwood~Lollar, Westgate, Ward, Slater, and Lacrampe-Couloume]{Sherwood2013}
B.~Sherwood~Lollar, T.~D. Westgate, J.~A. Ward, G.~F. Slater, and G.~Lacrampe-Couloume.
\newblock Abiogenic formation of alkanes in the {Earth}'s crust as a minor source for global hydrocarbon reservoirs.
\newblock \emph{Nature}, 416\penalty0 (6880):\penalty0 522--524, 2002.
\newblock \doi{10.1038/416522a}.

\bibitem[Shields et~al.(2016)Shields, Ballard, and Johnson]{Shields_2016}
A.~L. Shields, S.~Ballard, and J.~A. Johnson.
\newblock The habitability of planets orbiting m-dwarf stars.
\newblock \emph{Physics Reports}, 663:\penalty0 1--38, 2016.
\newblock \doi{10.1016/j.physrep.2016.10.003}.

\bibitem[Sotin and Tobie(2004)]{Sotin2004}
C.~Sotin and G.~Tobie.
\newblock Internal structure and dynamics of the large icy satellites.
\newblock \emph{Comptes Rendus Physique}, 5\penalty0 (7):\penalty0 769--780, 2004.
\newblock ISSN 1631-0705.
\newblock \doi{https://doi.org/10.1016/j.crhy.2004.08.001}.
\newblock URL \url{https://www.sciencedirect.com/science/article/pii/S163107050400146X}.
\newblock Ice: from dislocations to icy satellites.

\bibitem[Sotin et~al.(2007)Sotin, Grasset, and Mocquet]{Sotin2007}
C.~Sotin, O.~Grasset, and A.~Mocquet.
\newblock Mass-radius curve for extrasolar {Earth}-like planets and ocean planets.
\newblock \emph{Icarus}, 191\penalty0 (1):\penalty0 337--351, 2007.
\newblock \doi{10.1016/j.icarus.2007.04.006}.

\bibitem[Spohn et~al.(2026)Spohn, Roberge, Way, Duarte, Miozzi, Baumeister, Byrne, and Lineweaver]{Spohn2026}
T.~Spohn, A.~Roberge, M.~Way, J.~C. Duarte, F.~Miozzi, P.~Baumeister, P.~Byrne, and C.~H. Lineweaver.
\newblock Exo-geoscience perspectives beyond habitability.
\newblock \emph{Space Science Reviews}, 222\penalty0 (1):\penalty0 9, 2026.

\bibitem[Stacey and Davis(2008)]{Stacey2008}
F.~D. Stacey and P.~M. Davis.
\newblock \emph{Physics of the Earth}.
\newblock Cambridge University Press, 2008.

\bibitem[Stevenson(2015)]{Stevenson2015}
D.~S. Stevenson.
\newblock Life and its detection signatures in non-aqueous solvents.
\newblock \emph{International Journal of Astrobiology}, 14\penalty0 (4):\penalty0 545--557, 2015.
\newblock \doi{10.1017/S1473550414000561}.

\bibitem[Takai et~al.(2008)Takai, Nakamura, Toki, Tsunogai, Miyazaki, Miyazaki, Hirayama, Nakagawa, Nunoura, and Horikoshi]{Takai2008}
K.~Takai, K.~Nakamura, T.~Toki, U.~Tsunogai, M.~Miyazaki, J.~Miyazaki, H.~Hirayama, S.~Nakagawa, T.~Nunoura, and K.~Horikoshi.
\newblock Cell proliferation at 122 c and isotopically heavy ch4 production by a hyperthermophilic methanogen under high-pressure cultivation.
\newblock \emph{Proceedings of the National Academy of Sciences}, 105\penalty0 (31):\penalty0 10949--10954, 2008.

\bibitem[Tarnas et~al.(2018)Tarnas, Mustard, Sherwood~Lollar, Bramble, Cannon, Palumbo, and Plesa]{Tarnas2018}
J.~D. Tarnas, J.~F. Mustard, B.~Sherwood~Lollar, M.~S. Bramble, K.~M. Cannon, A.~M. Palumbo, and A.-C. Plesa.
\newblock Radiolytic {H}$_2$ production on {Noachian Mars}: Implications for habitability and atmospheric warming.
\newblock \emph{Earth and Planetary Science Letters}, 502:\penalty0 133--145, 2018.
\newblock \doi{10.1016/j.epsl.2018.09.001}.

\bibitem[Tenelanda-Osorio et~al.(2021)Tenelanda-Osorio, Parra, Cuartas-Restrepo, and Zuluaga]{tenelanda2021enceladus}
L.~I. Tenelanda-Osorio, J.~L. Parra, P.~Cuartas-Restrepo, and J.~I. Zuluaga.
\newblock Enceladus as a potential niche for methanogens and estimation of its biomass.
\newblock \emph{Life}, 11\penalty0 (11):\penalty0 1182, 2021.

\bibitem[Turcotte and Schubert(2014)]{Turcotte2014}
D.~L. Turcotte and G.~Schubert.
\newblock \emph{Geodynamics}.
\newblock Cambridge University Press, 3 edition, 2014.
\newblock ISBN 9781107069983.

\bibitem[Unterborn et~al.(2016)Unterborn, Dismukes, and Panero]{Unterborn2016}
C.~T. Unterborn, E.~E. Dismukes, and W.~R. Panero.
\newblock Scaling the earth: A sensitivity analysis of terrestrial exoplanetary interior models.
\newblock \emph{The Astrophysical Journal}, 819\penalty0 (1):\penalty0 32, feb 2016.
\newblock \doi{10.3847/0004-637X/819/1/32}.
\newblock URL \url{https://doi.org/10.3847/0004-637X/819/1/32}.

\bibitem[Unterborn et~al.(2023)Unterborn, Desch, Haldemann, Lorenzo, Schulze, Hinkel, and Panero]{Unterborn2023}
C.~T. Unterborn, S.~J. Desch, J.~Haldemann, A.~Lorenzo, J.~G. Schulze, N.~R. Hinkel, and W.~R. Panero.
\newblock The nominal ranges of rocky planet masses, radii, surface gravities, and bulk densities.
\newblock \emph{The Astrophysical Journal}, 944\penalty0 (1):\penalty0 42, 2023.
\newblock \doi{10.3847/1538-4357/acaa3b}.

\bibitem[Valencia et~al.(2006)Valencia, O'Connell, and Sasselov]{Valencia2006}
D.~Valencia, R.~J. O'Connell, and D.~Sasselov.
\newblock Internal structure of massive terrestrial planets.
\newblock \emph{Icarus}, 181\penalty0 (2):\penalty0 545--554, 2006.
\newblock ISSN 0019-1035.
\newblock \doi{https://doi.org/10.1016/j.icarus.2005.11.021}.
\newblock URL \url{https://www.sciencedirect.com/science/article/pii/S0019103505004574}.

\bibitem[Vance et~al.(2018)Vance, Panning, St{\"a}hler, Cammarano, Bills, Tobie, Kamata, Kedar, Sotin, Pike, et~al.]{Vance2018}
S.~D. Vance, M.~P. Panning, S.~St{\"a}hler, F.~Cammarano, B.~G. Bills, G.~Tobie, S.~Kamata, S.~Kedar, C.~Sotin, W.~T. Pike, et~al.
\newblock Geophysical investigations of habitability in ice-covered ocean worlds.
\newblock \emph{Journal of Geophysical Research: Planets}, 123\penalty0 (1):\penalty0 180--205, 2018.

\bibitem[Virtanen et~al.(2020)Virtanen, Gommers, Oliphant, et~al.]{virtanenSciPy10Fundamental2020}
P.~Virtanen, R.~Gommers, T.~E. Oliphant, et~al.
\newblock {{SciPy}} 1.0: Fundamental algorithms for scientific computing in {{Python}}.
\newblock \emph{Nature Methods}, 17\penalty0 (3):\penalty0 261--272, Mar. 2020.
\newblock \doi{10.1038/s41592-019-0686-2}.

\bibitem[Wagner and Pru{\ss}(2002)]{Wagner2002}
W.~Wagner and A.~Pru{\ss}.
\newblock The iapws formulation 1995 for the thermodynamic properties of ordinary water substance for general and scientific use.
\newblock \emph{Journal of physical and chemical reference data}, 31\penalty0 (2):\penalty0 387--535, 2002.

\bibitem[White(1984)]{White1984}
R.~H. White.
\newblock Hydrolytic stability of biomolecules at high temperatures and its implication for life at 250 °c.
\newblock \emph{Nature}, 310\penalty0 (5976):\penalty0 430--432, 1984.
\newblock \doi{10.1038/310430a0}.

\bibitem[Winter(2014)]{Winter2014}
J.~D. Winter.
\newblock \emph{Principles of igneous and metamorphic petrology}, volume~2.
\newblock Pearson education Harlow, UK, 2014.

\bibitem[Winter and Jeworrek(2009)]{Winter2009}
R.~Winter and C.~Jeworrek.
\newblock Effect of pressure on membranes.
\newblock \emph{Soft Matter}, 5\penalty0 (17):\penalty0 3157--3173, 2009.

\bibitem[Yardley(2009)]{Yardley2009}
B.~W.~D. Yardley.
\newblock The role of water in the evolution of the continental crust.
\newblock \emph{Journal of the Geological Society}, 166\penalty0 (4):\penalty0 585--600, 2009.
\newblock \doi{10.1144/0016-76492008-101}.

\bibitem[Yayanos et~al.(1981)Yayanos, Dietz, and Boxtel]{Yayanos1995}
A.~A. Yayanos, A.~S. Dietz, and R.~V. Boxtel.
\newblock Obligately barophilic bacterium from the mariana trench.
\newblock \emph{Proceedings of the National Academy of Sciences}, 78\penalty0 (8):\penalty0 5212--5215, 1981.
\newblock \doi{10.1073/pnas.78.8.5212}.
\newblock URL \url{https://www.pnas.org/doi/abs/10.1073/pnas.78.8.5212}.

\bibitem[{Zeng} et~al.(2016){Zeng}, {Sasselov}, and {Jacobsen}]{Zeng2016}
L.~{Zeng}, D.~D. {Sasselov}, and S.~B. {Jacobsen}.
\newblock {Mass-Radius Relation for Rocky Planets Based on PREM}.
\newblock \emph{\apj}, 819\penalty0 (2):\penalty0 127, Mar. 2016.
\newblock \doi{10.3847/0004-637X/819/2/127}.

\bibitem[Zuluaga et~al.(2013)Zuluaga, Bustamante, Cuartas, and Hoyos]{Zuluaga2013}
J.~I. Zuluaga, S.~Bustamante, P.~A. Cuartas, and J.~H. Hoyos.
\newblock The influence of thermal evolution in the magnetic protection of terrestrial planets.
\newblock \emph{The Astrophysical Journal}, 770\penalty0 (1):\penalty0 23, 2013.

\end{thebibliography}
\end{document}